\UseRawInputEncoding
\documentclass[twocolumn,           
               showpacs,            
               nopreprintnumbers,     
               aps,                 
               prd,          	    
               letterpaper,             
              groupeaddress,      
               nofootinbib,         
               tightenlines,        
               floats,floatfix,      
               showkeys
               ]{revtex4-1}
               
\usepackage[toc,page]{appendix}
\usepackage{graphicx}
\usepackage{dcolumn}
\usepackage{bm}
\usepackage{amsmath}
\usepackage{amsfonts,amssymb}
\usepackage{soul}
\usepackage{color}
\definecolor{v}{rgb}{0.6, 0.2, 0.8} 
\definecolor{MAGA}{rgb}{0.1, 0.43, 0.75}
\definecolor{jm}{rgb}{0.13, 0.48, 0.64}

\usepackage{xcolor}
\usepackage{enumerate}
\usepackage{float}
\usepackage{subfigure}

\begin{document}

\title{An exponential equation of state of dark energy in the light of 2018 CMB Planck data}

\author{M\'onica N. Castillo-Santos$^1$}
\email{mncastillo99@gmail.com}
\author{A.  Hern\'andez-Almada$^1$}
\email{ahalmada@uaq.mx, corresponding author}

\author{Miguel A. Garc\'ia-Aspeitia$^2$}
\email{angel.garcia@ibero.mx}
\author{Juan Maga\~na$^{3}$}
\email{juan.magana@ucentral.cl}

\affiliation{$^1$ Facultad de Ingenier\'ia, Universidad Aut\'onoma de
Quer\'etaro, Centro Universitario Cerro de las Campanas, 76010, Santiago de 
Quer\'etaro, M\'exico}
\affiliation{$^2$ Depto. de Física y Matemáticas, Universidad Iberoamericana Ciudad de México, Prolongación Paseo \\ de la Reforma 880, México D. F. 01219, México}
\affiliation{$^3$ Escuela de Ingenier\'ia, Universidad Central de Chile, Avenida Francisco de Aguirre 0405, 171-0164 La Serena, Coquimbo, Chile}

\begin{abstract}
The dynamics of the Universe is analyzed using an exponential function for the dark energy equation of state, known as Gong-Zhang parameterization. The phase space of the free parameters presented in the model is constrained using Cosmic Microwave Background radiation, Cosmic Chronometers, modulus distance from Hydrogen II Galaxies, Type Ia Supernovae and measurements from Baryon Acoustic Oscillations, together with a stronger bound from a Joint analysis. The cosmological model is confronted with $\Lambda$CDM, observing there is a strong evidence for $\Lambda$CDM in the Joint analysis although the exponential model is preferred when the data are separated. Based on the Joint analysis, a value of $\omega_0 = -1.202^{+0.027}_{-0.026}$ is found for the characteristic parameter presented in the equation of state. Additionally, the cosmographic parameters at current times are reported, having $q_0 = -0.789^{+0.034}_{-0.036}$, $j_0=1.779^{+0.130}_{-0.119}$, and a transition deceleration-acceleration redshift $z_T = 0.644^{+0.011}_{-0.012}$. Furthermore, the age of the Universe is estimated as $t_U = 13.788^{+0.019}_{-0.019}$ Gyrs. Finally, we open a discussion if this model could alleviate the $H_0$ and $S_8$ tensions.
\end{abstract}
\pacs{Dark energy, equation of state, cosmology}
\maketitle

\section{Introduction}

Since the discovery of the accelerated expansion of the Universe at current times through Ia Type Supernovae (SNIa) in 1998 \cite{Riess:1998,Perlmutter:1999}, its nature is one of the most elusive open question in cosmology. This phenomenon, associated to the called Dark Energy (DE), was later confirmed through the radiation coming from early times of the Universe, the well-known Cosmic Microwave Background radiation (CMB) \cite{Planck:2018}. There is no doubt the simplest way to model the DE is with a cosmological constant ($\Lambda$) introduced into Einstein field equations as a new field. By adding a matter component (baryons and Dark Matter (DM)) and a relativistic component (photons and neutrinos), has been constructed the standard cosmological model known as $\Lambda$-Cold Dark Matter ($\Lambda$CDM). Until now, this paradigm has been tested at different scales and data and it is successful to fit and be consistent with them. However $\Lambda$CDM is not free of problems, apart from the coincidence and fine-tuning problems \cite{Carroll:2000,Zeldovich,Weinberg} related to $\Lambda$, recent studies \cite{Zhao:2017cud,Yang:2018qmz,Akarsu:2019hmw,DiValentino:2020naf,Chudaykin:2020ghx,Yang:2021flj,Colgain:2021pmf,Escamilla:2021uoj,Sharma:2022ifr} indicate that the DE Equation of State (EoS) could be dynamical instead of a constant.

Furthermore, a recent problem in Cosmology is related to the Hubble constant ($H_0$) measurements. There is a deviation about $4\sigma$ between the local measurements of $H_0$ using Type Ia Supernovae (SNIa) \cite{Riess:2019cxk} corresponding at late times of the Universe and those performed using CMB radiation at early times \cite{Planck:2018}. This problem is also known as the $H_0$ tension in which $\Lambda$CDM does not give an answer.  Nevertheless Ref. \cite{Efstathiou:2021ocp} has recently suggested that this might be related with systematics of the distance ladder measurements instead of an “exotic new physics”. Finally a review of cosmological models that could resolve the $H_0$ tension is presented in \cite{DiValentino:2021izs,Motta:2021hvl, Bamba_2012,DiValentino:2020vhf,DiValentino:2020zio,DiValentino:2020vvd,DiValentino:2020srs,Perivolaropoulos:2021jda,Abdalla:2022yfr}.

Additionally to the $H_0$ problem, there is another tension in the amplitude of matter perturbations smoothed over $8h^{-1}\mathrm{Mpc}$, $\sigma_8$, where $h$ is the dimensionless Hubble parameter. Indeed, the tension is quantified through the quantity $S_8\equiv\sigma_8\sqrt{\Omega_{0m}/0.3}$, where $\Omega_{0m}$ is the density parameter of matter. The $\Lambda$CDM inferred value from CMB anisotropies is not consistent with those from weak gravitational lensing measurements \cite{DES:2021wwk,KiDS:2020suj, Heymans_2021}. Notice that Planck collaboration predicts $S_8=0.834\pm0.016$ while Kilo-Degree Survey (KiDS-1000) \cite{Heymans_2021} predicts $S_8=0.766^{+0.020}_{-0.014}$, which produce a tension of about $3\sigma$ between both results. See for instance \cite{intertwined:S8} for details.

In general, the efforts to search for alternatives to understand the DE comes from two perspectives. One approach considers that the cosmic acceleration comes from a modified gravity theory such as $f(R)$ theories, Einstein-Gauss-Bonnet gravity, extra dimensions, fractional calculus \cite{Garcia-Aspeitia:2018fvw, Garcia-Aspeitia:2022uxz}. Another considers there is an extra component (fluid or field) that drives the accelerated expansion of the Universe and some of them are Chaplygin gas \cite{chaplygin, Hernandez-Almada:2018osh}, viscous fluids \cite{Almada:2020, Herrera:2020} and emergent fluids \cite{Li_2019, Hernandez-Almada:2020uyr}, among others. In general, these dark energy fluids have an evolving EoS that can be parametrized as a function of redshift. A well-studied parametrization able to solve the $\Lambda$ problems is a linear function of the scale factor (Chevallier-Polarski and Linder (CPL) parametrization \cite{CPL:2001, Linder:2003}), however it present divergences at the future \cite{Ma:2011}.
Although some EoS parameterizations as function of the redshift or scale factor have been proposed, some of them present problems. For instance, an EoS represented by a linear function of the redshift \cite{Huterer:2001, Weller:2002} diverges at high redshift (in the past) and is in disagreement with CMB \cite{Caldwell:2004} and Big Bang Nucleosynthesis constraints \cite{Johri:2002}. Although, there are proposals to solve these conflicts, most of them involve two or more parameters in the EoS \cite{Upadhye:2005, JBP:2005,  Liu:2008, Barboza:2008, Ma:2011, Li:2011, Feng:2012, Magana:2014voa, Hu:2015ksa, Roman-Garza:2018cxf,Singh:2020, Perkovic:2020} that need to be determined by the observations. Moreover, in \cite{Bouhmadi-Lopez:2014cca,Yang:2018qmz,Poulin:2018cxd} it is possible to study models of one-parameter extension of $\Lambda$, similarly in \cite{Acquaviva:2021jov,Akarsu:2021fol,Poulin:2022sgp,Akarsu:2022typ} these models are able to relax $H_0$ or $S_8$ tensions.

In this work we explore a parameterization proposed by Gong and Zhang (GZ) for DE EoS \cite{Gong:2005}, which is formulated to avoid the divergences at the future or at the early stages of the Universe \footnote{ Other models that does not diverge in the past or in the future can be founded in \cite{Bassett:2002qu,Bassett:2004wz,Shafieloo:2009ti,DeFelice:2012vd,Wei:2013jya,Hazra:2013dsx,Akarsu:2015yea,Barboza:2008rh,Ma:2011nc}.}, only adding one extra parameter to the standard cosmology. 
In \cite{Yang:2018qmz}, the GZ parameterization is constrained using CMB, SNIa (JLA sample), baryon acoustic oscillation (BAO) and observational Hubble data (OHD, cosmic chronometers).  Their constraints for the current DE EoS ($w_0$) show a trend to the phantom region ($w<-1$). Also, the $H_0$ bounds are in agreement with those obtained with CMB, hence the GZ model could reduce the $H_0$ tension. Later on, \cite{Yang:2022klj} provides updated GZ constraints from recent CMB data, SNIa (Pantheon and JLA), BAO and OHD. They also obtained that the $w_0$ tends to the phantom regime, a similar result as in \cite{Yang:2018qmz}, whereas that point out that the $H_0$ problem remains unsolvable (see \cite{Yang:2022klj}). In this line, we confront the GZ cosmology with several cosmological datasets to provide  strong updated bounds on their parameters, in both background and perturbation levels. We analyze the behavior of the deceleration and jerk parameters and compare with the standard paradigm. Furthermore, we analyze if this model could alleviate the $H_0$ tension through the $\mathbf{\mathbb{H}}0(z)$ diagnostic. 

The organization of the paper is as follows: In Sec. \ref{sec:cosmology} we present the mathematical formalism for the Gong-Zhang parameterization in where it is presented the background and  perturbed cosmology. In Sec. \ref{sec:constraints} we show the data sets and constraints obtained through CMB, Cosmic Chronometers, Baryon Acoustic Oscillations, Hydrogen II Galaxies and SNIa observations. Sec. \ref{Results} presents our constraints and finally in \ref{SD} we summarize our results and present a discussion about the model and its consequences. We henceforth use units in which $c=\hbar=k_{B}=1$, unless explicitly stated otherwise.

\section{Gong-Zhang parameterization cosmology} \label{sec:cosmology}

\subsection{Background cosmology} \label{sec:background}

\begin{figure}
    \centering
    \includegraphics[width=0.45\textwidth]{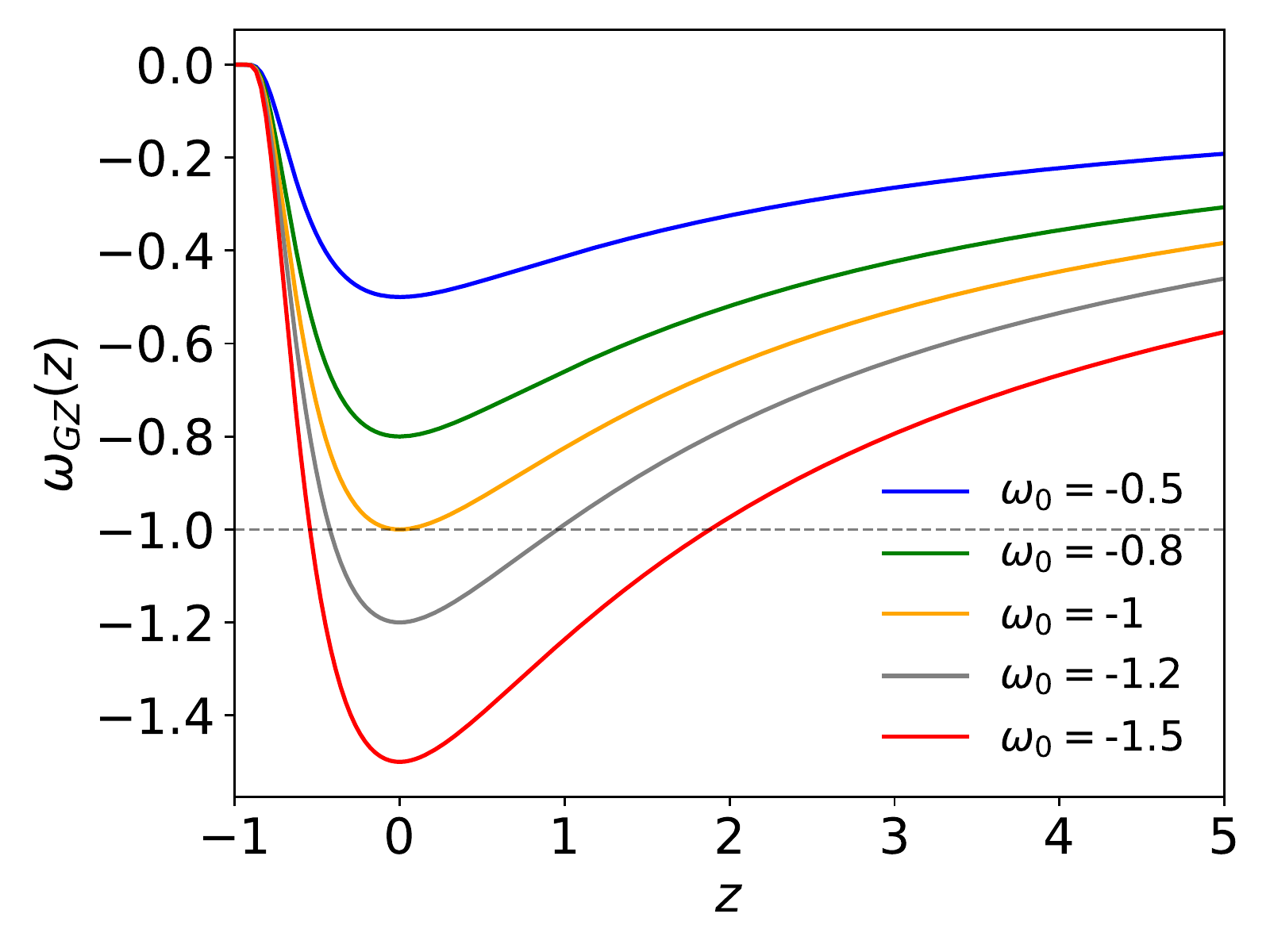}
    \caption{Evolution of the GZ EoS for several values of $\omega_0$, all of these EoS under the general relativity constraint of Universe acceleration $w_0<-1/3$.}
    \label{fig:wz_GZ}
\end{figure}

Under a flat Friedmann-Leamitre-Robertson-Walker (FLRW) spacetime, we consider a cosmological model composes by DM and baryon fluid with an EoS $\omega_{m}=0$, a radiation fluid with $\omega_r=1/3$, and a dynamical DE fluid whose EoS is parameterised \cite{Gong:2005}, as
\begin{equation}\label{eq:EoS_Gong}
    \omega_{GZ}(z) = \frac{w_0}{1+z} e^{z/(1+z)},
\end{equation}
where $z$ is the redshift and $w_0$ is a free parameter which represents the $\omega_{GZ}$ value at $z=0$. Notice that $w_{GZ}\to 0$ when $z\to \infty$ and when $z\to -1$. Notice that this model only adds one extra parameter over $\Lambda$CDM. Figure \ref{fig:wz_GZ} displays the evolution of $\omega_{GZ}(z)$ for several values of $\omega_0$.

At the background level, the cosmological system can be expressed as
\begin{eqnarray}
    H^2              &=& \frac{\kappa^2}{3}(\rho_m + \rho_r + \rho_{de}), \label{eq:H2}\\
    \dot{H}+H^2&=&-\frac{\kappa^2}{6}\sum_i(\rho_i+3p_i), \label{eq:2H2} \\
    \dot{\rho}_m     &=& -3H\rho_m \label{eq:rho_m},\\
    \dot{\rho}_r     &=& -4H\rho_r \label{eq:rho_r},\\
    \dot{\rho}_{de}  &=& - 3H(1+\omega_{GZ})\rho_{de}, \label{eq:rho_de}
\end{eqnarray}
where the dot symbol ($\,\cdot \,$) represents derivative with respect to cosmic time,  $\rho_{m}, \rho_{r}$ and $\rho_{de}$ refer to matter, radiation and dark energy densities respectively  and the sum is also by the three components, $H\equiv\dot{a}/a$ is the Hubble parameter, $a$ is the scale factor related with redshift through $a=(1+z)^{-1}$, $\kappa^2 \equiv 8\pi G$, and $G$ is the Newton gravitational constant. 

By integrating \eqref{eq:rho_m}-\eqref{eq:rho_de} and substituting into \eqref{eq:H2}, the Hubble parameter is given by
\begin{equation}\label{eq:E2}
    \left(\frac{H(z)}{H_0}\right)^2 = \Omega_{m0}(1+z)^3 + \Omega_{r0}(1+z)^4 + X_{de}(z),
\end{equation}
 where
\begin{equation} \label{eq:Ode}
    X_{de}(z) = \Omega_{de0}(1+z)^3\exp\left[3w_0\left( e^{\frac{z}{1+z}} -1\right)\right]\,,
\end{equation}
and $\Omega_{i0} = \kappa^2\rho_{i0}/3H^2_0$ with $i=m, r$ and $\Omega_{de0}=1-\Omega_{m0}-\Omega_{r0}$. For radiation density parameter we use $\Omega_{r0} = 2.469\times 10^{-5}h^{-2}(1+0.2271g_*)$, where $g_*=3.04$ is the effective number of relativistic species \cite{Komatsu_2011}.  Notice that the expression \eqref{eq:E2}, satisfies the flatness condition $H(0)=1$.  Additionally, notice that the DE energy density in eq. \eqref{eq:Ode} behaves as an effective DM density due to the cubic term implying a similar dust evolution. For instance, when $z\to-1$ we have $X_{de}\sim \Omega_{de0}e^{-3\omega_0}(1+z)^3$, and when $z\gg1$ the equation is $X_{de}\sim \Omega_{de0}e^{-3\omega_0(1-e)}(1+z)^3$, but at present it is a constant $X_{de}\sim \Omega_{de0}$.

On the other hand, the deceleration parameter $q(z)$ is estimated as
\begin{eqnarray}
    q(z) &=& -1 + \frac{H_0^2}{H^2(z)}\left[ \frac{3}{2}\Omega_{m0}(1+z)^3 + 2 \Omega_{r0}(1+z)^4 \right.\nonumber \\
    & & + \left. \frac{3}{2}\left( 1 + \omega_0- \omega_0 \frac{z}{1+z}\right)X_{de}(z) \right]\,, \label{eq:qz}
\end{eqnarray}
and the jerk parameter is obtained through the expression  
\begin{equation} \label{eq:jz}
    j(z) = q(2q+1) + (1+z)\frac{dq}{dz}\,.
\end{equation}
It is worth to mention that the jerk parameter for $\Lambda$CDM tends to $j\approx1$, and we have $j\rightarrow1$ as $z\rightarrow-1$ in GZ model.

\subsection{Perturbed cosmology} \label{sec:perturbed}

Within the EoS parameterization \eqref{eq:EoS_Gong}, the DE is dynamical and it plays an important role in the perturbation analysis.
In this sense, by perturbing the FLRW metric under the conformal Newtonian gauge as
\begin{equation}
    ds^2 = a^2(\eta)[-(1+2\psi)d\eta^2 + (1-2\phi)\delta_{ij}dx^idx^j]\,,
\end{equation}
where $\eta$ is the conformal time, $\psi(\eta,\vec{r})$ and $\phi(\eta,\vec{r})$ are scalar potentials. Thus the perturbed equations for the DE density and velocity are \cite{Bardeen:1980, Kodama:1984, Ma_1995}
\begin{eqnarray}
    \delta' &=& -(1+\omega_{GZ})(\theta-3\phi') - 3\mathcal{H}(c_s^2-\omega_{GZ})\delta\,, \label{deltaPert}\\
    \theta' &=& -\mathcal{H}(1-3\omega_{GZ}) - \frac{\omega_{GZ}'}{1+\omega_{GZ}}\theta + \frac{c_s^2}{1+\omega_{GZ}}k^2\delta \nonumber \\
    && + k^2\psi \label{thetaPert}\,,
\end{eqnarray}
respectively. In this notation, $\delta= \delta \rho/\rho  = (\rho -\bar{\rho})/\bar{\rho}$ is the density contrast between the density field $\rho$ and the spatial average density $\bar{\rho}$, $\theta = \nabla_i v^i$ is the velocity divergence and $k^2$ is the wave vector. The prime ($'$) denotes the derivatives with respect to $\eta$, the conformal Hubble parameter is $\mathcal{H}= a'/a$ and the square sound speed is defined in the form $c_s^2=\delta p/\delta \rho$.  It is worthy to note that for simplicity, the GZ parameterization is introduced as a fluid into the perturbation equations instead of a modification to GR. The another approach is that the GZ EoS is an effective EoS sourced by extra terms in the cosmological dynamical equations due to GR modifications. In this case, Eqs. \eqref{deltaPert}-\eqref{thetaPert}, could be not valid in general and they must be modified.

\section{Datasets and constraints} \label{sec:constraints}

The parameter phase-space of the GZ model is given by $\Theta=\{h, \Omega_{b0}, \Omega_{m0}, \omega_0, \log (10^{10}A_s), n_s, \tau{}_{reio}\}$ where $h$ is the dimensionless Hubble constant, $A_s$ is the amplitude of the initial power spectrum, $n_s$ is the scalar spectral index, $\tau{}_{reio}$ is the optical depth to re-ionization, $\Omega_{b0}=\kappa^2\rho_{b0}/3H^2_0$ with $\rho_{b0}$ is the current baryon density, and $\Omega_{m0}$ and $\omega_0$ were defined in the previous section. The aim is to bound them using CMB, CC, BAO, HIIG and SNIa datasets and Markov Chain Monte Carlo (MCMC) tools. Furthermore, a stronger bound on the phase-space is established by a joint analysis. Thus, a modified CLASS version \cite{Blas_2011} is used to implement the GZ parametrization and also a likelihood for HIIG sample is added to Montepython \cite{Brinckmann:2018cvx,Audren:2012wb} to perform the mentioned analysis. The convergence of the chains is verified through the Gelman-Rubin criterion.

\begin{table}
\centering
\resizebox{0.35\textwidth}{!}{%
\begin{tabular}{|lc|}
\hline
Parameter       &  Prior  \\ [0.9ex]
\hline
$h$  & Gauss($0.7324$,$0.0174$) \cite{Riess_2016}\\  [0.9ex]
$\Omega_{b0}$ & [$0.001, 0.15$]  \\ [0.9ex]
$\Omega_{m0}$  & [$0, 1$] \\ [0.9ex]
$\omega_0$     & [$-2,0$] \\ [0.9ex]
$\log (10^{10}A_s)$ & [$1.7,5.0$] \\ [0.9ex]
$n_s$  & [$0.7,1.3$]\\ [0.9ex]
$\tau{}_{reio}$  & [$0.004, 0.08$]\\ [0.9ex]
\hline

\hline
\end{tabular}
}
\caption{Priors considered over free parameters at background and perturbative level.
}
\label{tab:priors}
\end{table}

The configuration for the priors are Uniform distributions except for $h$ which is a Gaussian distribution as shown in Table \ref{tab:priors}.  We use a Gaussian prior over $h$ to obtain a largest deviation between our result and the Planck value instead of use a flat prior on $h$. In this sense, we would restrict the $h$ value to the SH0ES value \cite{Riess_2016} and leave to vary freely the rest of parameters, in particular, we are interested to observe the behaviour of the EoS parameter $\omega_0$.
Hence the figure-of-merit for the joint analysis is built through the a Gaussian log-likelihood given as $-2\ln(\mathcal{L}_{\rm data})\varpropto \chi^2_{\rm data}$ and
\begin{equation} \label{eq:chi2_joint}
\chi^2_{\rm data} = \chi_{\rm CMB}^2+\chi_{\rm CC}^2 + \chi_{\rm BAO}^2 + \chi_{\rm HIIG}^2 + \chi_{\rm SNIa}^2\,,
\end{equation}
where each terms refer to the $\chi^2$-function for every single dataset. Now, a description of all data is given in the rest of the section.

\subsection{Cosmic Microwave Background}

The latest CMB anisotropy data is provided by Planck 2018 legacy dataset release \cite{Planck:2018}. We use likelihoods of the high-$\ell$ for Temperature power spectrum (TT mode) covering a multipole interval $30\leq\ \ell<2508$, and for high TE multipole and polarization spectra EE modes in the range $30\leq \ell \leq 1996$. Additionally, we consider the low-$\ell$ TT-only likelihood which covers the multipole range $2\leq \ell \leq 29$.

\subsection{Cosmic Chronometers}

Cosmic chronometers consist of 31 Hubble parameter measurements which are considered as cosmological model independent because are obtained using differential age tools \cite{Moresco:2016mzx}. Furthermore, due these points are considered uncorrelated the $\chi^2$-function ca be written as
\begin{equation} \label{eq:chi2_OHD}
    \chi^2_{{\rm CC}}=\sum_{i=1}^{31}\left(\frac{H_{th}(z_i, \Theta)-H_{obs}(z_i)}{\sigma^i_{obs}}\right)^2,
\end{equation}
where the sum runs over the whole sample, and $H_{th}-H_{obs}$ is the difference between the theoretical and observational Hubble parameter at the redshift $z_i$ and $\sigma_{obs}$ is the uncertainty of $H_{obs}$.

\subsection{Baryon Acoustic Oscillations}

Baryon Acoustic Oscillations give an standard ruler evolving through the Universe since recombination epoch  produced from the interactions between baryons and photons. As BAO is a powerful tool for constraining cosmological models, we use several measurements of the BAO data coming from different tracers as galaxies, Ly$\alpha\times$ Ly$\alpha$ and Ly$\alpha\times$ QSO. Table \ref{tab:bao} summaries the BAO dataset used. BAO measurements in the transverse direction give an estimate of $D_M(z)/r_d$ and the BAO feature through the line of sight gives $D_H(z)/r_d = c/H(z)r_d$. Here,
\begin{equation}
    D_M(z) = c \int_0^z \frac{dz'}{H(z')}\,, \label{DM}
\end{equation}
is the comoving angular diameter for a flat cosmology at the redshift $z$, $r_d=r_s(z_d)$ is the size of the sound horizon at the drag epoch redshift $z_d$, and $c$ is the speed of light.
The datasets 6dFGS and SDSS DR7 report the measurement $D_V(z)r_d$ where the distance scale $D_V$ is defined as
\begin{equation}
    D_V(z) = [z\,D_H(z)\,D_M^2(z)]^{1/3}\,.
\end{equation}
The redshift $z_d$ is given by \cite{Eisenstein_1998}
\begin{equation}
    z_d = \frac{1291(\Omega_{m0}h^2)^{0.251}}{1+0.659(\Omega_{m0}h^2)^{0.828}}[1+b_1(\Omega_{b0}h^2)^{b_2}],
\end{equation}
and
\begin{eqnarray}
    b_1 &=& 0.313(\Omega_{m0}h^2)^{-0.419}[1+0.607(\Omega_{m0}h^2)^{0.674}], \\
    b_2 &=& 0.238(\Omega_{m0}h^2)^{0.223}\,.
\end{eqnarray}
Thus the $\chi^2$-function is built as
\begin{eqnarray}
    \chi^2_{{\rm BAO}} &=& \chi^2_{{\rm 6dFGS}} + \chi^2_{{\rm SDSS DR7}} + \chi^2_{{\rm SDSS DR12}} \nonumber \\
    & & + \chi^2_{{\rm SDSS DR14}}\,.
\end{eqnarray}
Due the 6dFGS \cite{Beutler_2011}, SDSS DR7 \cite{Ross_2015}, SDSS DR14 \cite{de_Sainte_Agathe_2019} datasets give uncorrelated points, each $\chi^2$ corresponds to the square of the difference between the observable measurement and the theoretical estimate divided by the uncertainty of the measurement. SDSS DR12 \cite{Alam_2017} dataset reports correlated measurements of the $D_M$ and $H$, thus the $\chi^2$ function is built using the covariance matrix, as $\chi^2= \vec{X}^T {\rm C}_{DR12}^{-1} \vec{X}$.

\begin{table*}
\centering
\resizebox{0.9\textwidth}{!}{%
\begin{tabular}{|lllccc|}
\hline
BAO measurement &  Dataset     &  Tracer & Parameter & $z_{\rm eff}$ & Reference  \\ [0.9ex]
\hline
6dFGS                             & 6dFGS     & galaxies &    $r_d/D_V(z_{\rm eff})$  & 0.106 & \cite{Beutler_2011} \\ [0.9ex]
SDSS MGS                          & SDSS DR7  & galaxies &    $D_V(z_{\rm eff})/r_d$  & 0.15  & \cite{Ross_2015}   \\ [0.9ex]
BOSS Gal                          & SDSS DR12 & galaxies &    $D_M(z_{\rm eff})/r_d$, $D_H(z_{\rm eff})/r_d$              & 0.38, 0.51, 0.61 & \cite{Alam_2017} \\ [0.9ex]
eBOSS Ly$\alpha\times$ Ly$\alpha$ & SDSS DR14 & Ly$\alpha\times$ Ly$\alpha$& $D_M(z_{\rm eff})/r_d$, $D_H(z_{\rm eff})/r_d$ & 2.34  & \cite{de_Sainte_Agathe_2019}\\ [0.9ex]
eBOSS Ly$\alpha\times$ QSO        & SDSS DR14 & Ly$\alpha\times$ QSO &  $D_M(z_{\rm eff})/r_d$, $D_H(z_{\rm eff})/r_d$  & 2.35  & \cite{Blomqvist_2019}\\ [0.9ex]
\hline
\end{tabular}
}
\caption{BAO measurements used in the cosmological parameter estimation.}
\label{tab:bao}
\end{table*}

\subsection{Hydrogen II Galaxies}

Hydrogen II galaxies are compact low mass galaxies which their luminosity is almost dominated by a young massive burst of star formation  and are useful to constrain cosmological parameters \cite{Chavez2014} because there is a correlation between the measured luminosity, $L$,  and the inferred velocity dispersion, $\sigma$, of the ionized gas. The largest sample reported by \cite{GonzalezMoran2019, Gonzalez-Moran:2021drc} consists of a full sample of 181 HIIG measurements in the redshift range $0.01<z<2.6$. The $\chi^2$-function is written as
\begin{equation}\label{eq:chi2_HIIG}
    \chi^2_{{\rm HIIG}}=\sum_i^{181}\frac{[\mu_{th}(z_i, {\Theta})-\mu_{obs}(z_i)]^2}{\epsilon_i^2},
\end{equation}
where $\mu_{obs}$ is the observational distance modulus given by
\begin{equation}
    \mu_{obs} = 2.5(\alpha + \beta\log \sigma -\log f - 40.08)\,,
\end{equation}
where $\alpha$ and $\beta$ correspond to the intercept and slope of the $L$-$\sigma$ relation and $f$ is the measured flux. The quantity $\epsilon_i$ is the uncertainty of $\mu_{obs}$ at $z_i$. It is worth to mention that \cite{Gonzalez-Moran:2021drc} reports measurements and uncertainties of $\log \alpha$, $\log\beta$ and $\log f$, hence it is necessary to propagate such errors to estimate $\epsilon_i$ (for more details see \cite{Gonzalez-Moran:2021drc}).

The theoretical counterpart is given as
\begin{equation}
    \mu_{th}(z, \Theta) = 5 \log_{10} \left [ \frac{d_L(z, \Theta)}{1\,{\rm Mpc}}\right] + 25,
\end{equation}
where $d_L$ is the luminosity distance measured in Mpc expressed as 
\begin{equation}
d_L(z, \Theta)=(1+z)D_M(z), \label{LD}
\end{equation}
being $D_M(z)$ defined by Eq. \eqref{DM}.

\subsection{Type Ia Supernovae}

The Pantheon sample \cite{Scolnic:2018} contains 1048 luminosity modulus measurements coming from Supernovae Ia. With this sample a redshift region $0.01<z<2.3$ is covered. In this case the measurements are correlated which is appropriate to use a $\chi^2$-function as
\begin{equation}\label{eq:chi2SnIa}
    \chi_{\rm SNIa}^{2}=\vec{X}^T {\rm C}_P^{-1} \vec{X},
\end{equation}
where $\vec{X}$ is the vector of residuals between the theoretical distance modulus and the observed one and the upper index $T$ represents the transpose of the vector, ${\rm C}_{P}$ is the covariance matrix formed by adding the systematic and statistic uncertainties, ${\rm C}_{P}={\rm C}_{P,sys}+{\rm C}_{P,stat}$.

The theoretical distance modulus is estimated by
\begin{equation}
    m_{th}=\mathcal{M}+5\log_{10}\left[\frac{d_L(z)}{10\, pc}\right],
\end{equation}
where $\mathcal{M}$ is a nuisance parameter and $d_L(z)$ is luminosity distance defined in Eq. \eqref{LD}.

\section{Results} \label{Results}

\begin{figure*}
    \centering
    \includegraphics[width=0.6\textwidth]{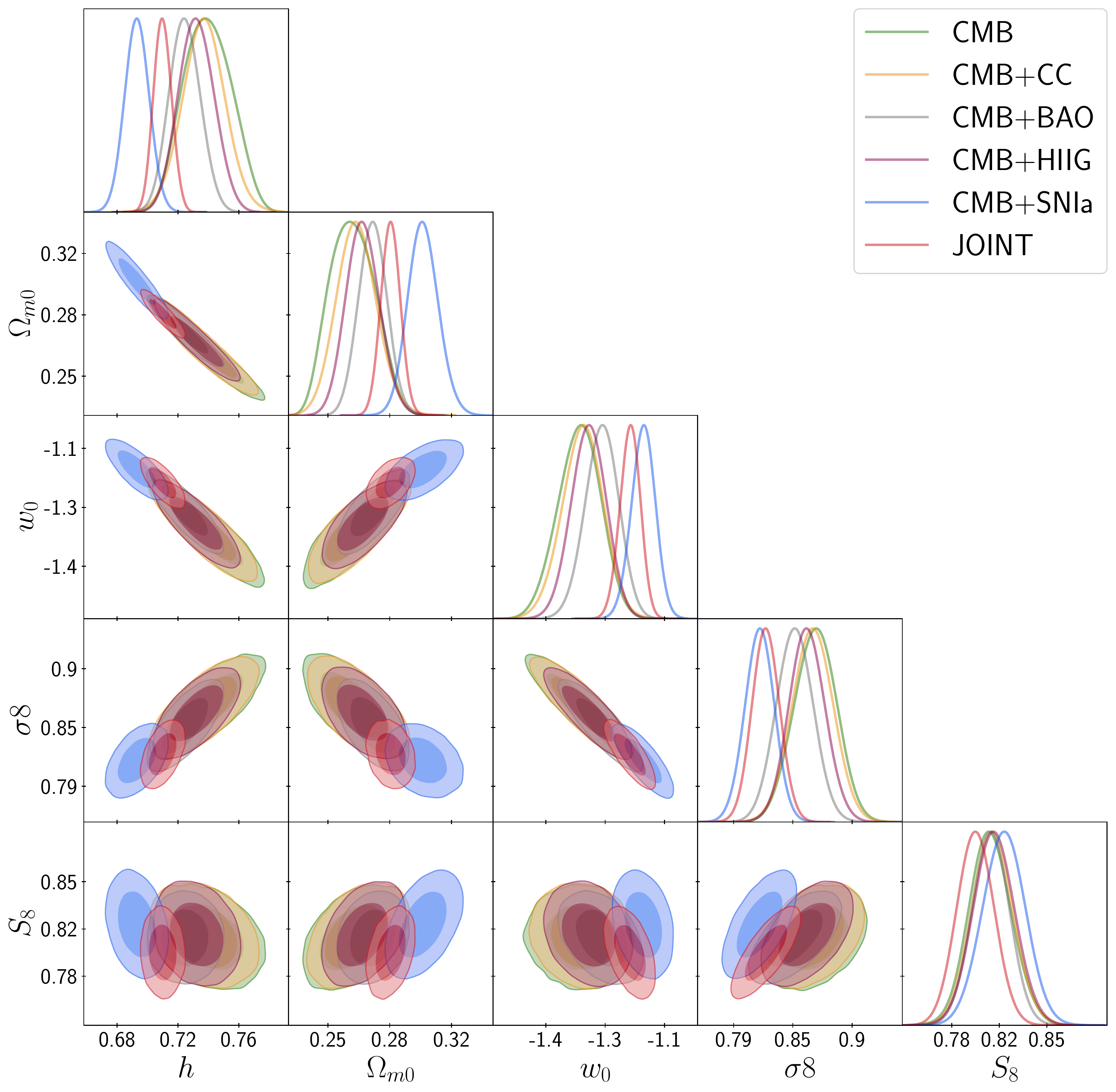}
    \caption{2D contours at $1\sigma$ (inner region) and $2\sigma$ (outermost region) CL for the GZ cosmology.}
    \label{fig:contours}
\end{figure*}

After confronting the cosmological model with CMB, CMB+CC, CMB+BAO, CMB+HIIG, CMB+SNIa data and CMB+CC+BAO+HIIG+SNIa (we will refer as Joint to this data combination),  the main values of the parameter space are summarized in the Table \ref{tab:bestfits}. Additionally, the 2D probability spaces at $68\%$ ($1\sigma$) and $95\%$ ($2\sigma$) and 1D posterior distributions are displayed in Figure \ref{fig:contours}. Furthermore, we also estimate the cosmological parameter $\sigma_8$, defined as the r.m.s. density variation when smoothed with a tophat-filter of radius of $8h^{-1}\mathrm{Mpc}$ \cite{peacock_1998}. Regarding this parameter, we find consistent results as those found using $\Lambda$CDM. For the Joint analysis we find our results in agreement within $2\sigma$ with those values reported in \cite{Yang:2019} (using Planck 2015+CC+BAO+JLA), and is deviated $1.2\sigma$ from the values reported in \cite{Gong:2005} (using WMAP+SNIa+SDSS) for the GZ cosmology; in particular, our yield $\omega_0=-1.202^{+0.027}_{-0.026}$ is within $1\sigma$ but about $50\%$ less of uncertainty. Based on this $\omega_0$ value, the DE behaves as phantom today and is deviated about $7.5\sigma$ of the quintessence behaviour instead of about $4\sigma$ reported in \cite{Yang:2019}. On the other hand, to compare statistically both GZ and $\Lambda$CDM models, we estimate their Bayesian evidences. In this test, a model $M_1$ is compared with respect to a model $M_2$ through the ratio of their posterior probabilities as
\begin{equation}
    \frac{P(M_1| D)}{P(M_2| D)} =  B_{12}\frac{P(M_1)}{P(M_2)}\,,
\end{equation}
where the factor $B_{12}$ is known as the Bayes' factor defined as the ratio of the evidence of the models, $B_{12}= P(D|M_1)/P(D| M_2)$. In the Jeffreys' scale \cite{BayesFactor:1995}, the strength of evidence is as follow. When $|\log B_{12}|<1$ there is a weak or inconclusive evidence, it is definite or positive evidence for $1\leq |\log B_{12}|<3$, a strong evidence for $3\leq |\log B_{12}|<5$, and a very strong evidence or decisive for $|\log B_{12}|>5$. Table \ref{tab:Bij} shows the difference of the $\log B_{12} = \log B_{\rm GZ} - \log B_{\Lambda \rm CDM}$ when GZ model is confronted with $\Lambda$CDM for the different samples used by using the publicly package MCEvidence \cite{Heavens:2017, Heavens:2017arxiv}. A negative (positive) value of $\log B_{ij}$ indicates a preference of the data over $\Lambda$CDM (GZ cosmology). We find an inconclusive evidence for CMB+CC and CMB+HIIG, a strong evidence of GZ model for CMB, CMB+BAO and CMB+SNIa, and a strong evidence of $\Lambda$CDM for the Joint analysis. 

\begin{table*}
\centering
\resizebox{0.8\textwidth}{!}{%
\begin{tabular}{|lccccccccc|}
\hline
Sample       &  $\chi^2_{\rm min}$     &  $h$ & $\Omega_{m0}$ & $\omega_0$ & $\log10^{10}A_{s}$ & $n_{s}$ & $\tau{}_{reio}$ & $\sigma_8$ & $S_8$\\ [0.9ex]
\hline
\hline
\multicolumn{10}{|c|}{GZ cosmology} \\ [0.9ex]
CMB & $2841$ & $0.742^{+0.017}_{-0.017}$ & $0.262^{+0.012}_{-0.014}$ & $-1.333^{+0.057}_{-0.054}$ & $3.048^{+0.015}_{-0.016}$ & $0.965^{+0.004}_{-0.004}$ & $0.055^{+0.007}_{-0.008}$ & $0.867^{+0.020}_{-0.020}$ & $0.810^{+0.015}_{-0.016}$ \\ [0.9ex]
CMB+CC & $2856$ & $0.739^{+0.016}_{-0.016}$ & $0.264^{+0.011}_{-0.013}$ & $-1.324^{+0.054}_{-0.053}$ & $3.049^{+0.015}_{-0.016}$ & $0.965^{+0.004}_{-0.004}$ & $0.055^{+0.007}_{-0.008}$ & $0.865^{+0.019}_{-0.019}$ & $0.811^{+0.015}_{-0.016}$ \\ [0.9ex]
CMB+BAO & $2861$ & $0.725^{+0.011}_{-0.012}$ & $0.274^{+0.008}_{-0.009}$ & $-1.275^{+0.045}_{-0.042}$ & $3.049^{+0.015}_{-0.016}$ & $0.966^{+0.003}_{-0.003}$ & $0.056^{+0.007}_{-0.008}$ & $0.848^{+0.017}_{-0.018}$ & $0.810^{+0.013}_{-0.014}$ \\ [0.9ex]
CMB+HIIG & $3278$ & $0.733^{+0.013}_{-0.013}$ & $0.268^{+0.009}_{-0.010}$ & $-1.309^{+0.047}_{-0.044}$ & $3.049^{+0.015}_{-0.016}$ & $0.965^{+0.004}_{-0.004}$ & $0.055^{+0.008}_{-0.008}$ & $0.860^{+0.017}_{-0.017}$ & $0.812^{+0.015}_{-0.016}$ \\ [0.9ex]
CMB+SNIa & $3883$ & $0.690^{+0.009}_{-0.009}$ & $0.303^{+0.009}_{-0.009}$ & $-1.168^{+0.032}_{-0.031}$ & $3.050^{+0.015}_{-0.016}$ & $0.965^{+0.004}_{-0.004}$ & $0.056^{+0.007}_{-0.008}$ & $0.815^{+0.014}_{-0.014}$ & $0.819^{+0.016}_{-0.016}$ \\ [0.9ex]
JOINT & $4367$ & $0.709^{+0.007}_{-0.007}$ & $0.284^{+0.006}_{-0.006}$ & $-1.202^{+0.027}_{-0.026}$ & $3.051^{+0.015}_{-0.017}$ & $0.968^{+0.003}_{-0.003}$ & $0.058^{+0.007}_{-0.009}$ & $0.821^{+0.013}_{-0.013}$ & $0.799^{+0.013}_{-0.013}$ \\ [0.9ex]
\hline
\multicolumn{10}{|c|}{$\Lambda$CDM} \\ [0.9ex]
CMB & $2849$ & $0.680^{+0.006}_{-0.006}$ & $0.307^{+0.008}_{-0.008}$ & --- & $3.048^{+0.015}_{-0.016}$ & $0.969^{+0.004}_{-0.004}$ & $0.057^{+0.007}_{-0.008}$ & $0.808^{+0.007}_{-0.008}$ & $0.817^{+0.016}_{-0.016}$ \\ [0.9ex]
CMB+CC & $2864$ & $0.680^{+0.006}_{-0.006}$ & $0.306^{+0.007}_{-0.008}$ & --- & $3.048^{+0.015}_{-0.016}$ & $0.969^{+0.004}_{-0.004}$ & $0.057^{+0.007}_{-0.009}$ & $0.808^{+0.007}_{-0.008}$ & $0.817^{+0.015}_{-0.016}$ \\ [0.9ex]
CMB+BAO & $2860$ & $0.681^{+0.004}_{-0.004}$ & $0.305^{+0.006}_{-0.006}$ & --- & $3.048^{+0.014}_{-0.016}$ & $0.970^{+0.003}_{-0.003}$ & $0.057^{+0.007}_{-0.008}$ & $0.808^{+0.007}_{-0.007}$ & $0.814^{+0.012}_{-0.013}$ \\ [0.9ex]
CMB+HIIG & $3289$ & $0.684^{+0.006}_{-0.005}$ & $0.302^{+0.007}_{-0.007}$ & --- & $3.049^{+0.015}_{-0.016}$ & $0.971^{+0.004}_{-0.004}$ & $0.058^{+0.007}_{-0.009}$ & $0.806^{+0.007}_{-0.008}$ & $0.809^{+0.014}_{-0.015}$ \\ [0.9ex]
CMB+SNIa & $3893$ & $0.688^{+0.005}_{-0.005}$ & $0.296^{+0.006}_{-0.007}$ & --- & $3.049^{+0.015}_{-0.017}$ & $0.973^{+0.003}_{-0.004}$ & $0.059^{+0.008}_{-0.009}$ & $0.803^{+0.007}_{-0.008}$ & $0.798^{+0.014}_{-0.014}$ \\ [0.9ex]
JOINT & $4358$ & $0.688^{+0.004}_{-0.004}$ & $0.296^{+0.005}_{-0.005}$ & --- & $3.049^{+0.015}_{-0.017}$ & $0.973^{+0.003}_{-0.003}$ & $0.059^{+0.008}_{-0.009}$ & $0.804^{+0.007}_{-0.007}$ & $0.799^{+0.011}_{-0.011}$ \\ [0.9ex]
\hline
\end{tabular}
}
\caption{First column is the data sample, second column is the minimum value of the $\chi^2$, and the rest of the columns are the bestfit values of the free parameters and their $1\sigma$ CL uncertainties for both GZ and $\Lambda$CDM cosmologies.}
\label{tab:bestfits}
\end{table*}

\begin{table}
\centering
\resizebox{0.2\textwidth}{!}{%
\begin{tabular}{|lc|}
\hline
Sample   & $\log B_{12}$ \\
\hline
CMB      & 18.99  \\
CMB+CC   & -0.29  \\
CMB+BAO  & 7.95   \\
CMB+HIIG & -1.69  \\
CMB+SNIa & 6.20   \\
JOINT    & -10.15 \\
\hline
\end{tabular}
}
\caption{Bayes factor difference, $\log B_{12}= \log B_{\rm GZ}-\log B_{\Lambda \rm CDM}$.}
\label{tab:Bij}
\end{table}

\begin{figure}
    \centering
    \includegraphics[width=0.5\textwidth]{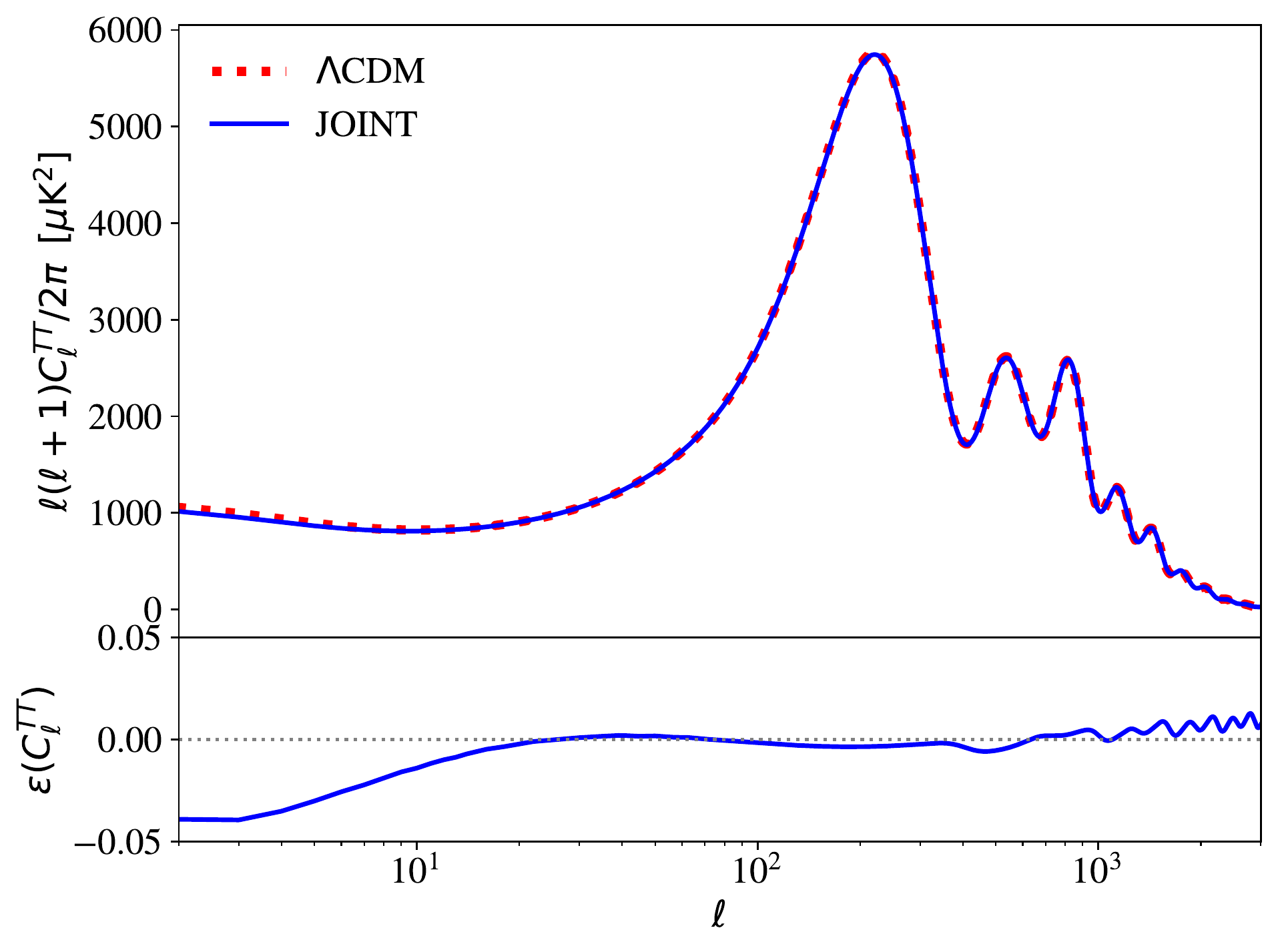}
    \caption{$C^{TT}_{\ell}$ CMB Temperature reconstruction for both GZ and $\Lambda$CDM cosmologies at top panel. Relative difference between both models, $\varepsilon = C^{TT}_{\ell}({\rm GZ})/C^{TT}_{\ell}(\Lambda{\rm CDM})-1$, at bottom panel.}
    \label{fig:cmb}
\end{figure}

Figure \ref{fig:cmb} displays the $C_\ell^{TT}$ CMB Temperature power spectrum for both GZ and $\Lambda$CDM cosmologies at the top panel and the relative error of GZ with respect to $\Lambda$CDM at the bottom panel. We find agreement between both models less than $5\%$. As it is expected, the largest contribution of cosmological models with variable EoS is at low $\ell$ as shown in the bottom panel \cite{Feng:2012}. 

\begin{figure*}
\centering
\includegraphics[width=0.31\textwidth]{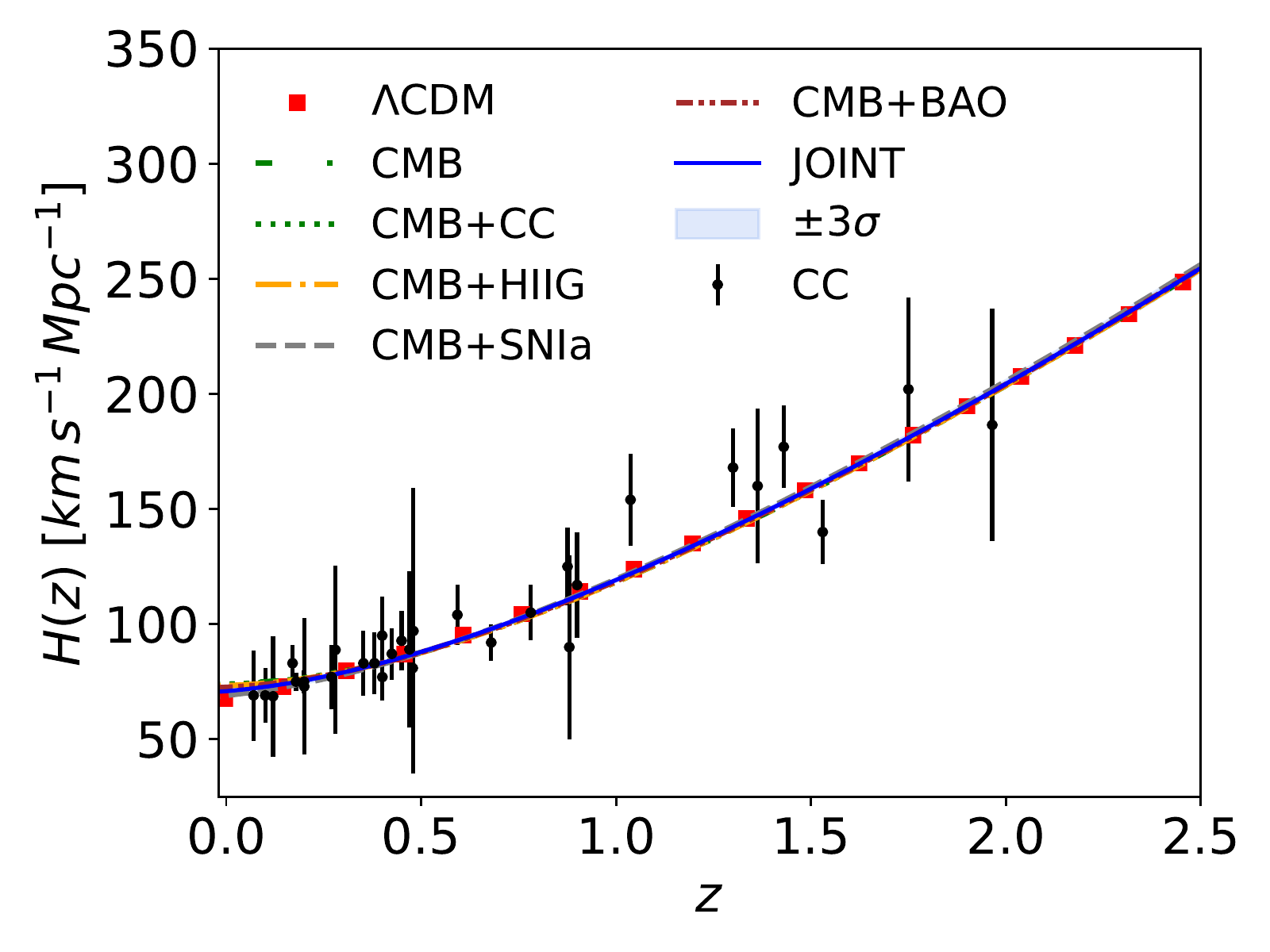}
\includegraphics[width=0.31\textwidth]{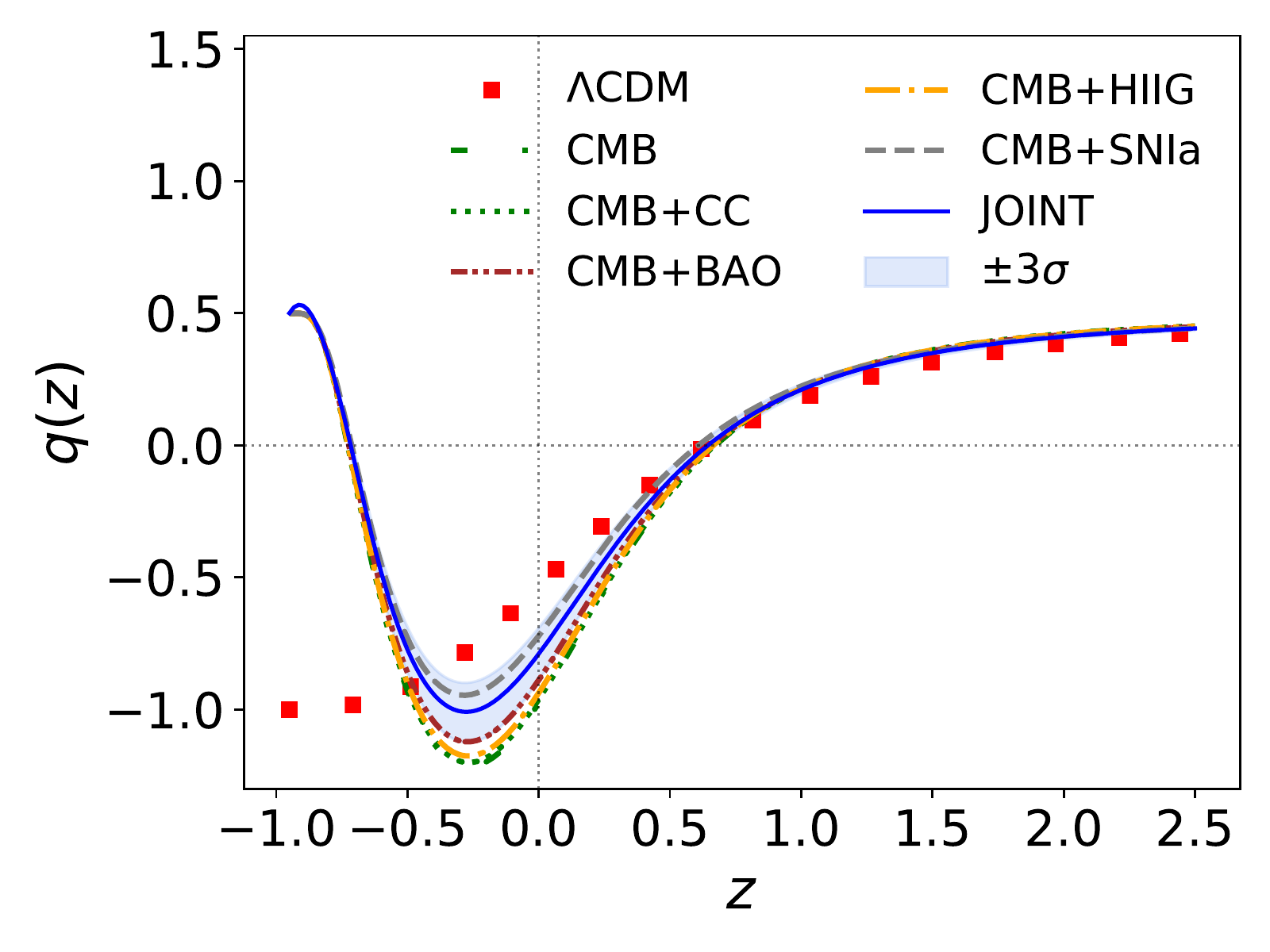}
\includegraphics[width=0.31\textwidth]{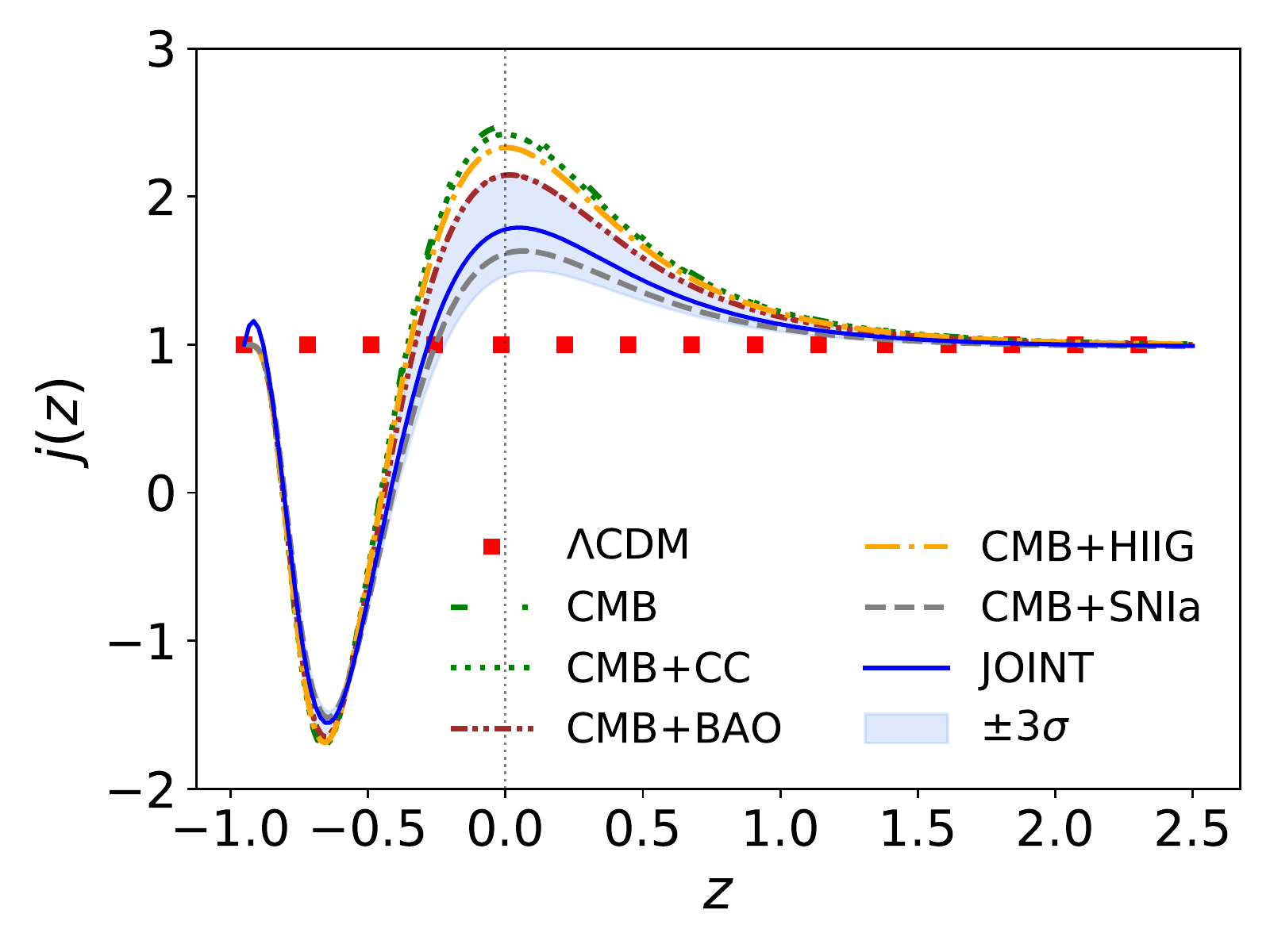}
 \caption{Left to right: reconstruction of the $H(z)$, $q(z)$ and $j(z)$ in GZ cosmology and $\Lambda$CDM for each dataset. The bands correspond to $3\sigma$ uncertainty for GZ model.
}
\label{fig:Hz_and_qz}
\end{figure*}

Figure \ref{fig:Hz_and_qz} shows the reconstruction of the Hubble (left panel), the deceleration (middle panel) and jerk (right panel) parameters for the GZ model as function of redshift. The blue shadow regions show the $3\sigma$ error for the joint constraints. For comparison, the $\Lambda$CDM prediction is also shown. Although the $H(z)$ reconstruction for GZ model is consistent with $\Lambda$CDM as shown in Fig. \ref{fig:Hz_and_qz}, the cosmographic parameters present different behaviours. Firstly,
an interesting feature is the slowing down of the cosmic acceleration observed in the $q(z)$ reconstruction at $z_{sd}\approx -0.3$ as shown in Fig. \ref{fig:Hz_and_qz}, where $z_{sd}$ is the slowing down redshift, being the accelerated epoch only a transitory effect ($-0.7\lesssim z\lesssim 0.7$), ending the evolution of the Universe similar as the matter dominated epoch. 
The slowing down of cosmic acceleration is also found in some dynamical dark energy models with parametric EoS \cite{Cardenas:2012mw,Cardenas:2013roa,Magana:2014voa,Hu:2015ksa,ElBoss,Zhang:2017jvo,Roman-Garza:2018cxf,Bolotin:2020qbx} but at current times. Nevertheless, this trend might not be present in these models if their constraints are updated from the recent data samples \cite{Zhang:2017jvo}. In general, not all the parametric models predict a future deceleration \cite{Yang:2019} and sometimes the results can be misleading when it is used old observational data to constrain them \cite{Zhang:2017jvo}. For the GZ model, a future deceleration is predicted instead of a slowing down of the acceleration at current times, this trend differs from the standard cosmological model. This phenomenon (i.e. the future deceleration) is also found in other models (see for instance \cite{Escobal:2023saa,Mamon}). On the other hand, the $j(z)$ presents also an oscillatory behaviour in the range $-1<z\lesssim 1$ and goes to $j\to1$ for $z\gg 1$ instead of a constant as the jerk for $\Lambda$CDM.

An estimated value of $q_0 = -0.789^{+0.034}_{-0.036}$ is found which is deviated more than $3\sigma$ from the $\Lambda$CDM value but the value of the deceleration-acceleration transition redshift of $z_T=0.644^{+0.011}_{-0.012}$ obtained is consistent within $1\sigma$ from the $\Lambda$CDM value. Additionally, a current value of the jerk parameter of $j_0=1.779^{+0.130}_{-0.119}$ is reported and is deviated more than $5\sigma$ from the $\Lambda$CDM value. The values of $q_0$ and $j_0$ are also in agreement with the regions $-1.4<q_0<-0.3$ and $-0.1<j_0<6.4$ reported in \cite{Riess_2004} and discussed by \cite{Visser_2004}. 

\begin{figure}
    \centering
    \includegraphics[width=0.5\textwidth]{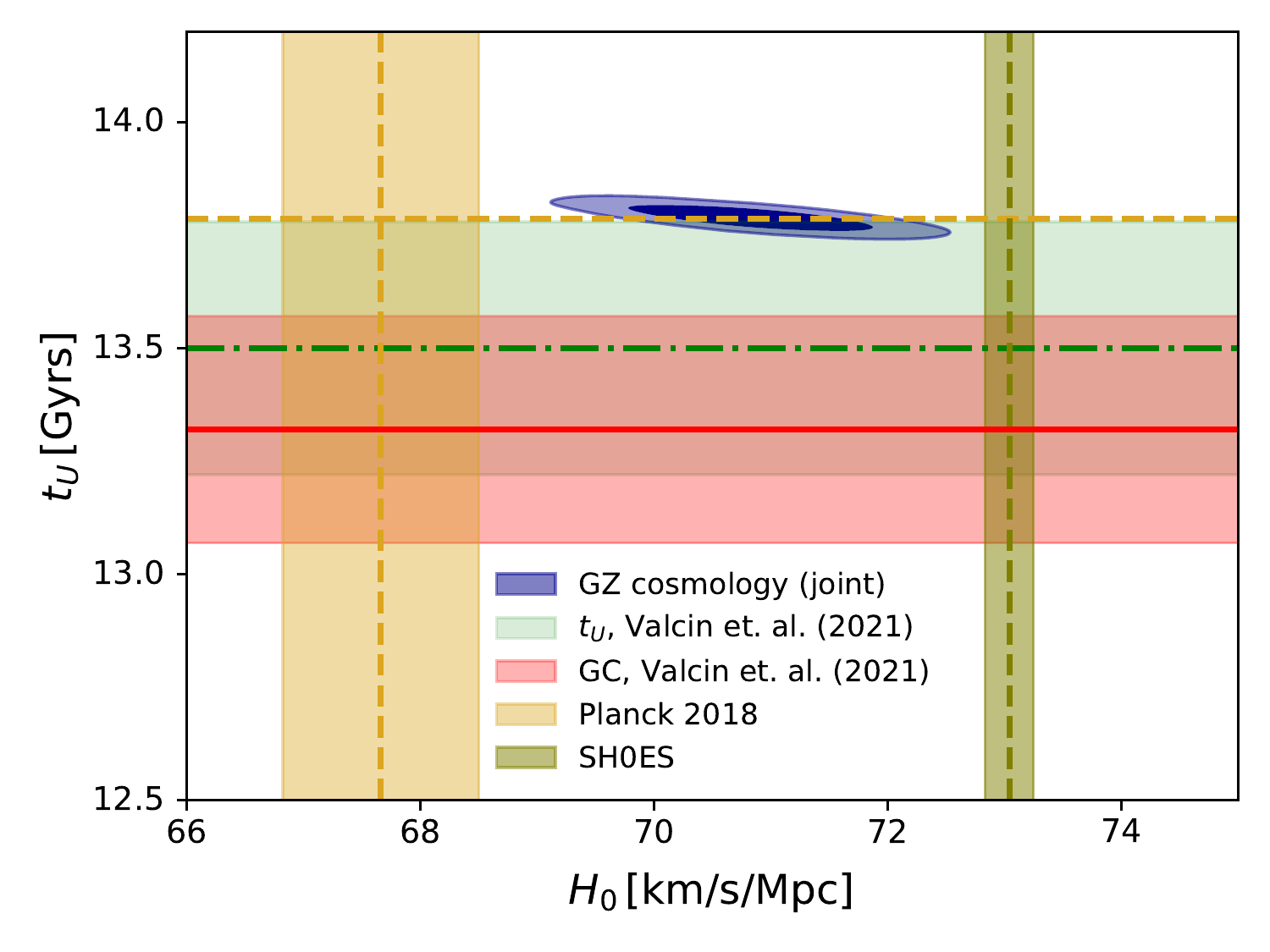}
    \caption{ Constraints at $1\sigma$ (inner region) and $3\sigma$ (outer region) in the $H_0$-$t_U$ plane for the GZ cosmology. Vertical bands are $2\sigma$ CL region centered at best fit values (dashed lines) for Planck 2018 and SH0ES respectively. Horizontal dashed line is the estimated value for the age of Universe by Planck 2018. Horizontal bands represent $1\sigma$ CL region centered at the bestfit values for the globular cluster age (solid line)  and the age of Universe (dashed dot line).}
    \label{fig:H0vstU}
\end{figure}

We estimate the age of the Universe as
$13.743^{+0.030}_{-0.030}$ Gyrs (CMB), 
$13.747^{+0.028}_{-0.027}$ Gyrs (CMB+CC),
$13.754^{+0.025}_{-0.025}$ Gyrs (CMB+HIIG),
$13.767^{+0.024}_{-0.023}$ Gyrs (CMB+BAO),
$13.834^{+0.024}_{-0.023}$ Gyrs (CMB+SNIa)
$13.788^{+0.019}_{-0.019}$ Gyrs (Joint). These values
are consistent within $1\sigma$ with the value reported by Planck \cite{Planck:2018}. Additionally they are in agreement within $1.02\sigma$ with the cosmological model-independent value of the age of the Universe ($t_{U}=13.5^{+0.16}_{-0.14}\mathrm{(stat.)} \pm 0.23 \mathrm{(sys.)}$) and deviated from the age of the oldest globular clusters (GC) ($t_{GC}=13.32\pm 0.1 \mathrm{(stat.)} \pm 0.23 \mathrm{(sys.)}$) in $2\sigma$ \cite{Valcin:2021} (see also \cite{Valcin_2020, Bernal:2021}).  Figure \ref{fig:H0vstU} shows the constrained contours at $1\sigma$ and $3\sigma$ in the plane $H_0$-$t_U$ for GZ cosmology using the joint analysis. Additionally, vertical lines with bands at $2\sigma$ represent the corresponding $H_0$ values for Planck 2018 and SH0ES. Horizontal lines is the age of GC (solid line) \cite{Valcin:2021}, age of Universe by Valcin 2021 \cite{Valcin:2021} (dashed dot green line) and Planck 2018 (dashed line) \cite{Planck:2018}.

 On the other hand, GZ model can relax the $H_0$ tension presented between the SH0ES measurements of $H_0$ and those by Planck collaboration. Our $h$ constraints (Table \ref{tab:bestfits}),
are deviated from the Planck value in $3.7\sigma$ (CMB), $3.8\sigma$ (CMB+CC), $3.8\sigma$ (CMB+BAO), $4.1\sigma$ (CMB+HIIG), $1.4 \sigma$ (CMB+SNIa) and $4.0\sigma$ (joint). Additionally, our $h$ values are deviated within $1\sigma$ for CMB, CMB+CC, CMB+BAO and CMB+HIIG combinations whereas it is  $4.45\sigma$ for CMB+SNIa and $3.02\sigma$ joint from SH0ES value. This means that reduce the $H_0$ tension from $\sim 4.8\sigma$ obtained between the Planck value and SH0ES value to $\sim 4\sigma$ for the joint analysis. 

 We find that GZ model could alleviate the $S_8$ tension. Our estimates report deviations of $1.72\sigma$ (CMB), $1.76\sigma$ (CMB+CC), $1.80\sigma$ (CMB+BAO), $1.80\sigma$ (CMB+HIIG), $2.07\sigma$ (CMB+SNIa) and  $1.38\sigma$ (Joint) from the value reported by KiDS-1000 ($S_8=0.766^{+0.020}_{-0.014}$) \cite{Heymans_2021}. This means a distance reduction of $2.65\sigma$ obtained between the Planck value and the KiDS-1000 value to $1.38\sigma$ for the joint analysis.

 Finally, we analyze the $\mathbf{\mathbb{H}}0(z)$ diagnostic \cite{H0diagnostic:2021} in which it is able to compare the Hubble constant from two different models, named model A and model B with an effective EoS $\omega_{eff}^{(A)}(z)$  and $\omega_{eff}^{(B)}(z)$ respectively through 
\begin{equation} \label{eq:H0diag}
    \frac{H_0^{(A)}}{H_0^{(B)}} = \exp \left( \frac{3}{2}\int_0^z \frac{\Delta w_{eff}(z')}{1+z'}dz'\right),
\end{equation}
where 
$\Delta w_{eff}(z)= \omega_{eff}^{(B)}(z)-\omega_{eff}^{(A)}(z)$. If $\Delta w_{eff}(z)=0$ for all redshift, the right hand side is a constant. In particular, if we confront GZ model and $\Lambda$CDM, the diagnostic $\mathbf{\mathbb{H}}0(z)$ is written as
\begin{equation}
    \mathbf{\mathbb{H}}0(z) = \frac{H_{GZ}(z)}{\sqrt{\Omega_{\Lambda}+\Omega_{m0}^{\mathrm{\Lambda CDM}}(1+z)^3 +\Omega_{r0}^{\mathrm{\Lambda CDM}}(1+z)^4}},
\end{equation}
where $\Omega_{\Lambda}=1-\Omega_{m0}^{\mathrm{\Lambda CDM}} -\Omega_{r0}^{\mathrm{\Lambda CDM}}$, and $\Omega_{m0}^{\mathrm{\Lambda CDM}}$, $\Omega_{r0}^{\mathrm{\Lambda CDM}}$ correspond to matter and radiation density parameter values  respectively for $\Lambda$CDM. We use the corresponding values reported in \cite{Planck:2018}. In this diagnostic, if $\mathbf{\mathbb{H}}0(z)$ is not a constant, then it suggests that the null model ($\Lambda$CDM) needs modification.

\begin{figure}
    \centering
    \includegraphics[width=0.45\textwidth]{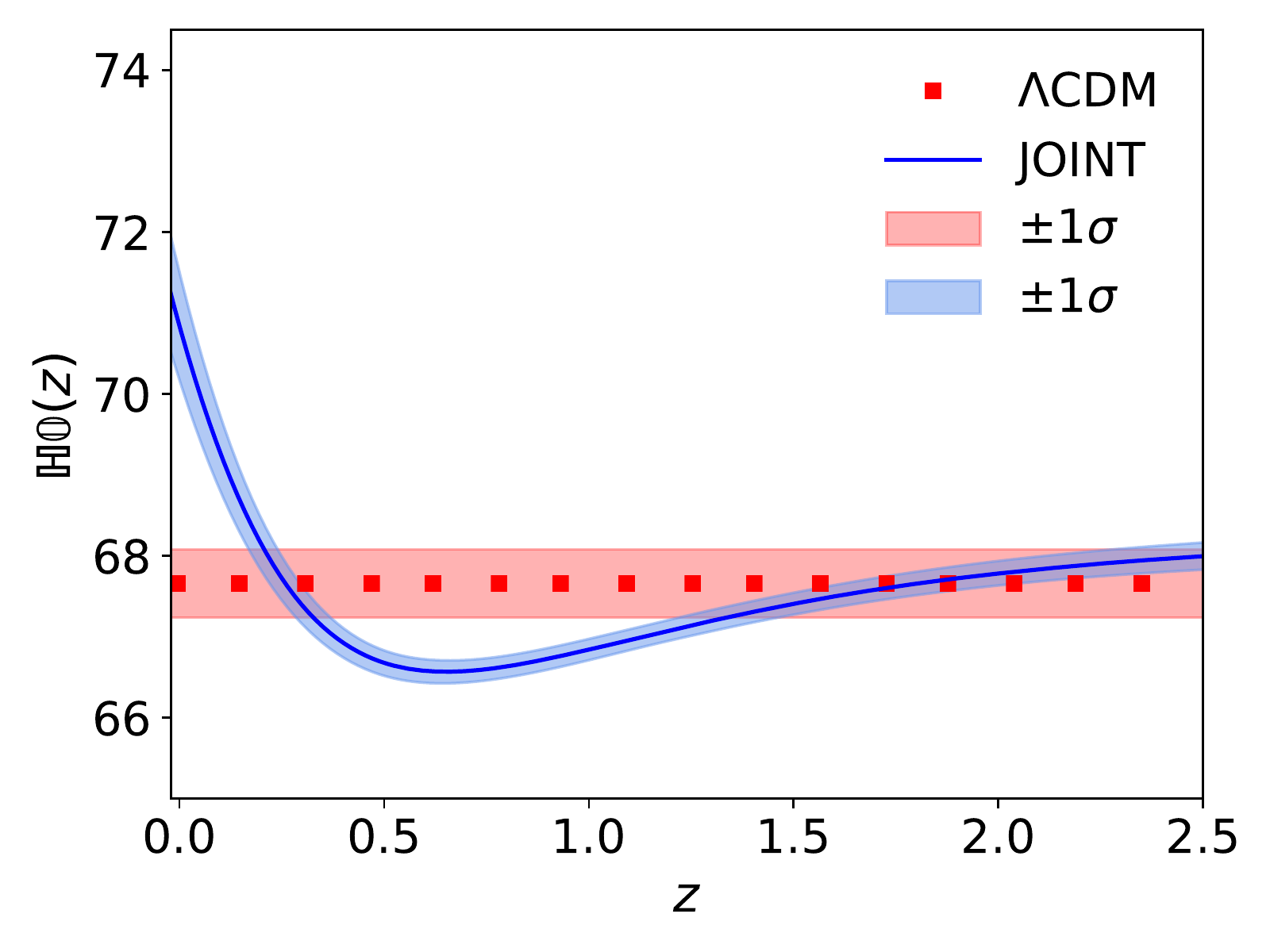}
    \caption{$\mathbf{\mathbb{H}}0(z)$ diagnostic for GZ cosmology in terms of the redshift $z$. Here we present the behavior of $\mathbb{H}0(z)$ under the joint analysis. The red squares represent the results for $\Lambda$CDM cosmology assuming $h=0.6766$ and $\Omega_{0m}=0.3111$ according to \cite{Planck:2018}. Notice that at high redshift, $\mathbf{\mathbb{H}}0(z)$ goes to Planck value whereas that at late times it presents a trend to go to our prior value of $H_0$ obtained by SH0ES collaboration \cite{Riess:2019cxk}.}
    \label{fig:H0diagnostic}
\end{figure}

Figure \ref{fig:H0diagnostic} shows the reconstruction of the $\mathbf{\mathbb{H}}0(z)$ diagnostic for the GZ cosmology and its error band at $1\sigma$ CL. Horizontal square points represents the CMB Planck value of $H_0$ for $\Lambda$CDM \cite{Planck:2018} and horizontal band is its uncertainty at $1\sigma$. It is interesting to observe that in the past (high redshift), $\mathbf{\mathbb{H}}0(z)$ goes to Planck value and  it presents a trend to go to our prior value of $H_0$ obtained by SH0ES collaboration \cite{Riess:2019cxk} at late times. This result is supported by the Bayesian evidence for CMB data in which the GZ model presents a strong evidence when a value of $h=0.742^{+0.017}_{-0.017}$ is obtained for CMB.

\section{Summary and Discussions} \label{SD}

We analyzed a model of the Universe which include a DE component characterized with an exponential EoS added to DM, baryons and radiation, namely GZ cosmology. This model was confronted with several cosmological datasets as CMB, CC, BAO, HIIG and SNIa, and also a joint analysis. Our results are consistent within $2\sigma$ with those reported in \cite{Gong:2005,Yang:2019}. Furthermore, the Bayes factor is estimated to compare statistically both GZ model and $\Lambda$CDM and shows that the latter was preferred by the Joint analysis but GZ cosmology is preferred when the data is separated, in particular, for CMB and the combinations CMB+BAO and CMB+SNIa. Based on the Joint constraint results, we reconstruct the deceleration and jerk parameters as shown in Fig. \ref{fig:Hz_and_qz} and we obtain current values $q_0=-0.789^{+0.034}_{-0.036}$ and $j_0=1.779^{+0.130}_{-0.119}$ respectively. Although both quantities are deviated more than $3\sigma$ from the $\Lambda$CDM values, the deceleration acceleration transition redshifts for both models are in agreement within $1\sigma$.  Additionally, an interesting difference between both models is also presented in these cosmographic parameters $q(z)$ and $j(z)$, observing oscillations at the future as shown in Fig. \ref{fig:Hz_and_qz} as happens in other models with parametric EoS presented in the literature \cite{Magana:2014voa}. In contrast to these models, GZ cosmology presents only an accelerated phase around $-0.7\lesssim z\lesssim 0.7$, going to a state similar to matter dominated era as $z\to-1$.  Our constraints show that $\omega_0=-1.202^{+0.027}_{-0.026}$ (joint) pointing a phantom DE behavior, however, in this case the final state of the Universe ($z\to-1$) is a decelerated state instead of an accelerated phase and even without a future singularity (Big Rip) commonly produced in the phantom DE scenarios. This kind of behavior is also presented in models as Parker Vacuum Metamorphosis \cite{Valentino:2008tsh0}.

On the other hand, we report the age of the Universe as $t_U=13.788^{+0.019}_{-0.019}$ Gyrs which is consistent with the standard cosmology within $1\sigma$ and is $2\sigma$ greater than the age of the oldest globular clusters \cite{Valcin:2021}.  Although we find slightly lower values of the Universe age with respect to the Planck 2018 value, they are in agreement within $1\sigma$.

We also confront GZ model and $\Lambda$CDM using the $\mathbf{\mathbb{H}}0(z)$ diagnostic and also based on results of the Bayesian statistical test for CMB data, we find GZ cosmology is an interesting candidate to alleviate the $H_0$ tension between early and late time estimation. Based on joint analysis, we find a $h$ value deviated $3.02\sigma$ from SH0ES and a distance of $4\sigma$ from the Planck 2018 value. However, when CMB and SNIa data are combined, we find a $1.4 \sigma$ from Planck value and about $4.5\sigma$ from SH0ES. For the other analyzed data, we find a distance up to $4.1\sigma$ from Planck value but $1\sigma$ from SH0ES value. The $H_0$ tension is reduced from $\sim 4.8\sigma$ obtained between the Planck value (based on $\Lambda$CDM) and SH0ES value to around $3\sigma-4\sigma$ for the joint analysis.

Moreover, we also estimate the quantities $\sigma_8$ and $S_8$.  We find that our results for $\sigma_8$ are consistent within $2\sigma$ with values reported by \cite{Yang:2019} when a combination of CMB Planck 2015, CC, BAO, JLA data are used and  are within $1.2\sigma$ from the values reported in a previous analysis for GZ model \cite{Gong:2005} using WMAP+SNIa+SDSS. Finally, we find $S_8$ values deviated up to $1.80\sigma$ with the value reported by KiDS-1000 \cite{Heymans_2021}. In particular we find that GZ model could alleviate the $S_8$ tension reducing from $2.65\sigma$ to $1.38\sigma$ for our joint analysis.

\begin{acknowledgements}
We thank anonymous referees for thoughtful remarks and suggestions. A.H.A. thank to the support from Luis Aguilar, 
Alejandro de Le\'on, Carlos Flores, and Jair Garc\'ia of the Laboratorio 
Nacional de Visualizaci\'on Cient\'ifica Avanzada. M.A.G.-A. acknowledges support from c\'atedra Marcos Moshinsky (MM), Universidad Iberoamericana for support with the SNI grant and the numerical analysis was also carried out by {\it Numerical Integration for Cosmological Theory and Experiments in High-energy Astrophysics} (Nicte Ha) cluster at IBERO University, acquired through c\'atedra MM support. J.M. acknowledges the support from  ANID REDES 190147. A.H.A, M.A.G.-A. and J.M. acknowledge partial support from project ANID Vinculaci\'on Internacional FOVI220144.
\end{acknowledgements}

\bibliography{main}

\begin{thebibliography}{111}%
\makeatletter
\providecommand \@ifxundefined [1]{%
 \@ifx{#1\undefined}
}%
\providecommand \@ifnum [1]{%
 \ifnum #1\expandafter \@firstoftwo
 \else \expandafter \@secondoftwo
 \fi
}%
\providecommand \@ifx [1]{%
 \ifx #1\expandafter \@firstoftwo
 \else \expandafter \@secondoftwo
 \fi
}%
\providecommand \natexlab [1]{#1}%
\providecommand \enquote  [1]{``#1''}%
\providecommand \bibnamefont  [1]{#1}%
\providecommand \bibfnamefont [1]{#1}%
\providecommand \citenamefont [1]{#1}%
\providecommand \href@noop [0]{\@secondoftwo}%
\providecommand \href [0]{\begingroup \@sanitize@url \@href}%
\providecommand \@href[1]{\@@startlink{#1}\@@href}%
\providecommand \@@href[1]{\endgroup#1\@@endlink}%
\providecommand \@sanitize@url [0]{\catcode `\\12\catcode `\$12\catcode
  `\&12\catcode `\#12\catcode `\^12\catcode `\_12\catcode `\%12\relax}%
\providecommand \@@startlink[1]{}%
\providecommand \@@endlink[0]{}%
\providecommand \url  [0]{\begingroup\@sanitize@url \@url }%
\providecommand \@url [1]{\endgroup\@href {#1}{\urlprefix }}%
\providecommand \urlprefix  [0]{URL }%
\providecommand \Eprint [0]{\href }%
\providecommand \doibase [0]{http://dx.doi.org/}%
\providecommand \selectlanguage [0]{\@gobble}%
\providecommand \bibinfo  [0]{\@secondoftwo}%
\providecommand \bibfield  [0]{\@secondoftwo}%
\providecommand \translation [1]{[#1]}%
\providecommand \BibitemOpen [0]{}%
\providecommand \bibitemStop [0]{}%
\providecommand \bibitemNoStop [0]{.\EOS\space}%
\providecommand \EOS [0]{\spacefactor3000\relax}%
\providecommand \BibitemShut  [1]{\csname bibitem#1\endcsname}%
\let\auto@bib@innerbib\@empty
\bibitem [{\citenamefont {Riess}\ \emph {et~al.}(1998)\citenamefont {Riess},
  \citenamefont {Filippenko}, \citenamefont {Challis}, \citenamefont
  {Clocchiatti}, \citenamefont {Diercks} \emph {et~al.}}]{Riess:1998}%
  \BibitemOpen
  \bibfield  {author} {\bibinfo {author} {\bibfnamefont {A.~G.}\ \bibnamefont
  {Riess}}, \bibinfo {author} {\bibfnamefont {A.~V.}\ \bibnamefont
  {Filippenko}}, \bibinfo {author} {\bibfnamefont {P.}~\bibnamefont {Challis}},
  \bibinfo {author} {\bibfnamefont {A.}~\bibnamefont {Clocchiatti}}, \bibinfo
  {author} {\bibfnamefont {A.}~\bibnamefont {Diercks}},  \emph {et~al.},\
  }\href {http://stacks.iop.org/1538-3881/116/i=3/a=1009} {\bibfield  {journal}
  {\bibinfo  {journal} {The Astronomical Journal}\ }\textbf {\bibinfo {volume}
  {116}},\ \bibinfo {pages} {1009} (\bibinfo {year} {1998})}\BibitemShut
  {NoStop}%
\bibitem [{\citenamefont {Perlmutter}\ \emph {et~al.}(1999)\citenamefont
  {Perlmutter}, \citenamefont {Aldering}, \citenamefont {Goldhaber},
  \citenamefont {Knop}, \citenamefont {Nugent}, \citenamefont {others},\ and\
  \citenamefont {Project}}]{Perlmutter:1999}%
  \BibitemOpen
  \bibfield  {author} {\bibinfo {author} {\bibfnamefont {S.}~\bibnamefont
  {Perlmutter}}, \bibinfo {author} {\bibfnamefont {G.}~\bibnamefont
  {Aldering}}, \bibinfo {author} {\bibfnamefont {G.}~\bibnamefont {Goldhaber}},
  \bibinfo {author} {\bibfnamefont {R.~A.}\ \bibnamefont {Knop}}, \bibinfo
  {author} {\bibfnamefont {P.}~\bibnamefont {Nugent}}, \bibinfo {author}
  {\bibnamefont {others}}, \ and\ \bibinfo {author} {\bibfnamefont {T.~S.~C.}\
  \bibnamefont {Project}},\ }\href
  {http://stacks.iop.org/0004-637X/517/i=2/a=565} {\bibfield  {journal}
  {\bibinfo  {journal} {The Astrophysical Journal}\ }\textbf {\bibinfo {volume}
  {517}},\ \bibinfo {pages} {565} (\bibinfo {year} {1999})}\BibitemShut
  {NoStop}%
\bibitem [{\citenamefont {Aghanim}\ and\ \citenamefont {et.
  al.}(2020)}]{Planck:2018}%
  \BibitemOpen
  \bibfield  {author} {\bibinfo {author} {\bibfnamefont {N.}~\bibnamefont
  {Aghanim}}\ and\ \bibinfo {author} {\bibnamefont {et. al.}},\ }\href
  {\doibase 10.1051/0004-6361/201833910} {\bibfield  {journal} {\bibinfo
  {journal} {A\&A}\ }\textbf {\bibinfo {volume} {641}},\ \bibinfo {pages} {A6}
  (\bibinfo {year} {2020})}\BibitemShut {NoStop}%
\bibitem [{\citenamefont {Carroll}(2001)}]{Carroll:2000}%
  \BibitemOpen
  \bibfield  {author} {\bibinfo {author} {\bibfnamefont {S.~M.}\ \bibnamefont
  {Carroll}},\ }\href {\doibase 10.12942/lrr-2001-1} {\bibfield  {journal}
  {\bibinfo  {journal} {Living Rev. Rel.}\ }\textbf {\bibinfo {volume} {4}},\
  \bibinfo {pages} {1} (\bibinfo {year} {2001})},\ \Eprint
  {http://arxiv.org/abs/astro-ph/0004075} {arXiv:astro-ph/0004075 [astro-ph]}
  \BibitemShut {NoStop}%
\bibitem [{\citenamefont {Zeldovich}(1968)}]{Zeldovich}%
  \BibitemOpen
  \bibfield  {author} {\bibinfo {author} {\bibfnamefont {Y.~B.}\ \bibnamefont
  {Zeldovich}},\ }\href@noop {} {\bibfield  {journal} {\bibinfo  {journal}
  {Soviet Physics Uspekhi}\ }\textbf {\bibinfo {volume} {11}} (\bibinfo {year}
  {1968})}\BibitemShut {NoStop}%
\bibitem [{\citenamefont {Weinberg}(1989)}]{Weinberg}%
  \BibitemOpen
  \bibfield  {author} {\bibinfo {author} {\bibfnamefont {S.}~\bibnamefont
  {Weinberg}},\ }\href@noop {} {\bibfield  {journal} {\bibinfo  {journal}
  {Reviews of Modern Physics}\ }\textbf {\bibinfo {volume} {61}} (\bibinfo
  {year} {1989})}\BibitemShut {NoStop}%
\bibitem [{\citenamefont {Zhao}\ \emph {et~al.}(2017)\citenamefont {Zhao} \emph
  {et~al.}}]{Zhao:2017cud}%
  \BibitemOpen
  \bibfield  {author} {\bibinfo {author} {\bibfnamefont {G.-B.}\ \bibnamefont
  {Zhao}} \emph {et~al.},\ }\href {\doibase 10.1038/s41550-017-0216-z}
  {\bibfield  {journal} {\bibinfo  {journal} {Nature Astron.}\ }\textbf
  {\bibinfo {volume} {1}},\ \bibinfo {pages} {627} (\bibinfo {year} {2017})},\
  \Eprint {http://arxiv.org/abs/1701.08165} {arXiv:1701.08165 [astro-ph.CO]}
  \BibitemShut {NoStop}%
\bibitem [{\citenamefont {Yang}\ \emph
  {et~al.}(2019{\natexlab{a}})\citenamefont {Yang}, \citenamefont {Pan},
  \citenamefont {Di~Valentino}, \citenamefont {Saridakis},\ and\ \citenamefont
  {Chakraborty}}]{Yang:2018qmz}%
  \BibitemOpen
  \bibfield  {author} {\bibinfo {author} {\bibfnamefont {W.}~\bibnamefont
  {Yang}}, \bibinfo {author} {\bibfnamefont {S.}~\bibnamefont {Pan}}, \bibinfo
  {author} {\bibfnamefont {E.}~\bibnamefont {Di~Valentino}}, \bibinfo {author}
  {\bibfnamefont {E.~N.}\ \bibnamefont {Saridakis}}, \ and\ \bibinfo {author}
  {\bibfnamefont {S.}~\bibnamefont {Chakraborty}},\ }\href {\doibase
  10.1103/PhysRevD.99.043543} {\bibfield  {journal} {\bibinfo  {journal} {Phys.
  Rev. D}\ }\textbf {\bibinfo {volume} {99}},\ \bibinfo {pages} {043543}
  (\bibinfo {year} {2019}{\natexlab{a}})},\ \Eprint
  {http://arxiv.org/abs/1810.05141} {arXiv:1810.05141 [astro-ph.CO]}
  \BibitemShut {NoStop}%
\bibitem [{\citenamefont {Akarsu}\ \emph {et~al.}(2020)\citenamefont {Akarsu},
  \citenamefont {Barrow}, \citenamefont {Escamilla},\ and\ \citenamefont
  {Vazquez}}]{Akarsu:2019hmw}%
  \BibitemOpen
  \bibfield  {author} {\bibinfo {author} {\bibfnamefont {O.}~\bibnamefont
  {Akarsu}}, \bibinfo {author} {\bibfnamefont {J.~D.}\ \bibnamefont {Barrow}},
  \bibinfo {author} {\bibfnamefont {L.~A.}\ \bibnamefont {Escamilla}}, \ and\
  \bibinfo {author} {\bibfnamefont {J.~A.}\ \bibnamefont {Vazquez}},\ }\href
  {\doibase 10.1103/PhysRevD.101.063528} {\bibfield  {journal} {\bibinfo
  {journal} {Phys. Rev. D}\ }\textbf {\bibinfo {volume} {101}},\ \bibinfo
  {pages} {063528} (\bibinfo {year} {2020})},\ \Eprint
  {http://arxiv.org/abs/1912.08751} {arXiv:1912.08751 [astro-ph.CO]}
  \BibitemShut {NoStop}%
\bibitem [{\citenamefont {Di~Valentino}\ \emph
  {et~al.}(2021{\natexlab{a}})\citenamefont {Di~Valentino}, \citenamefont
  {Mukherjee},\ and\ \citenamefont {Sen}}]{DiValentino:2020naf}%
  \BibitemOpen
  \bibfield  {author} {\bibinfo {author} {\bibfnamefont {E.}~\bibnamefont
  {Di~Valentino}}, \bibinfo {author} {\bibfnamefont {A.}~\bibnamefont
  {Mukherjee}}, \ and\ \bibinfo {author} {\bibfnamefont {A.~A.}\ \bibnamefont
  {Sen}},\ }\href {\doibase 10.3390/e23040404} {\bibfield  {journal} {\bibinfo
  {journal} {Entropy}\ }\textbf {\bibinfo {volume} {23}},\ \bibinfo {pages}
  {404} (\bibinfo {year} {2021}{\natexlab{a}})},\ \Eprint
  {http://arxiv.org/abs/2005.12587} {arXiv:2005.12587 [astro-ph.CO]}
  \BibitemShut {NoStop}%
\bibitem [{\citenamefont {Chudaykin}\ \emph {et~al.}(2021)\citenamefont
  {Chudaykin}, \citenamefont {Dolgikh},\ and\ \citenamefont
  {Ivanov}}]{Chudaykin:2020ghx}%
  \BibitemOpen
  \bibfield  {author} {\bibinfo {author} {\bibfnamefont {A.}~\bibnamefont
  {Chudaykin}}, \bibinfo {author} {\bibfnamefont {K.}~\bibnamefont {Dolgikh}},
  \ and\ \bibinfo {author} {\bibfnamefont {M.~M.}\ \bibnamefont {Ivanov}},\
  }\href {\doibase 10.1103/PhysRevD.103.023507} {\bibfield  {journal} {\bibinfo
   {journal} {Phys. Rev. D}\ }\textbf {\bibinfo {volume} {103}},\ \bibinfo
  {pages} {023507} (\bibinfo {year} {2021})},\ \Eprint
  {http://arxiv.org/abs/2009.10106} {arXiv:2009.10106 [astro-ph.CO]}
  \BibitemShut {NoStop}%
\bibitem [{\citenamefont {Yang}\ \emph {et~al.}(2021)\citenamefont {Yang},
  \citenamefont {Di~Valentino}, \citenamefont {Pan}, \citenamefont {Wu},\ and\
  \citenamefont {Lu}}]{Yang:2021flj}%
  \BibitemOpen
  \bibfield  {author} {\bibinfo {author} {\bibfnamefont {W.}~\bibnamefont
  {Yang}}, \bibinfo {author} {\bibfnamefont {E.}~\bibnamefont {Di~Valentino}},
  \bibinfo {author} {\bibfnamefont {S.}~\bibnamefont {Pan}}, \bibinfo {author}
  {\bibfnamefont {Y.}~\bibnamefont {Wu}}, \ and\ \bibinfo {author}
  {\bibfnamefont {J.}~\bibnamefont {Lu}},\ }\href {\doibase
  10.1093/mnras/staa3914} {\bibfield  {journal} {\bibinfo  {journal} {Mon. Not.
  Roy. Astron. Soc.}\ }\textbf {\bibinfo {volume} {501}},\ \bibinfo {pages}
  {5845} (\bibinfo {year} {2021})},\ \Eprint {http://arxiv.org/abs/2101.02168}
  {arXiv:2101.02168 [astro-ph.CO]} \BibitemShut {NoStop}%
\bibitem [{\citenamefont {Colg\'ain}\ \emph {et~al.}(2021)\citenamefont
  {Colg\'ain}, \citenamefont {Sheikh-Jabbari},\ and\ \citenamefont
  {Yin}}]{Colgain:2021pmf}%
  \BibitemOpen
  \bibfield  {author} {\bibinfo {author} {\bibfnamefont {E.~O.}\ \bibnamefont
  {Colg\'ain}}, \bibinfo {author} {\bibfnamefont {M.~M.}\ \bibnamefont
  {Sheikh-Jabbari}}, \ and\ \bibinfo {author} {\bibfnamefont {L.}~\bibnamefont
  {Yin}},\ }\href {\doibase 10.1103/PhysRevD.104.023510} {\bibfield  {journal}
  {\bibinfo  {journal} {Phys. Rev. D}\ }\textbf {\bibinfo {volume} {104}},\
  \bibinfo {pages} {023510} (\bibinfo {year} {2021})},\ \Eprint
  {http://arxiv.org/abs/2104.01930} {arXiv:2104.01930 [astro-ph.CO]}
  \BibitemShut {NoStop}%
\bibitem [{\citenamefont {Escamilla}\ and\ \citenamefont
  {Vazquez}(2021)}]{Escamilla:2021uoj}%
  \BibitemOpen
  \bibfield  {author} {\bibinfo {author} {\bibfnamefont {L.~A.}\ \bibnamefont
  {Escamilla}}\ and\ \bibinfo {author} {\bibfnamefont {J.~A.}\ \bibnamefont
  {Vazquez}},\ }\href@noop {} {\  (\bibinfo {year} {2021})},\ \Eprint
  {http://arxiv.org/abs/2111.10457} {arXiv:2111.10457 [astro-ph.CO]}
  \BibitemShut {NoStop}%
\bibitem [{\citenamefont {Sharma}\ \emph {et~al.}(2022)\citenamefont {Sharma},
  \citenamefont {Pandey},\ and\ \citenamefont {Das}}]{Sharma:2022ifr}%
  \BibitemOpen
  \bibfield  {author} {\bibinfo {author} {\bibfnamefont {R.~K.}\ \bibnamefont
  {Sharma}}, \bibinfo {author} {\bibfnamefont {K.~L.}\ \bibnamefont {Pandey}},
  \ and\ \bibinfo {author} {\bibfnamefont {S.}~\bibnamefont {Das}},\ }\href
  {\doibase 10.3847/1538-4357/ac7a33} {\bibfield  {journal} {\bibinfo
  {journal} {Astrophys. J.}\ }\textbf {\bibinfo {volume} {934}},\ \bibinfo
  {pages} {113} (\bibinfo {year} {2022})},\ \Eprint
  {http://arxiv.org/abs/2202.01749} {arXiv:2202.01749 [astro-ph.CO]}
  \BibitemShut {NoStop}%
\bibitem [{\citenamefont {Riess}\ \emph {et~al.}(2019)\citenamefont {Riess},
  \citenamefont {Casertano}, \citenamefont {Yuan}, \citenamefont {Macri},\ and\
  \citenamefont {Scolnic}}]{Riess:2019cxk}%
  \BibitemOpen
  \bibfield  {author} {\bibinfo {author} {\bibfnamefont {A.~G.}\ \bibnamefont
  {Riess}}, \bibinfo {author} {\bibfnamefont {S.}~\bibnamefont {Casertano}},
  \bibinfo {author} {\bibfnamefont {W.}~\bibnamefont {Yuan}}, \bibinfo {author}
  {\bibfnamefont {L.~M.}\ \bibnamefont {Macri}}, \ and\ \bibinfo {author}
  {\bibfnamefont {D.}~\bibnamefont {Scolnic}},\ }\href {\doibase
  10.3847/1538-4357/ab1422} {\bibfield  {journal} {\bibinfo  {journal}
  {Astrophys. J.}\ }\textbf {\bibinfo {volume} {876}},\ \bibinfo {pages} {85}
  (\bibinfo {year} {2019})},\ \Eprint {http://arxiv.org/abs/1903.07603}
  {arXiv:1903.07603 [astro-ph.CO]} \BibitemShut {NoStop}%
\bibitem [{\citenamefont {Efstathiou}(2021)}]{Efstathiou:2021ocp}%
  \BibitemOpen
  \bibfield  {author} {\bibinfo {author} {\bibfnamefont {G.}~\bibnamefont
  {Efstathiou}},\ }\href {\doibase 10.1093/mnras/stab1588} {\bibfield
  {journal} {\bibinfo  {journal} {Mon. Not. Roy. Astron. Soc.}\ }\textbf
  {\bibinfo {volume} {505}},\ \bibinfo {pages} {3866} (\bibinfo {year}
  {2021})},\ \Eprint {http://arxiv.org/abs/2103.08723} {arXiv:2103.08723
  [astro-ph.CO]} \BibitemShut {NoStop}%
\bibitem [{\citenamefont {Di~Valentino}\ \emph
  {et~al.}(2021{\natexlab{b}})\citenamefont {Di~Valentino}, \citenamefont
  {Mena}, \citenamefont {Pan}, \citenamefont {Visinelli}, \citenamefont {Yang},
  \citenamefont {Melchiorri}, \citenamefont {Mota}, \citenamefont {Riess},\
  and\ \citenamefont {Silk}}]{DiValentino:2021izs}%
  \BibitemOpen
  \bibfield  {author} {\bibinfo {author} {\bibfnamefont {E.}~\bibnamefont
  {Di~Valentino}}, \bibinfo {author} {\bibfnamefont {O.}~\bibnamefont {Mena}},
  \bibinfo {author} {\bibfnamefont {S.}~\bibnamefont {Pan}}, \bibinfo {author}
  {\bibfnamefont {L.}~\bibnamefont {Visinelli}}, \bibinfo {author}
  {\bibfnamefont {W.}~\bibnamefont {Yang}}, \bibinfo {author} {\bibfnamefont
  {A.}~\bibnamefont {Melchiorri}}, \bibinfo {author} {\bibfnamefont {D.~F.}\
  \bibnamefont {Mota}}, \bibinfo {author} {\bibfnamefont {A.~G.}\ \bibnamefont
  {Riess}}, \ and\ \bibinfo {author} {\bibfnamefont {J.}~\bibnamefont {Silk}},\
  }\href {\doibase 10.1088/1361-6382/ac086d} {\bibfield  {journal} {\bibinfo
  {journal} {Class. Quant. Grav.}\ }\textbf {\bibinfo {volume} {38}},\ \bibinfo
  {pages} {153001} (\bibinfo {year} {2021}{\natexlab{b}})},\ \Eprint
  {http://arxiv.org/abs/2103.01183} {arXiv:2103.01183 [astro-ph.CO]}
  \BibitemShut {NoStop}%
\bibitem [{\citenamefont {Motta}\ \emph {et~al.}(2021)\citenamefont {Motta},
  \citenamefont {Garc\'\i{}a-Aspeitia}, \citenamefont {Hern\'andez-Almada},
  \citenamefont {Maga\~na},\ and\ \citenamefont {Verdugo}}]{Motta:2021hvl}%
  \BibitemOpen
  \bibfield  {author} {\bibinfo {author} {\bibfnamefont {V.}~\bibnamefont
  {Motta}}, \bibinfo {author} {\bibfnamefont {M.~A.}\ \bibnamefont
  {Garc\'\i{}a-Aspeitia}}, \bibinfo {author} {\bibfnamefont {A.}~\bibnamefont
  {Hern\'andez-Almada}}, \bibinfo {author} {\bibfnamefont {J.}~\bibnamefont
  {Maga\~na}}, \ and\ \bibinfo {author} {\bibfnamefont {T.}~\bibnamefont
  {Verdugo}},\ }\href {\doibase 10.3390/universe7060163} {\bibfield  {journal}
  {\bibinfo  {journal} {Universe}\ }\textbf {\bibinfo {volume} {7}},\ \bibinfo
  {pages} {163} (\bibinfo {year} {2021})},\ \Eprint
  {http://arxiv.org/abs/2104.04642} {arXiv:2104.04642 [astro-ph.CO]}
  \BibitemShut {NoStop}%
\bibitem [{\citenamefont {Bamba}\ \emph {et~al.}(2012)\citenamefont {Bamba},
  \citenamefont {Capozziello}, \citenamefont {Nojiri},\ and\ \citenamefont
  {Odintsov}}]{Bamba_2012}%
  \BibitemOpen
  \bibfield  {author} {\bibinfo {author} {\bibfnamefont {K.}~\bibnamefont
  {Bamba}}, \bibinfo {author} {\bibfnamefont {S.}~\bibnamefont {Capozziello}},
  \bibinfo {author} {\bibfnamefont {S.}~\bibnamefont {Nojiri}}, \ and\ \bibinfo
  {author} {\bibfnamefont {S.~D.}\ \bibnamefont {Odintsov}},\ }\href {\doibase
  10.1007/s10509-012-1181-8} {\bibfield  {journal} {\bibinfo  {journal}
  {Astrophysics and Space Science}\ }\textbf {\bibinfo {volume} {342}},\
  \bibinfo {pages} {155} (\bibinfo {year} {2012})}\BibitemShut {NoStop}%
\bibitem [{\citenamefont {Di~Valentino}\ \emph
  {et~al.}(2021{\natexlab{c}})\citenamefont {Di~Valentino} \emph
  {et~al.}}]{DiValentino:2020vhf}%
  \BibitemOpen
  \bibfield  {author} {\bibinfo {author} {\bibfnamefont {E.}~\bibnamefont
  {Di~Valentino}} \emph {et~al.},\ }\href {\doibase
  10.1016/j.astropartphys.2021.102606} {\bibfield  {journal} {\bibinfo
  {journal} {Astropart. Phys.}\ }\textbf {\bibinfo {volume} {131}},\ \bibinfo
  {pages} {102606} (\bibinfo {year} {2021}{\natexlab{c}})},\ \Eprint
  {http://arxiv.org/abs/2008.11283} {arXiv:2008.11283 [astro-ph.CO]}
  \BibitemShut {NoStop}%
\bibitem [{\citenamefont {Di~Valentino}\ \emph
  {et~al.}(2021{\natexlab{d}})\citenamefont {Di~Valentino} \emph
  {et~al.}}]{DiValentino:2020zio}%
  \BibitemOpen
  \bibfield  {author} {\bibinfo {author} {\bibfnamefont {E.}~\bibnamefont
  {Di~Valentino}} \emph {et~al.},\ }\href {\doibase
  10.1016/j.astropartphys.2021.102605} {\bibfield  {journal} {\bibinfo
  {journal} {Astropart. Phys.}\ }\textbf {\bibinfo {volume} {131}},\ \bibinfo
  {pages} {102605} (\bibinfo {year} {2021}{\natexlab{d}})},\ \Eprint
  {http://arxiv.org/abs/2008.11284} {arXiv:2008.11284 [astro-ph.CO]}
  \BibitemShut {NoStop}%
\bibitem [{\citenamefont {Di~Valentino}\ \emph
  {et~al.}(2021{\natexlab{e}})\citenamefont {Di~Valentino} \emph
  {et~al.}}]{DiValentino:2020vvd}%
  \BibitemOpen
  \bibfield  {author} {\bibinfo {author} {\bibfnamefont {E.}~\bibnamefont
  {Di~Valentino}} \emph {et~al.},\ }\href {\doibase
  10.1016/j.astropartphys.2021.102604} {\bibfield  {journal} {\bibinfo
  {journal} {Astropart. Phys.}\ }\textbf {\bibinfo {volume} {131}},\ \bibinfo
  {pages} {102604} (\bibinfo {year} {2021}{\natexlab{e}})},\ \Eprint
  {http://arxiv.org/abs/2008.11285} {arXiv:2008.11285 [astro-ph.CO]}
  \BibitemShut {NoStop}%
\bibitem [{\citenamefont {Di~Valentino}\ \emph
  {et~al.}(2021{\natexlab{f}})\citenamefont {Di~Valentino} \emph
  {et~al.}}]{DiValentino:2020srs}%
  \BibitemOpen
  \bibfield  {author} {\bibinfo {author} {\bibfnamefont {E.}~\bibnamefont
  {Di~Valentino}} \emph {et~al.},\ }\href {\doibase
  10.1016/j.astropartphys.2021.102607} {\bibfield  {journal} {\bibinfo
  {journal} {Astropart. Phys.}\ }\textbf {\bibinfo {volume} {131}},\ \bibinfo
  {pages} {102607} (\bibinfo {year} {2021}{\natexlab{f}})},\ \Eprint
  {http://arxiv.org/abs/2008.11286} {arXiv:2008.11286 [astro-ph.CO]}
  \BibitemShut {NoStop}%
\bibitem [{\citenamefont {Perivolaropoulos}\ and\ \citenamefont
  {Skara}(2022)}]{Perivolaropoulos:2021jda}%
  \BibitemOpen
  \bibfield  {author} {\bibinfo {author} {\bibfnamefont {L.}~\bibnamefont
  {Perivolaropoulos}}\ and\ \bibinfo {author} {\bibfnamefont {F.}~\bibnamefont
  {Skara}},\ }\href {\doibase 10.1016/j.newar.2022.101659} {\bibfield
  {journal} {\bibinfo  {journal} {New Astron. Rev.}\ }\textbf {\bibinfo
  {volume} {95}},\ \bibinfo {pages} {101659} (\bibinfo {year} {2022})},\
  \Eprint {http://arxiv.org/abs/2105.05208} {arXiv:2105.05208 [astro-ph.CO]}
  \BibitemShut {NoStop}%
\bibitem [{\citenamefont {Abdalla}\ \emph {et~al.}(2022)\citenamefont {Abdalla}
  \emph {et~al.}}]{Abdalla:2022yfr}%
  \BibitemOpen
  \bibfield  {author} {\bibinfo {author} {\bibfnamefont {E.}~\bibnamefont
  {Abdalla}} \emph {et~al.},\ }\href {\doibase 10.1016/j.jheap.2022.04.002}
  {\bibfield  {journal} {\bibinfo  {journal} {JHEAp}\ }\textbf {\bibinfo
  {volume} {34}},\ \bibinfo {pages} {49} (\bibinfo {year} {2022})},\ \Eprint
  {http://arxiv.org/abs/2203.06142} {arXiv:2203.06142 [astro-ph.CO]}
  \BibitemShut {NoStop}%
\bibitem [{\citenamefont {Abbott}\ \emph {et~al.}(2022)\citenamefont {Abbott}
  \emph {et~al.}}]{DES:2021wwk}%
  \BibitemOpen
  \bibfield  {author} {\bibinfo {author} {\bibfnamefont {T.~M.~C.}\
  \bibnamefont {Abbott}} \emph {et~al.} (\bibinfo {collaboration} {DES}),\
  }\href {\doibase 10.1103/PhysRevD.105.023520} {\bibfield  {journal} {\bibinfo
   {journal} {Phys. Rev. D}\ }\textbf {\bibinfo {volume} {105}},\ \bibinfo
  {pages} {023520} (\bibinfo {year} {2022})},\ \Eprint
  {http://arxiv.org/abs/2105.13549} {arXiv:2105.13549 [astro-ph.CO]}
  \BibitemShut {NoStop}%
\bibitem [{\citenamefont {Asgari}\ \emph {et~al.}(2021)\citenamefont {Asgari}
  \emph {et~al.}}]{KiDS:2020suj}%
  \BibitemOpen
  \bibfield  {author} {\bibinfo {author} {\bibfnamefont {M.}~\bibnamefont
  {Asgari}} \emph {et~al.} (\bibinfo {collaboration} {KiDS}),\ }\href {\doibase
  10.1051/0004-6361/202039070} {\bibfield  {journal} {\bibinfo  {journal}
  {Astron. Astrophys.}\ }\textbf {\bibinfo {volume} {645}},\ \bibinfo {pages}
  {A104} (\bibinfo {year} {2021})},\ \Eprint {http://arxiv.org/abs/2007.15633}
  {arXiv:2007.15633 [astro-ph.CO]} \BibitemShut {NoStop}%
\bibitem [{\citenamefont {Heymans}\ \emph {et~al.}(2021)\citenamefont
  {Heymans}, \citenamefont {Tröster}, \citenamefont {Asgari}, \citenamefont
  {Blake}, \citenamefont {Hildebrandt}, \citenamefont {Joachimi}, \citenamefont
  {Kuijken}, \citenamefont {Lin}, \citenamefont {S{\'{a}}nchez}, \citenamefont
  {van~den Busch}, \citenamefont {Wright}, \citenamefont {Amon}, \citenamefont
  {Bilicki}, \citenamefont {de~Jong}, \citenamefont {Crocce}, \citenamefont
  {Dvornik}, \citenamefont {Erben}, \citenamefont {Fortuna}, \citenamefont
  {Getman}, \citenamefont {Giblin}, \citenamefont {Glazebrook}, \citenamefont
  {Hoekstra}, \citenamefont {Joudaki}, \citenamefont {Kannawadi}, \citenamefont
  {Köhlinger}, \citenamefont {Lidman}, \citenamefont {Miller}, \citenamefont
  {Napolitano}, \citenamefont {Parkinson}, \citenamefont {Schneider},
  \citenamefont {Shan}, \citenamefont {Valentijn}, \citenamefont {Kleijn},\
  and\ \citenamefont {Wolf}}]{Heymans_2021}%
  \BibitemOpen
  \bibfield  {author} {\bibinfo {author} {\bibfnamefont {C.}~\bibnamefont
  {Heymans}}, \bibinfo {author} {\bibfnamefont {T.}~\bibnamefont {Tröster}},
  \bibinfo {author} {\bibfnamefont {M.}~\bibnamefont {Asgari}}, \bibinfo
  {author} {\bibfnamefont {C.}~\bibnamefont {Blake}}, \bibinfo {author}
  {\bibfnamefont {H.}~\bibnamefont {Hildebrandt}}, \bibinfo {author}
  {\bibfnamefont {B.}~\bibnamefont {Joachimi}}, \bibinfo {author}
  {\bibfnamefont {K.}~\bibnamefont {Kuijken}}, \bibinfo {author} {\bibfnamefont
  {C.-A.}\ \bibnamefont {Lin}}, \bibinfo {author} {\bibfnamefont {A.~G.}\
  \bibnamefont {S{\'{a}}nchez}}, \bibinfo {author} {\bibfnamefont {J.~L.}\
  \bibnamefont {van~den Busch}}, \bibinfo {author} {\bibfnamefont {A.~H.}\
  \bibnamefont {Wright}}, \bibinfo {author} {\bibfnamefont {A.}~\bibnamefont
  {Amon}}, \bibinfo {author} {\bibfnamefont {M.}~\bibnamefont {Bilicki}},
  \bibinfo {author} {\bibfnamefont {J.}~\bibnamefont {de~Jong}}, \bibinfo
  {author} {\bibfnamefont {M.}~\bibnamefont {Crocce}}, \bibinfo {author}
  {\bibfnamefont {A.}~\bibnamefont {Dvornik}}, \bibinfo {author} {\bibfnamefont
  {T.}~\bibnamefont {Erben}}, \bibinfo {author} {\bibfnamefont {M.~C.}\
  \bibnamefont {Fortuna}}, \bibinfo {author} {\bibfnamefont {F.}~\bibnamefont
  {Getman}}, \bibinfo {author} {\bibfnamefont {B.}~\bibnamefont {Giblin}},
  \bibinfo {author} {\bibfnamefont {K.}~\bibnamefont {Glazebrook}}, \bibinfo
  {author} {\bibfnamefont {H.}~\bibnamefont {Hoekstra}}, \bibinfo {author}
  {\bibfnamefont {S.}~\bibnamefont {Joudaki}}, \bibinfo {author} {\bibfnamefont
  {A.}~\bibnamefont {Kannawadi}}, \bibinfo {author} {\bibfnamefont
  {F.}~\bibnamefont {Köhlinger}}, \bibinfo {author} {\bibfnamefont
  {C.}~\bibnamefont {Lidman}}, \bibinfo {author} {\bibfnamefont
  {L.}~\bibnamefont {Miller}}, \bibinfo {author} {\bibfnamefont {N.~R.}\
  \bibnamefont {Napolitano}}, \bibinfo {author} {\bibfnamefont
  {D.}~\bibnamefont {Parkinson}}, \bibinfo {author} {\bibfnamefont
  {P.}~\bibnamefont {Schneider}}, \bibinfo {author} {\bibfnamefont
  {H.}~\bibnamefont {Shan}}, \bibinfo {author} {\bibfnamefont {E.~A.}\
  \bibnamefont {Valentijn}}, \bibinfo {author} {\bibfnamefont {G.~V.}\
  \bibnamefont {Kleijn}}, \ and\ \bibinfo {author} {\bibfnamefont
  {C.}~\bibnamefont {Wolf}},\ }\href {\doibase 10.1051/0004-6361/202039063}
  {\bibfield  {journal} {\bibinfo  {journal} {Astronomy {\&} Astrophysics}\
  }\textbf {\bibinfo {volume} {646}},\ \bibinfo {pages} {A140} (\bibinfo {year}
  {2021})}\BibitemShut {NoStop}%
\bibitem [{\citenamefont {{Di Valentino}}\ \emph {et~al.}(2021)\citenamefont
  {{Di Valentino}}, \citenamefont {Anchordoqui}, \citenamefont {Özgür
  Akarsu}, \citenamefont {Ali-Haimoud}, \citenamefont {Amendola}, \citenamefont
  {Arendse}, \citenamefont {Asgari}, \citenamefont {Ballardini}, \citenamefont
  {Basilakos}, \citenamefont {Battistelli}, \citenamefont {Benetti},
  \citenamefont {Birrer}, \citenamefont {Bouchet}, \citenamefont {Bruni},
  \citenamefont {Calabrese}, \citenamefont {Camarena}, \citenamefont
  {Capozziello}, \citenamefont {Chen}, \citenamefont {Chluba}, \citenamefont
  {Chudaykin}, \citenamefont {Colgáin}, \citenamefont {Cyr-Racine},
  \citenamefont {{de Bernardis}}, \citenamefont {{de Cruz Pérez}},
  \citenamefont {Delabrouille}, \citenamefont {Dunkley}, \citenamefont
  {Escamilla-Rivera}, \citenamefont {Ferté}, \citenamefont {Finelli},
  \citenamefont {Freedman}, \citenamefont {Frusciante}, \citenamefont
  {Giusarma}, \citenamefont {Gómez-Valent}, \citenamefont {Handley},
  \citenamefont {Harrison}, \citenamefont {Hart}, \citenamefont {Heavens},
  \citenamefont {Hildebrandt}, \citenamefont {Holz}, \citenamefont {Huterer},
  \citenamefont {Ivanov}, \citenamefont {Joudaki}, \citenamefont
  {Kamionkowski}, \citenamefont {Karwal}, \citenamefont {Knox}, \citenamefont
  {Kumar}, \citenamefont {Lamagna}, \citenamefont {Lesgourgues}, \citenamefont
  {Lucca}, \citenamefont {Marra}, \citenamefont {Masi}, \citenamefont
  {Matarrese}, \citenamefont {Mazumdar}, \citenamefont {Melchiorri},
  \citenamefont {Mena}, \citenamefont {Mersini-Houghton}, \citenamefont
  {Miranda}, \citenamefont {Moreno-Pulido}, \citenamefont {Mota}, \citenamefont
  {Muir}, \citenamefont {Mukherjee}, \citenamefont {Niedermann}, \citenamefont
  {Notari}, \citenamefont {Nunes}, \citenamefont {Pace}, \citenamefont
  {Paliathanasis}, \citenamefont {Palmese}, \citenamefont {Pan}, \citenamefont
  {Paoletti}, \citenamefont {Pettorino}, \citenamefont {Piacentini},
  \citenamefont {Poulin}, \citenamefont {Raveri}, \citenamefont {Riess},
  \citenamefont {Salzano}, \citenamefont {Saridakis}, \citenamefont {Sen},
  \citenamefont {Shafieloo}, \citenamefont {Shajib}, \citenamefont {Silk},
  \citenamefont {Silvestri}, \citenamefont {Sloth}, \citenamefont {Smith},
  \citenamefont {{Solà Peracaula}}, \citenamefont {{van de Bruck}},
  \citenamefont {Verde}, \citenamefont {Visinelli}, \citenamefont {Wandelt},
  \citenamefont {Wang}, \citenamefont {Wang}, \citenamefont {Yadav},\ and\
  \citenamefont {Yang}}]{intertwined:S8}%
  \BibitemOpen
  \bibfield  {author} {\bibinfo {author} {\bibfnamefont {E.}~\bibnamefont {{Di
  Valentino}}}, \bibinfo {author} {\bibfnamefont {L.~A.}\ \bibnamefont
  {Anchordoqui}}, \bibinfo {author} {\bibnamefont {Özgür Akarsu}}, \bibinfo
  {author} {\bibfnamefont {Y.}~\bibnamefont {Ali-Haimoud}}, \bibinfo {author}
  {\bibfnamefont {L.}~\bibnamefont {Amendola}}, \bibinfo {author}
  {\bibfnamefont {N.}~\bibnamefont {Arendse}}, \bibinfo {author} {\bibfnamefont
  {M.}~\bibnamefont {Asgari}}, \bibinfo {author} {\bibfnamefont
  {M.}~\bibnamefont {Ballardini}}, \bibinfo {author} {\bibfnamefont
  {S.}~\bibnamefont {Basilakos}}, \bibinfo {author} {\bibfnamefont
  {E.}~\bibnamefont {Battistelli}}, \bibinfo {author} {\bibfnamefont
  {M.}~\bibnamefont {Benetti}}, \bibinfo {author} {\bibfnamefont
  {S.}~\bibnamefont {Birrer}}, \bibinfo {author} {\bibfnamefont {F.~R.}\
  \bibnamefont {Bouchet}}, \bibinfo {author} {\bibfnamefont {M.}~\bibnamefont
  {Bruni}}, \bibinfo {author} {\bibfnamefont {E.}~\bibnamefont {Calabrese}},
  \bibinfo {author} {\bibfnamefont {D.}~\bibnamefont {Camarena}}, \bibinfo
  {author} {\bibfnamefont {S.}~\bibnamefont {Capozziello}}, \bibinfo {author}
  {\bibfnamefont {A.}~\bibnamefont {Chen}}, \bibinfo {author} {\bibfnamefont
  {J.}~\bibnamefont {Chluba}}, \bibinfo {author} {\bibfnamefont
  {A.}~\bibnamefont {Chudaykin}}, \bibinfo {author} {\bibfnamefont {E.~Ã.}\
  \bibnamefont {Colgáin}}, \bibinfo {author} {\bibfnamefont {F.-Y.}\
  \bibnamefont {Cyr-Racine}}, \bibinfo {author} {\bibfnamefont
  {P.}~\bibnamefont {{de Bernardis}}}, \bibinfo {author} {\bibfnamefont
  {J.}~\bibnamefont {{de Cruz Pérez}}}, \bibinfo {author} {\bibfnamefont
  {J.}~\bibnamefont {Delabrouille}}, \bibinfo {author} {\bibfnamefont
  {J.}~\bibnamefont {Dunkley}}, \bibinfo {author} {\bibfnamefont
  {C.}~\bibnamefont {Escamilla-Rivera}}, \bibinfo {author} {\bibfnamefont
  {A.}~\bibnamefont {Ferté}}, \bibinfo {author} {\bibfnamefont
  {F.}~\bibnamefont {Finelli}}, \bibinfo {author} {\bibfnamefont
  {W.}~\bibnamefont {Freedman}}, \bibinfo {author} {\bibfnamefont
  {N.}~\bibnamefont {Frusciante}}, \bibinfo {author} {\bibfnamefont
  {E.}~\bibnamefont {Giusarma}}, \bibinfo {author} {\bibfnamefont
  {A.}~\bibnamefont {Gómez-Valent}}, \bibinfo {author} {\bibfnamefont
  {W.}~\bibnamefont {Handley}}, \bibinfo {author} {\bibfnamefont
  {I.}~\bibnamefont {Harrison}}, \bibinfo {author} {\bibfnamefont
  {L.}~\bibnamefont {Hart}}, \bibinfo {author} {\bibfnamefont {A.}~\bibnamefont
  {Heavens}}, \bibinfo {author} {\bibfnamefont {H.}~\bibnamefont
  {Hildebrandt}}, \bibinfo {author} {\bibfnamefont {D.}~\bibnamefont {Holz}},
  \bibinfo {author} {\bibfnamefont {D.}~\bibnamefont {Huterer}}, \bibinfo
  {author} {\bibfnamefont {M.~M.}\ \bibnamefont {Ivanov}}, \bibinfo {author}
  {\bibfnamefont {S.}~\bibnamefont {Joudaki}}, \bibinfo {author} {\bibfnamefont
  {M.}~\bibnamefont {Kamionkowski}}, \bibinfo {author} {\bibfnamefont
  {T.}~\bibnamefont {Karwal}}, \bibinfo {author} {\bibfnamefont
  {L.}~\bibnamefont {Knox}}, \bibinfo {author} {\bibfnamefont {S.}~\bibnamefont
  {Kumar}}, \bibinfo {author} {\bibfnamefont {L.}~\bibnamefont {Lamagna}},
  \bibinfo {author} {\bibfnamefont {J.}~\bibnamefont {Lesgourgues}}, \bibinfo
  {author} {\bibfnamefont {M.}~\bibnamefont {Lucca}}, \bibinfo {author}
  {\bibfnamefont {V.}~\bibnamefont {Marra}}, \bibinfo {author} {\bibfnamefont
  {S.}~\bibnamefont {Masi}}, \bibinfo {author} {\bibfnamefont {S.}~\bibnamefont
  {Matarrese}}, \bibinfo {author} {\bibfnamefont {A.}~\bibnamefont {Mazumdar}},
  \bibinfo {author} {\bibfnamefont {A.}~\bibnamefont {Melchiorri}}, \bibinfo
  {author} {\bibfnamefont {O.}~\bibnamefont {Mena}}, \bibinfo {author}
  {\bibfnamefont {L.}~\bibnamefont {Mersini-Houghton}}, \bibinfo {author}
  {\bibfnamefont {V.}~\bibnamefont {Miranda}}, \bibinfo {author} {\bibfnamefont
  {C.}~\bibnamefont {Moreno-Pulido}}, \bibinfo {author} {\bibfnamefont {D.~F.}\
  \bibnamefont {Mota}}, \bibinfo {author} {\bibfnamefont {J.}~\bibnamefont
  {Muir}}, \bibinfo {author} {\bibfnamefont {A.}~\bibnamefont {Mukherjee}},
  \bibinfo {author} {\bibfnamefont {F.}~\bibnamefont {Niedermann}}, \bibinfo
  {author} {\bibfnamefont {A.}~\bibnamefont {Notari}}, \bibinfo {author}
  {\bibfnamefont {R.~C.}\ \bibnamefont {Nunes}}, \bibinfo {author}
  {\bibfnamefont {F.}~\bibnamefont {Pace}}, \bibinfo {author} {\bibfnamefont
  {A.}~\bibnamefont {Paliathanasis}}, \bibinfo {author} {\bibfnamefont
  {A.}~\bibnamefont {Palmese}}, \bibinfo {author} {\bibfnamefont
  {S.}~\bibnamefont {Pan}}, \bibinfo {author} {\bibfnamefont {D.}~\bibnamefont
  {Paoletti}}, \bibinfo {author} {\bibfnamefont {V.}~\bibnamefont {Pettorino}},
  \bibinfo {author} {\bibfnamefont {F.}~\bibnamefont {Piacentini}}, \bibinfo
  {author} {\bibfnamefont {V.}~\bibnamefont {Poulin}}, \bibinfo {author}
  {\bibfnamefont {M.}~\bibnamefont {Raveri}}, \bibinfo {author} {\bibfnamefont
  {A.~G.}\ \bibnamefont {Riess}}, \bibinfo {author} {\bibfnamefont
  {V.}~\bibnamefont {Salzano}}, \bibinfo {author} {\bibfnamefont {E.~N.}\
  \bibnamefont {Saridakis}}, \bibinfo {author} {\bibfnamefont {A.~A.}\
  \bibnamefont {Sen}}, \bibinfo {author} {\bibfnamefont {A.}~\bibnamefont
  {Shafieloo}}, \bibinfo {author} {\bibfnamefont {A.~J.}\ \bibnamefont
  {Shajib}}, \bibinfo {author} {\bibfnamefont {J.}~\bibnamefont {Silk}},
  \bibinfo {author} {\bibfnamefont {A.}~\bibnamefont {Silvestri}}, \bibinfo
  {author} {\bibfnamefont {M.~S.}\ \bibnamefont {Sloth}}, \bibinfo {author}
  {\bibfnamefont {T.~L.}\ \bibnamefont {Smith}}, \bibinfo {author}
  {\bibfnamefont {J.}~\bibnamefont {{Solà Peracaula}}}, \bibinfo {author}
  {\bibfnamefont {C.}~\bibnamefont {{van de Bruck}}}, \bibinfo {author}
  {\bibfnamefont {L.}~\bibnamefont {Verde}}, \bibinfo {author} {\bibfnamefont
  {L.}~\bibnamefont {Visinelli}}, \bibinfo {author} {\bibfnamefont {B.~D.}\
  \bibnamefont {Wandelt}}, \bibinfo {author} {\bibfnamefont {D.}~\bibnamefont
  {Wang}}, \bibinfo {author} {\bibfnamefont {J.-M.}\ \bibnamefont {Wang}},
  \bibinfo {author} {\bibfnamefont {A.~K.}\ \bibnamefont {Yadav}}, \ and\
  \bibinfo {author} {\bibfnamefont {W.}~\bibnamefont {Yang}},\ }\href {\doibase
  https://doi.org/10.1016/j.astropartphys.2021.102604} {\bibfield  {journal}
  {\bibinfo  {journal} {Astroparticle Physics}\ }\textbf {\bibinfo {volume}
  {131}},\ \bibinfo {pages} {102604} (\bibinfo {year} {2021})}\BibitemShut
  {NoStop}%
\bibitem [{\citenamefont {Garcia-Aspeitia}\ \emph {et~al.}(2018)\citenamefont
  {Garcia-Aspeitia}, \citenamefont {Hernandez-Almada}, \citenamefont
  {Maga\~na}, \citenamefont {Amante}, \citenamefont {Motta},\ and\
  \citenamefont {Mart\'\i{}nez-Robles}}]{Garcia-Aspeitia:2018fvw}%
  \BibitemOpen
  \bibfield  {author} {\bibinfo {author} {\bibfnamefont {M.~A.}\ \bibnamefont
  {Garcia-Aspeitia}}, \bibinfo {author} {\bibfnamefont {A.}~\bibnamefont
  {Hernandez-Almada}}, \bibinfo {author} {\bibfnamefont {J.}~\bibnamefont
  {Maga\~na}}, \bibinfo {author} {\bibfnamefont {M.~H.}\ \bibnamefont
  {Amante}}, \bibinfo {author} {\bibfnamefont {V.}~\bibnamefont {Motta}}, \
  and\ \bibinfo {author} {\bibfnamefont {C.}~\bibnamefont
  {Mart\'\i{}nez-Robles}},\ }\href {\doibase 10.1103/PhysRevD.97.101301}
  {\bibfield  {journal} {\bibinfo  {journal} {Phys. Rev. D}\ }\textbf {\bibinfo
  {volume} {97}},\ \bibinfo {pages} {101301} (\bibinfo {year} {2018})},\
  \Eprint {http://arxiv.org/abs/1804.05085} {arXiv:1804.05085 [gr-qc]}
  \BibitemShut {NoStop}%
\bibitem [{\citenamefont {Garc\'\i{}a-Aspeitia}\ \emph
  {et~al.}(2022)\citenamefont {Garc\'\i{}a-Aspeitia}, \citenamefont
  {Fernandez-Anaya}, \citenamefont {Hern\'andez-Almada}, \citenamefont {Leon},\
  and\ \citenamefont {Maga\~na}}]{Garcia-Aspeitia:2022uxz}%
  \BibitemOpen
  \bibfield  {author} {\bibinfo {author} {\bibfnamefont {M.~A.}\ \bibnamefont
  {Garc\'\i{}a-Aspeitia}}, \bibinfo {author} {\bibfnamefont {G.}~\bibnamefont
  {Fernandez-Anaya}}, \bibinfo {author} {\bibfnamefont {A.}~\bibnamefont
  {Hern\'andez-Almada}}, \bibinfo {author} {\bibfnamefont {G.}~\bibnamefont
  {Leon}}, \ and\ \bibinfo {author} {\bibfnamefont {J.}~\bibnamefont
  {Maga\~na}},\ }\href {\doibase 10.1093/mnras/stac3006} {\  (\bibinfo {year}
  {2022}),\ 10.1093/mnras/stac3006},\ \Eprint {http://arxiv.org/abs/2207.00878}
  {arXiv:2207.00878 [gr-qc]} \BibitemShut {NoStop}%
\bibitem [{\citenamefont {Chaplygin}(1904)}]{chaplygin}%
  \BibitemOpen
  \bibfield  {author} {\bibinfo {author} {\bibfnamefont {S.~A.}\ \bibnamefont
  {Chaplygin}},\ }\href@noop {} {\bibfield  {journal} {\bibinfo  {journal}
  {Sci. Mem. Moscow Univ. Math. Phys.}\ }\textbf {\bibinfo {volume} {21}}
  (\bibinfo {year} {1904})}\BibitemShut {NoStop}%
\bibitem [{\citenamefont {Hernandez-Almada}\ \emph {et~al.}(2019)\citenamefont
  {Hernandez-Almada}, \citenamefont {Magana}, \citenamefont {Garcia-Aspeitia},\
  and\ \citenamefont {Motta}}]{Hernandez-Almada:2018osh}%
  \BibitemOpen
  \bibfield  {author} {\bibinfo {author} {\bibfnamefont {A.}~\bibnamefont
  {Hernandez-Almada}}, \bibinfo {author} {\bibfnamefont {J.}~\bibnamefont
  {Magana}}, \bibinfo {author} {\bibfnamefont {M.~A.}\ \bibnamefont
  {Garcia-Aspeitia}}, \ and\ \bibinfo {author} {\bibfnamefont {V.}~\bibnamefont
  {Motta}},\ }\href {\doibase 10.1140/epjc/s10052-018-6521-6} {\bibfield
  {journal} {\bibinfo  {journal} {Eur. Phys. J. C}\ }\textbf {\bibinfo {volume}
  {79}},\ \bibinfo {pages} {12} (\bibinfo {year} {2019})},\ \Eprint
  {http://arxiv.org/abs/1805.07895} {arXiv:1805.07895 [astro-ph.CO]}
  \BibitemShut {NoStop}%
\bibitem [{\citenamefont {Hern\'andez-Almada}\ \emph
  {et~al.}(2020{\natexlab{a}})\citenamefont {Hern\'andez-Almada}, \citenamefont
  {Garc\'{\i}a-Aspeitia}, \citenamefont {Maga\~na},\ and\ \citenamefont
  {Motta}}]{Almada:2020}%
  \BibitemOpen
  \bibfield  {author} {\bibinfo {author} {\bibfnamefont {A.}~\bibnamefont
  {Hern\'andez-Almada}}, \bibinfo {author} {\bibfnamefont {M.~A.}\ \bibnamefont
  {Garc\'{\i}a-Aspeitia}}, \bibinfo {author} {\bibfnamefont {J.}~\bibnamefont
  {Maga\~na}}, \ and\ \bibinfo {author} {\bibfnamefont {V.}~\bibnamefont
  {Motta}},\ }\href {\doibase 10.1103/PhysRevD.101.063516} {\bibfield
  {journal} {\bibinfo  {journal} {Phys. Rev. D}\ }\textbf {\bibinfo {volume}
  {101}},\ \bibinfo {pages} {063516} (\bibinfo {year}
  {2020}{\natexlab{a}})}\BibitemShut {NoStop}%
\bibitem [{\citenamefont {Herrera-Zamorano}\ \emph {et~al.}(2020)\citenamefont
  {Herrera-Zamorano}, \citenamefont {Hern\'andez-Almada},\ and\ \citenamefont
  {Garc\'ia-Aspeitia}}]{Herrera:2020}%
  \BibitemOpen
  \bibfield  {author} {\bibinfo {author} {\bibfnamefont {L.}~\bibnamefont
  {Herrera-Zamorano}}, \bibinfo {author} {\bibfnamefont {A.}~\bibnamefont
  {Hern\'andez-Almada}}, \ and\ \bibinfo {author} {\bibfnamefont
  {M.}~\bibnamefont {Garc\'ia-Aspeitia}},\ }\href {\doibase
  10.1140/epjc/s10052-020-8225-y} {\bibfield  {journal} {\bibinfo  {journal}
  {Eur. Phys. J. C}\ }\textbf {\bibinfo {volume} {80}},\ \bibinfo {pages} {637}
  (\bibinfo {year} {2020})},\ \Eprint {http://arxiv.org/abs/2007.04507}
  {arXiv:2007.04507} \BibitemShut {NoStop}%
\bibitem [{\citenamefont {Li}\ and\ \citenamefont {Shafieloo}(2019)}]{Li_2019}%
  \BibitemOpen
  \bibfield  {author} {\bibinfo {author} {\bibfnamefont {X.}~\bibnamefont
  {Li}}\ and\ \bibinfo {author} {\bibfnamefont {A.}~\bibnamefont {Shafieloo}},\
  }\href {\doibase 10.3847/2041-8213/ab3e09} {\bibfield  {journal} {\bibinfo
  {journal} {The Astrophysical Journal}\ }\textbf {\bibinfo {volume} {883}},\
  \bibinfo {pages} {L3} (\bibinfo {year} {2019})}\BibitemShut {NoStop}%
\bibitem [{\citenamefont {Hern\'andez-Almada}\ \emph
  {et~al.}(2020{\natexlab{b}})\citenamefont {Hern\'andez-Almada}, \citenamefont
  {Leon}, \citenamefont {Maga\~na}, \citenamefont {Garc\'\i{}a-Aspeitia},\ and\
  \citenamefont {Motta}}]{Hernandez-Almada:2020uyr}%
  \BibitemOpen
  \bibfield  {author} {\bibinfo {author} {\bibfnamefont {A.}~\bibnamefont
  {Hern\'andez-Almada}}, \bibinfo {author} {\bibfnamefont {G.}~\bibnamefont
  {Leon}}, \bibinfo {author} {\bibfnamefont {J.}~\bibnamefont {Maga\~na}},
  \bibinfo {author} {\bibfnamefont {M.~A.}\ \bibnamefont
  {Garc\'\i{}a-Aspeitia}}, \ and\ \bibinfo {author} {\bibfnamefont
  {V.}~\bibnamefont {Motta}},\ }\href {\doibase 10.1093/mnras/staa2052}
  {\bibfield  {journal} {\bibinfo  {journal} {Mon. Not. Roy. Astron. Soc.}\
  }\textbf {\bibinfo {volume} {497}},\ \bibinfo {pages} {1590} (\bibinfo {year}
  {2020}{\natexlab{b}})},\ \Eprint {http://arxiv.org/abs/2002.12881}
  {arXiv:2002.12881 [astro-ph.CO]} \BibitemShut {NoStop}%
\bibitem [{\citenamefont {Chevallier}\ and\ \citenamefont
  {Polarski}(2001)}]{CPL:2001}%
  \BibitemOpen
  \bibfield  {author} {\bibinfo {author} {\bibfnamefont {M.}~\bibnamefont
  {Chevallier}}\ and\ \bibinfo {author} {\bibfnamefont {D.}~\bibnamefont
  {Polarski}},\ }\href {\doibase 10.1142/s0218271801000822} {\bibfield
  {journal} {\bibinfo  {journal} {International Journal of Modern Physics D}\
  }\textbf {\bibinfo {volume} {10}},\ \bibinfo {pages} {213} (\bibinfo {year}
  {2001})}\BibitemShut {NoStop}%
\bibitem [{\citenamefont {Linder}(2003)}]{Linder:2003}%
  \BibitemOpen
  \bibfield  {author} {\bibinfo {author} {\bibfnamefont {E.~V.}\ \bibnamefont
  {Linder}},\ }\href {\doibase 10.1103/physrevlett.90.091301} {\bibfield
  {journal} {\bibinfo  {journal} {Physical Review Letters}\ }\textbf {\bibinfo
  {volume} {90}} (\bibinfo {year} {2003}),\
  10.1103/physrevlett.90.091301}\BibitemShut {NoStop}%
\bibitem [{\citenamefont {Ma}\ and\ \citenamefont
  {Zhang}(2011{\natexlab{a}})}]{Ma:2011}%
  \BibitemOpen
  \bibfield  {author} {\bibinfo {author} {\bibfnamefont {J.-Z.}\ \bibnamefont
  {Ma}}\ and\ \bibinfo {author} {\bibfnamefont {X.}~\bibnamefont {Zhang}},\
  }\href {\doibase 10.1016/j.physletb.2011.04.013} {\bibfield  {journal}
  {\bibinfo  {journal} {Physics Letters B}\ }\textbf {\bibinfo {volume}
  {699}},\ \bibinfo {pages} {233} (\bibinfo {year}
  {2011}{\natexlab{a}})}\BibitemShut {NoStop}%
\bibitem [{\citenamefont {Huterer}\ and\ \citenamefont
  {Turner}(2001)}]{Huterer:2001}%
  \BibitemOpen
  \bibfield  {author} {\bibinfo {author} {\bibfnamefont {D.}~\bibnamefont
  {Huterer}}\ and\ \bibinfo {author} {\bibfnamefont {M.~S.}\ \bibnamefont
  {Turner}},\ }\href {\doibase 10.1103/PhysRevD.64.123527} {\bibfield
  {journal} {\bibinfo  {journal} {Phys. Rev. D}\ }\textbf {\bibinfo {volume}
  {64}},\ \bibinfo {pages} {123527} (\bibinfo {year} {2001})}\BibitemShut
  {NoStop}%
\bibitem [{\citenamefont {Weller}\ and\ \citenamefont
  {Albrecht}(2002)}]{Weller:2002}%
  \BibitemOpen
  \bibfield  {author} {\bibinfo {author} {\bibfnamefont {J.}~\bibnamefont
  {Weller}}\ and\ \bibinfo {author} {\bibfnamefont {A.}~\bibnamefont
  {Albrecht}},\ }\href {\doibase 10.1103/PhysRevD.65.103512} {\bibfield
  {journal} {\bibinfo  {journal} {Phys. Rev. D}\ }\textbf {\bibinfo {volume}
  {65}},\ \bibinfo {pages} {103512} (\bibinfo {year} {2002})}\BibitemShut
  {NoStop}%
\bibitem [{\citenamefont {Caldwell}\ and\ \citenamefont
  {Doran}(2004)}]{Caldwell:2004}%
  \BibitemOpen
  \bibfield  {author} {\bibinfo {author} {\bibfnamefont {R.~R.}\ \bibnamefont
  {Caldwell}}\ and\ \bibinfo {author} {\bibfnamefont {M.}~\bibnamefont
  {Doran}},\ }\href {\doibase 10.1103/PhysRevD.69.103517} {\bibfield  {journal}
  {\bibinfo  {journal} {Phys. Rev. D}\ }\textbf {\bibinfo {volume} {69}},\
  \bibinfo {pages} {103517} (\bibinfo {year} {2004})}\BibitemShut {NoStop}%
\bibitem [{\citenamefont {Johri}(2002)}]{Johri:2002}%
  \BibitemOpen
  \bibfield  {author} {\bibinfo {author} {\bibfnamefont {V.~B.}\ \bibnamefont
  {Johri}},\ }\href@noop {} {\bibfield  {journal} {\bibinfo  {journal} {Pramana
  - J Phys}\ }\textbf {\bibinfo {volume} {59}},\ \bibinfo {pages} {L553}
  (\bibinfo {year} {2002})}\BibitemShut {NoStop}%
\bibitem [{\citenamefont {Upadhye}\ \emph {et~al.}(2005)\citenamefont
  {Upadhye}, \citenamefont {Ishak},\ and\ \citenamefont
  {Steinhardt}}]{Upadhye:2005}%
  \BibitemOpen
  \bibfield  {author} {\bibinfo {author} {\bibfnamefont {A.}~\bibnamefont
  {Upadhye}}, \bibinfo {author} {\bibfnamefont {M.}~\bibnamefont {Ishak}}, \
  and\ \bibinfo {author} {\bibfnamefont {P.~J.}\ \bibnamefont {Steinhardt}},\
  }\href {\doibase 10.1103/PhysRevD.72.063501} {\bibfield  {journal} {\bibinfo
  {journal} {Phys. Rev. D}\ }\textbf {\bibinfo {volume} {72}},\ \bibinfo
  {pages} {063501} (\bibinfo {year} {2005})}\BibitemShut {NoStop}%
\bibitem [{\citenamefont {Jassal}\ \emph {et~al.}(2005)\citenamefont {Jassal},
  \citenamefont {Bagla},\ and\ \citenamefont {Padmanabhan}}]{JBP:2005}%
  \BibitemOpen
  \bibfield  {author} {\bibinfo {author} {\bibfnamefont {H.~K.}\ \bibnamefont
  {Jassal}}, \bibinfo {author} {\bibfnamefont {J.~S.}\ \bibnamefont {Bagla}}, \
  and\ \bibinfo {author} {\bibfnamefont {T.}~\bibnamefont {Padmanabhan}},\
  }\href {\doibase 10.1111/j.1745-3933.2005.08577.x} {\bibfield  {journal}
  {\bibinfo  {journal} {Monthly Notices of the Royal Astronomical Society:
  Letters}\ }\textbf {\bibinfo {volume} {356}},\ \bibinfo {pages} {L11}
  (\bibinfo {year} {2005})},\ \Eprint
  {http://arxiv.org/abs/https://academic.oup.com/mnrasl/article-pdf/356/1/L11/6278031/356-1-L11.pdf}
  {https://academic.oup.com/mnrasl/article-pdf/356/1/L11/6278031/356-1-L11.pdf}
  \BibitemShut {NoStop}%
\bibitem [{\citenamefont {Liu}\ \emph {et~al.}(2008)\citenamefont {Liu},
  \citenamefont {Li}, \citenamefont {Hao},\ and\ \citenamefont
  {Jin}}]{Liu:2008}%
  \BibitemOpen
  \bibfield  {author} {\bibinfo {author} {\bibfnamefont {D.-J.}\ \bibnamefont
  {Liu}}, \bibinfo {author} {\bibfnamefont {X.-Z.}\ \bibnamefont {Li}},
  \bibinfo {author} {\bibfnamefont {J.}~\bibnamefont {Hao}}, \ and\ \bibinfo
  {author} {\bibfnamefont {X.-H.}\ \bibnamefont {Jin}},\ }\href {\doibase
  https://doi.org/10.1111/j.1365-2966.2008.13380.x} {\bibfield  {journal}
  {\bibinfo  {journal} {Monthly Notices of the Royal Astronomical Society}\
  }\textbf {\bibinfo {volume} {388}},\ \bibinfo {pages} {275} (\bibinfo {year}
  {2008})},\ \Eprint
  {http://arxiv.org/abs/https://onlinelibrary.wiley.com/doi/pdf/10.1111/j.1365-2966.2008.13380.x}
  {https://onlinelibrary.wiley.com/doi/pdf/10.1111/j.1365-2966.2008.13380.x}
  \BibitemShut {NoStop}%
\bibitem [{\citenamefont {Barboza}\ and\ \citenamefont
  {Alcaniz}(2008{\natexlab{a}})}]{Barboza:2008}%
  \BibitemOpen
  \bibfield  {author} {\bibinfo {author} {\bibfnamefont {E.}~\bibnamefont
  {Barboza}}\ and\ \bibinfo {author} {\bibfnamefont {J.}~\bibnamefont
  {Alcaniz}},\ }\href {\doibase https://doi.org/10.1016/j.physletb.2008.08.012}
  {\bibfield  {journal} {\bibinfo  {journal} {Physics Letters B}\ }\textbf
  {\bibinfo {volume} {666}},\ \bibinfo {pages} {415} (\bibinfo {year}
  {2008}{\natexlab{a}})}\BibitemShut {NoStop}%
\bibitem [{\citenamefont {Li}\ and\ \citenamefont {Zhang}(2011)}]{Li:2011}%
  \BibitemOpen
  \bibfield  {author} {\bibinfo {author} {\bibfnamefont {H.}~\bibnamefont
  {Li}}\ and\ \bibinfo {author} {\bibfnamefont {X.}~\bibnamefont {Zhang}},\
  }\href {\doibase https://doi.org/10.1016/j.physletb.2011.07.069} {\bibfield
  {journal} {\bibinfo  {journal} {Physics Letters B}\ }\textbf {\bibinfo
  {volume} {703}},\ \bibinfo {pages} {119} (\bibinfo {year}
  {2011})}\BibitemShut {NoStop}%
\bibitem [{\citenamefont {Feng}\ \emph {et~al.}(2012)\citenamefont {Feng},
  \citenamefont {Shen}, \citenamefont {Li},\ and\ \citenamefont
  {Li}}]{Feng:2012}%
  \BibitemOpen
  \bibfield  {author} {\bibinfo {author} {\bibfnamefont {C.-J.}\ \bibnamefont
  {Feng}}, \bibinfo {author} {\bibfnamefont {X.-Y.}\ \bibnamefont {Shen}},
  \bibinfo {author} {\bibfnamefont {P.}~\bibnamefont {Li}}, \ and\ \bibinfo
  {author} {\bibfnamefont {X.-Z.}\ \bibnamefont {Li}},\ }\href@noop {}
  {\bibfield  {journal} {\bibinfo  {journal} {J. Cosmol. Astropart. Phys.}\
  }\textbf {\bibinfo {volume} {2012}},\ \bibinfo {pages} {023} (\bibinfo {year}
  {2012})}\BibitemShut {NoStop}%
\bibitem [{\citenamefont {Maga\~na}\ \emph {et~al.}(2014)\citenamefont
  {Maga\~na}, \citenamefont {C\'ardenas},\ and\ \citenamefont
  {Motta}}]{Magana:2014voa}%
  \BibitemOpen
  \bibfield  {author} {\bibinfo {author} {\bibfnamefont {J.}~\bibnamefont
  {Maga\~na}}, \bibinfo {author} {\bibfnamefont {V.~H.}\ \bibnamefont
  {C\'ardenas}}, \ and\ \bibinfo {author} {\bibfnamefont {V.}~\bibnamefont
  {Motta}},\ }\href {\doibase 10.1088/1475-7516/2014/10/017} {\bibfield
  {journal} {\bibinfo  {journal} {JCAP}\ }\textbf {\bibinfo {volume} {10}},\
  \bibinfo {pages} {017} (\bibinfo {year} {2014})},\ \Eprint
  {http://arxiv.org/abs/1407.1632} {arXiv:1407.1632 [astro-ph.CO]} \BibitemShut
  {NoStop}%
\bibitem [{\citenamefont {Hu}\ \emph {et~al.}(2016)\citenamefont {Hu},
  \citenamefont {Li}, \citenamefont {Li},\ and\ \citenamefont
  {Wang}}]{Hu:2015ksa}%
  \BibitemOpen
  \bibfield  {author} {\bibinfo {author} {\bibfnamefont {Y.}~\bibnamefont
  {Hu}}, \bibinfo {author} {\bibfnamefont {M.}~\bibnamefont {Li}}, \bibinfo
  {author} {\bibfnamefont {N.}~\bibnamefont {Li}}, \ and\ \bibinfo {author}
  {\bibfnamefont {S.}~\bibnamefont {Wang}},\ }\href {\doibase
  10.3847/0004-637X/821/1/60} {\bibfield  {journal} {\bibinfo  {journal}
  {Astrophys. J.}\ }\textbf {\bibinfo {volume} {821}},\ \bibinfo {pages} {60}
  (\bibinfo {year} {2016})},\ \Eprint {http://arxiv.org/abs/1509.03461}
  {arXiv:1509.03461 [astro-ph.CO]} \BibitemShut {NoStop}%
\bibitem [{\citenamefont {Roman-Garza}\ \emph {et~al.}(2019)\citenamefont
  {Roman-Garza}, \citenamefont {Verdugo}, \citenamefont {Magana},\ and\
  \citenamefont {Motta}}]{Roman-Garza:2018cxf}%
  \BibitemOpen
  \bibfield  {author} {\bibinfo {author} {\bibfnamefont {J.}~\bibnamefont
  {Roman-Garza}}, \bibinfo {author} {\bibfnamefont {T.}~\bibnamefont
  {Verdugo}}, \bibinfo {author} {\bibfnamefont {J.}~\bibnamefont {Magana}}, \
  and\ \bibinfo {author} {\bibfnamefont {V.}~\bibnamefont {Motta}},\ }\href
  {\doibase 10.1140/epjc/s10052-019-7390-3} {\bibfield  {journal} {\bibinfo
  {journal} {Eur. Phys. J. C}\ }\textbf {\bibinfo {volume} {79}},\ \bibinfo
  {pages} {890} (\bibinfo {year} {2019})},\ \Eprint
  {http://arxiv.org/abs/1806.03538} {arXiv:1806.03538 [astro-ph.CO]}
  \BibitemShut {NoStop}%
\bibitem [{\citenamefont {Singh}\ and\ \citenamefont
  {Nagpal}(2020)}]{Singh:2020}%
  \BibitemOpen
  \bibfield  {author} {\bibinfo {author} {\bibfnamefont {J.~K.}\ \bibnamefont
  {Singh}}\ and\ \bibinfo {author} {\bibfnamefont {R.}~\bibnamefont {Nagpal}},\
  }\href@noop {} {\bibfield  {journal} {\bibinfo  {journal} {Eur. Phys. J. C
  Part. Fields}\ }\textbf {\bibinfo {volume} {80}} (\bibinfo {year}
  {2020})}\BibitemShut {NoStop}%
\bibitem [{\citenamefont {Perkovi{\'c}}\ and\ \citenamefont {{\v S}tefan{\v
  c}i{\'c}}(2020)}]{Perkovic:2020}%
  \BibitemOpen
  \bibfield  {author} {\bibinfo {author} {\bibfnamefont {D.}~\bibnamefont
  {Perkovi{\'c}}}\ and\ \bibinfo {author} {\bibfnamefont {H.}~\bibnamefont {{\v
  S}tefan{\v c}i{\'c}}},\ }\href@noop {} {\bibfield  {journal} {\bibinfo
  {journal} {Eur. Phys. J. C Part. Fields}\ }\textbf {\bibinfo {volume} {80}}
  (\bibinfo {year} {2020})}\BibitemShut {NoStop}%
\bibitem [{\citenamefont {Bouhmadi-Lopez}\ \emph {et~al.}(2015)\citenamefont
  {Bouhmadi-Lopez}, \citenamefont {Errahmani}, \citenamefont {Martin-Moruno},
  \citenamefont {Ouali},\ and\ \citenamefont
  {Tavakoli}}]{Bouhmadi-Lopez:2014cca}%
  \BibitemOpen
  \bibfield  {author} {\bibinfo {author} {\bibfnamefont {M.}~\bibnamefont
  {Bouhmadi-Lopez}}, \bibinfo {author} {\bibfnamefont {A.}~\bibnamefont
  {Errahmani}}, \bibinfo {author} {\bibfnamefont {P.}~\bibnamefont
  {Martin-Moruno}}, \bibinfo {author} {\bibfnamefont {T.}~\bibnamefont
  {Ouali}}, \ and\ \bibinfo {author} {\bibfnamefont {Y.}~\bibnamefont
  {Tavakoli}},\ }\href {\doibase 10.1142/S0218271815500789} {\bibfield
  {journal} {\bibinfo  {journal} {Int. J. Mod. Phys. D}\ }\textbf {\bibinfo
  {volume} {24}},\ \bibinfo {pages} {1550078} (\bibinfo {year} {2015})},\
  \Eprint {http://arxiv.org/abs/1407.2446} {arXiv:1407.2446 [gr-qc]}
  \BibitemShut {NoStop}%
\bibitem [{\citenamefont {Poulin}\ \emph {et~al.}(2019)\citenamefont {Poulin},
  \citenamefont {Smith}, \citenamefont {Karwal},\ and\ \citenamefont
  {Kamionkowski}}]{Poulin:2018cxd}%
  \BibitemOpen
  \bibfield  {author} {\bibinfo {author} {\bibfnamefont {V.}~\bibnamefont
  {Poulin}}, \bibinfo {author} {\bibfnamefont {T.~L.}\ \bibnamefont {Smith}},
  \bibinfo {author} {\bibfnamefont {T.}~\bibnamefont {Karwal}}, \ and\ \bibinfo
  {author} {\bibfnamefont {M.}~\bibnamefont {Kamionkowski}},\ }\href {\doibase
  10.1103/PhysRevLett.122.221301} {\bibfield  {journal} {\bibinfo  {journal}
  {Phys. Rev. Lett.}\ }\textbf {\bibinfo {volume} {122}},\ \bibinfo {pages}
  {221301} (\bibinfo {year} {2019})},\ \Eprint
  {http://arxiv.org/abs/1811.04083} {arXiv:1811.04083 [astro-ph.CO]}
  \BibitemShut {NoStop}%
\bibitem [{\citenamefont {Acquaviva}\ \emph {et~al.}(2021)\citenamefont
  {Acquaviva}, \citenamefont {Akarsu}, \citenamefont {Katirci},\ and\
  \citenamefont {Vazquez}}]{Acquaviva:2021jov}%
  \BibitemOpen
  \bibfield  {author} {\bibinfo {author} {\bibfnamefont {G.}~\bibnamefont
  {Acquaviva}}, \bibinfo {author} {\bibfnamefont {O.}~\bibnamefont {Akarsu}},
  \bibinfo {author} {\bibfnamefont {N.}~\bibnamefont {Katirci}}, \ and\
  \bibinfo {author} {\bibfnamefont {J.~A.}\ \bibnamefont {Vazquez}},\ }\href
  {\doibase 10.1103/PhysRevD.104.023505} {\bibfield  {journal} {\bibinfo
  {journal} {Phys. Rev. D}\ }\textbf {\bibinfo {volume} {104}},\ \bibinfo
  {pages} {023505} (\bibinfo {year} {2021})},\ \Eprint
  {http://arxiv.org/abs/2104.02623} {arXiv:2104.02623 [astro-ph.CO]}
  \BibitemShut {NoStop}%
\bibitem [{\citenamefont {Akarsu}\ \emph {et~al.}(2021)\citenamefont {Akarsu},
  \citenamefont {Kumar}, \citenamefont {\"Oz\"ulker},\ and\ \citenamefont
  {Vazquez}}]{Akarsu:2021fol}%
  \BibitemOpen
  \bibfield  {author} {\bibinfo {author} {\bibfnamefont {O.}~\bibnamefont
  {Akarsu}}, \bibinfo {author} {\bibfnamefont {S.}~\bibnamefont {Kumar}},
  \bibinfo {author} {\bibfnamefont {E.}~\bibnamefont {\"Oz\"ulker}}, \ and\
  \bibinfo {author} {\bibfnamefont {J.~A.}\ \bibnamefont {Vazquez}},\ }\href
  {\doibase 10.1103/PhysRevD.104.123512} {\bibfield  {journal} {\bibinfo
  {journal} {Phys. Rev. D}\ }\textbf {\bibinfo {volume} {104}},\ \bibinfo
  {pages} {123512} (\bibinfo {year} {2021})},\ \Eprint
  {http://arxiv.org/abs/2108.09239} {arXiv:2108.09239 [astro-ph.CO]}
  \BibitemShut {NoStop}%
\bibitem [{\citenamefont {Poulin}\ \emph {et~al.}(2022)\citenamefont {Poulin},
  \citenamefont {Bernal}, \citenamefont {Kovetz},\ and\ \citenamefont
  {Kamionkowski}}]{Poulin:2022sgp}%
  \BibitemOpen
  \bibfield  {author} {\bibinfo {author} {\bibfnamefont {V.}~\bibnamefont
  {Poulin}}, \bibinfo {author} {\bibfnamefont {J.~L.}\ \bibnamefont {Bernal}},
  \bibinfo {author} {\bibfnamefont {E.}~\bibnamefont {Kovetz}}, \ and\ \bibinfo
  {author} {\bibfnamefont {M.}~\bibnamefont {Kamionkowski}},\ }\href@noop {} {\
   (\bibinfo {year} {2022})},\ \Eprint {http://arxiv.org/abs/2209.06217}
  {arXiv:2209.06217 [astro-ph.CO]} \BibitemShut {NoStop}%
\bibitem [{\citenamefont {Akarsu}\ \emph {et~al.}(2022)\citenamefont {Akarsu},
  \citenamefont {Kumar}, \citenamefont {\"Oz\"ulker}, \citenamefont {Vazquez},\
  and\ \citenamefont {Yadav}}]{Akarsu:2022typ}%
  \BibitemOpen
  \bibfield  {author} {\bibinfo {author} {\bibfnamefont {O.}~\bibnamefont
  {Akarsu}}, \bibinfo {author} {\bibfnamefont {S.}~\bibnamefont {Kumar}},
  \bibinfo {author} {\bibfnamefont {E.}~\bibnamefont {\"Oz\"ulker}}, \bibinfo
  {author} {\bibfnamefont {J.~A.}\ \bibnamefont {Vazquez}}, \ and\ \bibinfo
  {author} {\bibfnamefont {A.}~\bibnamefont {Yadav}},\ }\href@noop {} {\
  (\bibinfo {year} {2022})},\ \Eprint {http://arxiv.org/abs/2211.05742}
  {arXiv:2211.05742 [astro-ph.CO]} \BibitemShut {NoStop}%
\bibitem [{\citenamefont {Gong}\ and\ \citenamefont {Zhang}(2005)}]{Gong:2005}%
  \BibitemOpen
  \bibfield  {author} {\bibinfo {author} {\bibfnamefont {Y.}~\bibnamefont
  {Gong}}\ and\ \bibinfo {author} {\bibfnamefont {Y.-Z.}\ \bibnamefont
  {Zhang}},\ }\href {\doibase 10.1103/PhysRevD.72.043518} {\bibfield  {journal}
  {\bibinfo  {journal} {Phys. Rev. D}\ }\textbf {\bibinfo {volume} {72}},\
  \bibinfo {pages} {043518} (\bibinfo {year} {2005})}\BibitemShut {NoStop}%
\bibitem [{\citenamefont {Bassett}\ \emph {et~al.}(2002)\citenamefont
  {Bassett}, \citenamefont {Kunz}, \citenamefont {Silk},\ and\ \citenamefont
  {Ungarelli}}]{Bassett:2002qu}%
  \BibitemOpen
  \bibfield  {author} {\bibinfo {author} {\bibfnamefont {B.~A.}\ \bibnamefont
  {Bassett}}, \bibinfo {author} {\bibfnamefont {M.}~\bibnamefont {Kunz}},
  \bibinfo {author} {\bibfnamefont {J.}~\bibnamefont {Silk}}, \ and\ \bibinfo
  {author} {\bibfnamefont {C.}~\bibnamefont {Ungarelli}},\ }\href {\doibase
  10.1046/j.1365-8711.2002.05887.x} {\bibfield  {journal} {\bibinfo  {journal}
  {Mon. Not. Roy. Astron. Soc.}\ }\textbf {\bibinfo {volume} {336}},\ \bibinfo
  {pages} {1217} (\bibinfo {year} {2002})},\ \Eprint
  {http://arxiv.org/abs/astro-ph/0203383} {arXiv:astro-ph/0203383} \BibitemShut
  {NoStop}%
\bibitem [{\citenamefont {Bassett}\ \emph {et~al.}(2004)\citenamefont
  {Bassett}, \citenamefont {Corasaniti},\ and\ \citenamefont
  {Kunz}}]{Bassett:2004wz}%
  \BibitemOpen
  \bibfield  {author} {\bibinfo {author} {\bibfnamefont {B.~A.}\ \bibnamefont
  {Bassett}}, \bibinfo {author} {\bibfnamefont {P.~S.}\ \bibnamefont
  {Corasaniti}}, \ and\ \bibinfo {author} {\bibfnamefont {M.}~\bibnamefont
  {Kunz}},\ }\href {\doibase 10.1086/427023} {\bibfield  {journal} {\bibinfo
  {journal} {Astrophys. J. Lett.}\ }\textbf {\bibinfo {volume} {617}},\
  \bibinfo {pages} {L1} (\bibinfo {year} {2004})},\ \Eprint
  {http://arxiv.org/abs/astro-ph/0407364} {arXiv:astro-ph/0407364} \BibitemShut
  {NoStop}%
\bibitem [{\citenamefont {Shafieloo}\ \emph {et~al.}(2009)\citenamefont
  {Shafieloo}, \citenamefont {Sahni},\ and\ \citenamefont
  {Starobinsky}}]{Shafieloo:2009ti}%
  \BibitemOpen
  \bibfield  {author} {\bibinfo {author} {\bibfnamefont {A.}~\bibnamefont
  {Shafieloo}}, \bibinfo {author} {\bibfnamefont {V.}~\bibnamefont {Sahni}}, \
  and\ \bibinfo {author} {\bibfnamefont {A.~A.}\ \bibnamefont {Starobinsky}},\
  }\href {\doibase 10.1103/PhysRevD.80.101301} {\bibfield  {journal} {\bibinfo
  {journal} {Phys. Rev. D}\ }\textbf {\bibinfo {volume} {80}},\ \bibinfo
  {pages} {101301} (\bibinfo {year} {2009})},\ \Eprint
  {http://arxiv.org/abs/0903.5141} {arXiv:0903.5141 [astro-ph.CO]} \BibitemShut
  {NoStop}%
\bibitem [{\citenamefont {De~Felice}\ \emph {et~al.}(2012)\citenamefont
  {De~Felice}, \citenamefont {Nesseris},\ and\ \citenamefont
  {Tsujikawa}}]{DeFelice:2012vd}%
  \BibitemOpen
  \bibfield  {author} {\bibinfo {author} {\bibfnamefont {A.}~\bibnamefont
  {De~Felice}}, \bibinfo {author} {\bibfnamefont {S.}~\bibnamefont {Nesseris}},
  \ and\ \bibinfo {author} {\bibfnamefont {S.}~\bibnamefont {Tsujikawa}},\
  }\href {\doibase 10.1088/1475-7516/2012/05/029} {\bibfield  {journal}
  {\bibinfo  {journal} {JCAP}\ }\textbf {\bibinfo {volume} {05}},\ \bibinfo
  {pages} {029} (\bibinfo {year} {2012})},\ \Eprint
  {http://arxiv.org/abs/1203.6760} {arXiv:1203.6760 [astro-ph.CO]} \BibitemShut
  {NoStop}%
\bibitem [{\citenamefont {Wei}\ \emph {et~al.}(2014)\citenamefont {Wei},
  \citenamefont {Yan},\ and\ \citenamefont {Zhou}}]{Wei:2013jya}%
  \BibitemOpen
  \bibfield  {author} {\bibinfo {author} {\bibfnamefont {H.}~\bibnamefont
  {Wei}}, \bibinfo {author} {\bibfnamefont {X.-P.}\ \bibnamefont {Yan}}, \ and\
  \bibinfo {author} {\bibfnamefont {Y.-N.}\ \bibnamefont {Zhou}},\ }\href
  {\doibase 10.1088/1475-7516/2014/01/045} {\bibfield  {journal} {\bibinfo
  {journal} {JCAP}\ }\textbf {\bibinfo {volume} {01}},\ \bibinfo {pages} {045}
  (\bibinfo {year} {2014})},\ \Eprint {http://arxiv.org/abs/1312.1117}
  {arXiv:1312.1117 [astro-ph.CO]} \BibitemShut {NoStop}%
\bibitem [{\citenamefont {Hazra}\ \emph {et~al.}(2015)\citenamefont {Hazra},
  \citenamefont {Majumdar}, \citenamefont {Pal}, \citenamefont {Panda},\ and\
  \citenamefont {Sen}}]{Hazra:2013dsx}%
  \BibitemOpen
  \bibfield  {author} {\bibinfo {author} {\bibfnamefont {D.~K.}\ \bibnamefont
  {Hazra}}, \bibinfo {author} {\bibfnamefont {S.}~\bibnamefont {Majumdar}},
  \bibinfo {author} {\bibfnamefont {S.}~\bibnamefont {Pal}}, \bibinfo {author}
  {\bibfnamefont {S.}~\bibnamefont {Panda}}, \ and\ \bibinfo {author}
  {\bibfnamefont {A.~A.}\ \bibnamefont {Sen}},\ }\href {\doibase
  10.1103/PhysRevD.91.083005} {\bibfield  {journal} {\bibinfo  {journal} {Phys.
  Rev. D}\ }\textbf {\bibinfo {volume} {91}},\ \bibinfo {pages} {083005}
  (\bibinfo {year} {2015})},\ \Eprint {http://arxiv.org/abs/1310.6161}
  {arXiv:1310.6161 [astro-ph.CO]} \BibitemShut {NoStop}%
\bibitem [{\citenamefont {Akarsu}\ \emph {et~al.}(2015)\citenamefont {Akarsu},
  \citenamefont {Dereli},\ and\ \citenamefont {Vazquez}}]{Akarsu:2015yea}%
  \BibitemOpen
  \bibfield  {author} {\bibinfo {author} {\bibfnamefont {O.}~\bibnamefont
  {Akarsu}}, \bibinfo {author} {\bibfnamefont {T.}~\bibnamefont {Dereli}}, \
  and\ \bibinfo {author} {\bibfnamefont {J.~A.}\ \bibnamefont {Vazquez}},\
  }\href {\doibase 10.1088/1475-7516/2015/06/049} {\bibfield  {journal}
  {\bibinfo  {journal} {JCAP}\ }\textbf {\bibinfo {volume} {06}},\ \bibinfo
  {pages} {049} (\bibinfo {year} {2015})},\ \Eprint
  {http://arxiv.org/abs/1501.07598} {arXiv:1501.07598 [astro-ph.CO]}
  \BibitemShut {NoStop}%
\bibitem [{\citenamefont {Barboza}\ and\ \citenamefont
  {Alcaniz}(2008{\natexlab{b}})}]{Barboza:2008rh}%
  \BibitemOpen
  \bibfield  {author} {\bibinfo {author} {\bibfnamefont {E.~M.}\ \bibnamefont
  {Barboza}, \bibfnamefont {Jr.}}\ and\ \bibinfo {author} {\bibfnamefont
  {J.~S.}\ \bibnamefont {Alcaniz}},\ }\href {\doibase
  10.1016/j.physletb.2008.08.012} {\bibfield  {journal} {\bibinfo  {journal}
  {Phys. Lett. B}\ }\textbf {\bibinfo {volume} {666}},\ \bibinfo {pages} {415}
  (\bibinfo {year} {2008}{\natexlab{b}})},\ \Eprint
  {http://arxiv.org/abs/0805.1713} {arXiv:0805.1713 [astro-ph]} \BibitemShut
  {NoStop}%
\bibitem [{\citenamefont {Ma}\ and\ \citenamefont
  {Zhang}(2011{\natexlab{b}})}]{Ma:2011nc}%
  \BibitemOpen
  \bibfield  {author} {\bibinfo {author} {\bibfnamefont {J.-Z.}\ \bibnamefont
  {Ma}}\ and\ \bibinfo {author} {\bibfnamefont {X.}~\bibnamefont {Zhang}},\
  }\href {\doibase 10.1016/j.physletb.2011.04.013} {\bibfield  {journal}
  {\bibinfo  {journal} {Phys. Lett. B}\ }\textbf {\bibinfo {volume} {699}},\
  \bibinfo {pages} {233} (\bibinfo {year} {2011}{\natexlab{b}})},\ \Eprint
  {http://arxiv.org/abs/1102.2671} {arXiv:1102.2671 [astro-ph.CO]} \BibitemShut
  {NoStop}%
\bibitem [{\citenamefont {Yang}\ \emph {et~al.}(2023)\citenamefont {Yang},
  \citenamefont {Fan}, \citenamefont {Feng},\ and\ \citenamefont
  {Zhai}}]{Yang:2022klj}%
  \BibitemOpen
  \bibfield  {author} {\bibinfo {author} {\bibfnamefont {J.}~\bibnamefont
  {Yang}}, \bibinfo {author} {\bibfnamefont {X.-Y.}\ \bibnamefont {Fan}},
  \bibinfo {author} {\bibfnamefont {C.-J.}\ \bibnamefont {Feng}}, \ and\
  \bibinfo {author} {\bibfnamefont {X.-H.}\ \bibnamefont {Zhai}},\ }\href
  {\doibase 10.1088/0256-307X/40/1/019801} {\bibfield  {journal} {\bibinfo
  {journal} {Chin. Phys. Lett.}\ }\textbf {\bibinfo {volume} {40}},\ \bibinfo
  {pages} {019801} (\bibinfo {year} {2023})},\ \Eprint
  {http://arxiv.org/abs/2211.15881} {arXiv:2211.15881 [astro-ph.CO]}
  \BibitemShut {NoStop}%
\bibitem [{\citenamefont {Komatsu}\ \emph {et~al.}(2011)\citenamefont
  {Komatsu}, \citenamefont {Smith}, \citenamefont {Dunkley}, \citenamefont
  {Bennett}, \citenamefont {Gold}, \citenamefont {Hinshaw}, \citenamefont
  {Jarosik}, \citenamefont {Larson}, \citenamefont {Nolta}, \citenamefont
  {Page}, \citenamefont {Spergel}, \citenamefont {Halpern}, \citenamefont
  {Hill}, \citenamefont {Kogut}, \citenamefont {Limon}, \citenamefont {Meyer},
  \citenamefont {Odegard}, \citenamefont {Tucker}, \citenamefont {Weiland},
  \citenamefont {Wollack},\ and\ \citenamefont {Wright}}]{Komatsu_2011}%
  \BibitemOpen
  \bibfield  {author} {\bibinfo {author} {\bibfnamefont {E.}~\bibnamefont
  {Komatsu}}, \bibinfo {author} {\bibfnamefont {K.~M.}\ \bibnamefont {Smith}},
  \bibinfo {author} {\bibfnamefont {J.}~\bibnamefont {Dunkley}}, \bibinfo
  {author} {\bibfnamefont {C.~L.}\ \bibnamefont {Bennett}}, \bibinfo {author}
  {\bibfnamefont {B.}~\bibnamefont {Gold}}, \bibinfo {author} {\bibfnamefont
  {G.}~\bibnamefont {Hinshaw}}, \bibinfo {author} {\bibfnamefont
  {N.}~\bibnamefont {Jarosik}}, \bibinfo {author} {\bibfnamefont
  {D.}~\bibnamefont {Larson}}, \bibinfo {author} {\bibfnamefont {M.~R.}\
  \bibnamefont {Nolta}}, \bibinfo {author} {\bibfnamefont {L.}~\bibnamefont
  {Page}}, \bibinfo {author} {\bibfnamefont {D.~N.}\ \bibnamefont {Spergel}},
  \bibinfo {author} {\bibfnamefont {M.}~\bibnamefont {Halpern}}, \bibinfo
  {author} {\bibfnamefont {R.~S.}\ \bibnamefont {Hill}}, \bibinfo {author}
  {\bibfnamefont {A.}~\bibnamefont {Kogut}}, \bibinfo {author} {\bibfnamefont
  {M.}~\bibnamefont {Limon}}, \bibinfo {author} {\bibfnamefont {S.~S.}\
  \bibnamefont {Meyer}}, \bibinfo {author} {\bibfnamefont {N.}~\bibnamefont
  {Odegard}}, \bibinfo {author} {\bibfnamefont {G.~S.}\ \bibnamefont {Tucker}},
  \bibinfo {author} {\bibfnamefont {J.~L.}\ \bibnamefont {Weiland}}, \bibinfo
  {author} {\bibfnamefont {E.}~\bibnamefont {Wollack}}, \ and\ \bibinfo
  {author} {\bibfnamefont {E.~L.}\ \bibnamefont {Wright}},\ }\href {\doibase
  10.1088/0067-0049/192/2/18} {\bibfield  {journal} {\bibinfo  {journal} {The
  Astrophysical Journal Supplement Series}\ }\textbf {\bibinfo {volume}
  {192}},\ \bibinfo {pages} {18} (\bibinfo {year} {2011})}\BibitemShut
  {NoStop}%
\bibitem [{\citenamefont {Bardeen}(1980)}]{Bardeen:1980}%
  \BibitemOpen
  \bibfield  {author} {\bibinfo {author} {\bibfnamefont {J.~M.}\ \bibnamefont
  {Bardeen}},\ }\href {\doibase 10.1103/PhysRevD.22.1882} {\bibfield  {journal}
  {\bibinfo  {journal} {Phys. Rev. D}\ }\textbf {\bibinfo {volume} {22}},\
  \bibinfo {pages} {1882} (\bibinfo {year} {1980})}\BibitemShut {NoStop}%
\bibitem [{\citenamefont {Kodama}\ and\ \citenamefont
  {Sasaki}(1984)}]{Kodama:1984}%
  \BibitemOpen
  \bibfield  {author} {\bibinfo {author} {\bibfnamefont {H.}~\bibnamefont
  {Kodama}}\ and\ \bibinfo {author} {\bibfnamefont {M.}~\bibnamefont
  {Sasaki}},\ }\href {\doibase 10.1143/PTPS.78.1} {\bibfield  {journal}
  {\bibinfo  {journal} {Progress of Theoretical Physics Supplement}\ }\textbf
  {\bibinfo {volume} {78}},\ \bibinfo {pages} {1} (\bibinfo {year} {1984})},\
  \Eprint
  {http://arxiv.org/abs/https://academic.oup.com/ptps/article-pdf/doi/10.1143/PTPS.78.1/5321391/78-1.pdf}
  {https://academic.oup.com/ptps/article-pdf/doi/10.1143/PTPS.78.1/5321391/78-1.pdf}
  \BibitemShut {NoStop}%
\bibitem [{\citenamefont {Ma}\ and\ \citenamefont
  {Bertschinger}(1995)}]{Ma_1995}%
  \BibitemOpen
  \bibfield  {author} {\bibinfo {author} {\bibfnamefont {C.-P.}\ \bibnamefont
  {Ma}}\ and\ \bibinfo {author} {\bibfnamefont {E.}~\bibnamefont
  {Bertschinger}},\ }\href {\doibase 10.1086/176550} {\bibfield  {journal}
  {\bibinfo  {journal} {The Astrophysical Journal}\ }\textbf {\bibinfo {volume}
  {455}},\ \bibinfo {pages} {7} (\bibinfo {year} {1995})}\BibitemShut {NoStop}%
\bibitem [{\citenamefont {Blas}\ \emph {et~al.}(2011)\citenamefont {Blas},
  \citenamefont {Lesgourgues},\ and\ \citenamefont {Tram}}]{Blas_2011}%
  \BibitemOpen
  \bibfield  {author} {\bibinfo {author} {\bibfnamefont {D.}~\bibnamefont
  {Blas}}, \bibinfo {author} {\bibfnamefont {J.}~\bibnamefont {Lesgourgues}}, \
  and\ \bibinfo {author} {\bibfnamefont {T.}~\bibnamefont {Tram}},\ }\href
  {\doibase 10.1088/1475-7516/2011/07/034} {\bibfield  {journal} {\bibinfo
  {journal} {Journal of Cosmology and Astroparticle Physics}\ }\textbf
  {\bibinfo {volume} {2011}},\ \bibinfo {pages} {034} (\bibinfo {year}
  {2011})}\BibitemShut {NoStop}%
\bibitem [{\citenamefont {Brinckmann}\ and\ \citenamefont
  {Lesgourgues}(2018)}]{Brinckmann:2018cvx}%
  \BibitemOpen
  \bibfield  {author} {\bibinfo {author} {\bibfnamefont {T.}~\bibnamefont
  {Brinckmann}}\ and\ \bibinfo {author} {\bibfnamefont {J.}~\bibnamefont
  {Lesgourgues}},\ }\href@noop {} {\  (\bibinfo {year} {2018})},\ \Eprint
  {http://arxiv.org/abs/1804.07261} {arXiv:1804.07261 [astro-ph.CO]}
  \BibitemShut {NoStop}%
\bibitem [{\citenamefont {Audren}\ \emph {et~al.}(2013)\citenamefont {Audren},
  \citenamefont {Lesgourgues}, \citenamefont {Benabed},\ and\ \citenamefont
  {Prunet}}]{Audren:2012wb}%
  \BibitemOpen
  \bibfield  {author} {\bibinfo {author} {\bibfnamefont {B.}~\bibnamefont
  {Audren}}, \bibinfo {author} {\bibfnamefont {J.}~\bibnamefont {Lesgourgues}},
  \bibinfo {author} {\bibfnamefont {K.}~\bibnamefont {Benabed}}, \ and\
  \bibinfo {author} {\bibfnamefont {S.}~\bibnamefont {Prunet}},\ }\href
  {\doibase 10.1088/1475-7516/2013/02/001} {\bibfield  {journal} {\bibinfo
  {journal} {JCAP}\ }\textbf {\bibinfo {volume} {1302}},\ \bibinfo {pages}
  {001} (\bibinfo {year} {2013})},\ \Eprint {http://arxiv.org/abs/1210.7183}
  {arXiv:1210.7183 [astro-ph.CO]} \BibitemShut {NoStop}%
\bibitem [{\citenamefont {Riess}\ \emph {et~al.}(2016)\citenamefont {Riess},
  \citenamefont {Macri}, \citenamefont {Hoffmann}, \citenamefont {Scolnic},
  \citenamefont {Casertano}, \citenamefont {Filippenko}, \citenamefont
  {Tucker}, \citenamefont {Reid}, \citenamefont {Jones}, \citenamefont
  {Silverman}, \citenamefont {Chornock}, \citenamefont {Challis}, \citenamefont
  {Yuan}, \citenamefont {Brown},\ and\ \citenamefont {Foley}}]{Riess_2016}%
  \BibitemOpen
  \bibfield  {author} {\bibinfo {author} {\bibfnamefont {A.~G.}\ \bibnamefont
  {Riess}}, \bibinfo {author} {\bibfnamefont {L.~M.}\ \bibnamefont {Macri}},
  \bibinfo {author} {\bibfnamefont {S.~L.}\ \bibnamefont {Hoffmann}}, \bibinfo
  {author} {\bibfnamefont {D.}~\bibnamefont {Scolnic}}, \bibinfo {author}
  {\bibfnamefont {S.}~\bibnamefont {Casertano}}, \bibinfo {author}
  {\bibfnamefont {A.~V.}\ \bibnamefont {Filippenko}}, \bibinfo {author}
  {\bibfnamefont {B.~E.}\ \bibnamefont {Tucker}}, \bibinfo {author}
  {\bibfnamefont {M.~J.}\ \bibnamefont {Reid}}, \bibinfo {author}
  {\bibfnamefont {D.~O.}\ \bibnamefont {Jones}}, \bibinfo {author}
  {\bibfnamefont {J.~M.}\ \bibnamefont {Silverman}}, \bibinfo {author}
  {\bibfnamefont {R.}~\bibnamefont {Chornock}}, \bibinfo {author}
  {\bibfnamefont {P.}~\bibnamefont {Challis}}, \bibinfo {author} {\bibfnamefont
  {W.}~\bibnamefont {Yuan}}, \bibinfo {author} {\bibfnamefont {P.~J.}\
  \bibnamefont {Brown}}, \ and\ \bibinfo {author} {\bibfnamefont {R.~J.}\
  \bibnamefont {Foley}},\ }\href {\doibase 10.3847/0004-637x/826/1/56}
  {\bibfield  {journal} {\bibinfo  {journal} {The Astrophysical Journal}\
  }\textbf {\bibinfo {volume} {826}},\ \bibinfo {pages} {56} (\bibinfo {year}
  {2016})}\BibitemShut {NoStop}%
\bibitem [{\citenamefont {Moresco}\ \emph {et~al.}(2016)\citenamefont
  {Moresco}, \citenamefont {Pozzetti}, \citenamefont {Cimatti}, \citenamefont
  {Jimenez}, \citenamefont {Maraston}, \citenamefont {Verde}, \citenamefont
  {Thomas}, \citenamefont {Citro}, \citenamefont {Tojeiro},\ and\ \citenamefont
  {Wilkinson}}]{Moresco:2016mzx}%
  \BibitemOpen
  \bibfield  {author} {\bibinfo {author} {\bibfnamefont {M.}~\bibnamefont
  {Moresco}}, \bibinfo {author} {\bibfnamefont {L.}~\bibnamefont {Pozzetti}},
  \bibinfo {author} {\bibfnamefont {A.}~\bibnamefont {Cimatti}}, \bibinfo
  {author} {\bibfnamefont {R.}~\bibnamefont {Jimenez}}, \bibinfo {author}
  {\bibfnamefont {C.}~\bibnamefont {Maraston}}, \bibinfo {author}
  {\bibfnamefont {L.}~\bibnamefont {Verde}}, \bibinfo {author} {\bibfnamefont
  {D.}~\bibnamefont {Thomas}}, \bibinfo {author} {\bibfnamefont
  {A.}~\bibnamefont {Citro}}, \bibinfo {author} {\bibfnamefont
  {R.}~\bibnamefont {Tojeiro}}, \ and\ \bibinfo {author} {\bibfnamefont
  {D.}~\bibnamefont {Wilkinson}},\ }\href {\doibase
  10.1088/1475-7516/2016/05/014} {\bibfield  {journal} {\bibinfo  {journal}
  {JCAP}\ }\textbf {\bibinfo {volume} {1605}},\ \bibinfo {pages} {014}
  (\bibinfo {year} {2016})},\ \Eprint {http://arxiv.org/abs/1601.01701}
  {arXiv:1601.01701 [astro-ph.CO]} \BibitemShut {NoStop}%
\bibitem [{\citenamefont {Eisenstein}\ and\ \citenamefont
  {Hu}(1998)}]{Eisenstein_1998}%
  \BibitemOpen
  \bibfield  {author} {\bibinfo {author} {\bibfnamefont {D.~J.}\ \bibnamefont
  {Eisenstein}}\ and\ \bibinfo {author} {\bibfnamefont {W.}~\bibnamefont
  {Hu}},\ }\href {\doibase 10.1086/305424} {\bibfield  {journal} {\bibinfo
  {journal} {The Astrophysical Journal}\ }\textbf {\bibinfo {volume} {496}},\
  \bibinfo {pages} {605} (\bibinfo {year} {1998})}\BibitemShut {NoStop}%
\bibitem [{\citenamefont {Beutler}\ \emph {et~al.}(2011)\citenamefont
  {Beutler}, \citenamefont {Blake}, \citenamefont {Colless}, \citenamefont
  {Jones}, \citenamefont {Staveley-Smith}, \citenamefont {Campbell},
  \citenamefont {Parker}, \citenamefont {Saunders},\ and\ \citenamefont
  {Watson}}]{Beutler_2011}%
  \BibitemOpen
  \bibfield  {author} {\bibinfo {author} {\bibfnamefont {F.}~\bibnamefont
  {Beutler}}, \bibinfo {author} {\bibfnamefont {C.}~\bibnamefont {Blake}},
  \bibinfo {author} {\bibfnamefont {M.}~\bibnamefont {Colless}}, \bibinfo
  {author} {\bibfnamefont {D.~H.}\ \bibnamefont {Jones}}, \bibinfo {author}
  {\bibfnamefont {L.}~\bibnamefont {Staveley-Smith}}, \bibinfo {author}
  {\bibfnamefont {L.}~\bibnamefont {Campbell}}, \bibinfo {author}
  {\bibfnamefont {Q.}~\bibnamefont {Parker}}, \bibinfo {author} {\bibfnamefont
  {W.}~\bibnamefont {Saunders}}, \ and\ \bibinfo {author} {\bibfnamefont
  {F.}~\bibnamefont {Watson}},\ }\href {\doibase
  10.1111/j.1365-2966.2011.19250.x} {\bibfield  {journal} {\bibinfo  {journal}
  {Monthly Notices of the Royal Astronomical Society}\ }\textbf {\bibinfo
  {volume} {416}},\ \bibinfo {pages} {3017} (\bibinfo {year}
  {2011})}\BibitemShut {NoStop}%
\bibitem [{\citenamefont {Ross}\ \emph {et~al.}(2015)\citenamefont {Ross},
  \citenamefont {Samushia}, \citenamefont {Howlett}, \citenamefont {Percival},
  \citenamefont {Burden},\ and\ \citenamefont {Manera}}]{Ross_2015}%
  \BibitemOpen
  \bibfield  {author} {\bibinfo {author} {\bibfnamefont {A.~J.}\ \bibnamefont
  {Ross}}, \bibinfo {author} {\bibfnamefont {L.}~\bibnamefont {Samushia}},
  \bibinfo {author} {\bibfnamefont {C.}~\bibnamefont {Howlett}}, \bibinfo
  {author} {\bibfnamefont {W.~J.}\ \bibnamefont {Percival}}, \bibinfo {author}
  {\bibfnamefont {A.}~\bibnamefont {Burden}}, \ and\ \bibinfo {author}
  {\bibfnamefont {M.}~\bibnamefont {Manera}},\ }\href {\doibase
  10.1093/mnras/stv154} {\bibfield  {journal} {\bibinfo  {journal} {Monthly
  Notices of the Royal Astronomical Society}\ }\textbf {\bibinfo {volume}
  {449}},\ \bibinfo {pages} {835} (\bibinfo {year} {2015})}\BibitemShut
  {NoStop}%
\bibitem [{\citenamefont {de~Sainte~Agathe}\ \emph {et~al.}(2019)\citenamefont
  {de~Sainte~Agathe}, \citenamefont {Balland}, \citenamefont {du~Mas~des
  Bourboux}, \citenamefont {Busca}, \citenamefont {Blomqvist}, \citenamefont
  {Guy}, \citenamefont {Rich}, \citenamefont {Font-Ribera}, \citenamefont
  {Pieri}, \citenamefont {Bautista}, \citenamefont {Dawson}, \citenamefont
  {Goff}, \citenamefont {de~la Macorra}, \citenamefont {Palanque-Delabrouille},
  \citenamefont {Percival}, \citenamefont {P{\'{e}}rez-R{\`{a}}fols},
  \citenamefont {Schneider}, \citenamefont {Slosar},\ and\ \citenamefont
  {Y{\`{e}}che}}]{de_Sainte_Agathe_2019}%
  \BibitemOpen
  \bibfield  {author} {\bibinfo {author} {\bibfnamefont {V.}~\bibnamefont
  {de~Sainte~Agathe}}, \bibinfo {author} {\bibfnamefont {C.}~\bibnamefont
  {Balland}}, \bibinfo {author} {\bibfnamefont {H.}~\bibnamefont {du~Mas~des
  Bourboux}}, \bibinfo {author} {\bibfnamefont {N.~G.}\ \bibnamefont {Busca}},
  \bibinfo {author} {\bibfnamefont {M.}~\bibnamefont {Blomqvist}}, \bibinfo
  {author} {\bibfnamefont {J.}~\bibnamefont {Guy}}, \bibinfo {author}
  {\bibfnamefont {J.}~\bibnamefont {Rich}}, \bibinfo {author} {\bibfnamefont
  {A.}~\bibnamefont {Font-Ribera}}, \bibinfo {author} {\bibfnamefont {M.~M.}\
  \bibnamefont {Pieri}}, \bibinfo {author} {\bibfnamefont {J.~E.}\ \bibnamefont
  {Bautista}}, \bibinfo {author} {\bibfnamefont {K.}~\bibnamefont {Dawson}},
  \bibinfo {author} {\bibfnamefont {J.-M.~L.}\ \bibnamefont {Goff}}, \bibinfo
  {author} {\bibfnamefont {A.}~\bibnamefont {de~la Macorra}}, \bibinfo {author}
  {\bibfnamefont {N.}~\bibnamefont {Palanque-Delabrouille}}, \bibinfo {author}
  {\bibfnamefont {W.~J.}\ \bibnamefont {Percival}}, \bibinfo {author}
  {\bibfnamefont {I.}~\bibnamefont {P{\'{e}}rez-R{\`{a}}fols}}, \bibinfo
  {author} {\bibfnamefont {D.~P.}\ \bibnamefont {Schneider}}, \bibinfo {author}
  {\bibfnamefont {A.}~\bibnamefont {Slosar}}, \ and\ \bibinfo {author}
  {\bibfnamefont {C.}~\bibnamefont {Y{\`{e}}che}},\ }\href {\doibase
  10.1051/0004-6361/201935638} {\bibfield  {journal} {\bibinfo  {journal}
  {Astronomy {\&} Astrophysics}\ }\textbf {\bibinfo {volume} {629}},\ \bibinfo
  {pages} {A85} (\bibinfo {year} {2019})}\BibitemShut {NoStop}%
\bibitem [{\citenamefont {Alam}\ \emph {et~al.}(2017)\citenamefont {Alam},
  \citenamefont {Ata}, \citenamefont {Bailey}, \citenamefont {Beutler},
  \citenamefont {Bizyaev}, \citenamefont {Blazek}, \citenamefont {Bolton},
  \citenamefont {Brownstein}, \citenamefont {Burden}, \citenamefont {Chuang},
  \citenamefont {Comparat}, \citenamefont {Cuesta}, \citenamefont {Dawson},
  \citenamefont {Eisenstein}, \citenamefont {Escoffier}, \citenamefont
  {Gil-Mar{\'{\i}}n}, \citenamefont {Grieb}, \citenamefont {Hand},
  \citenamefont {Ho}, \citenamefont {Kinemuchi}, \citenamefont {Kirkby},
  \citenamefont {Kitaura}, \citenamefont {Malanushenko}, \citenamefont
  {Malanushenko}, \citenamefont {Maraston}, \citenamefont {McBride},
  \citenamefont {Nichol}, \citenamefont {Olmstead}, \citenamefont {Oravetz},
  \citenamefont {Padmanabhan}, \citenamefont {Palanque-Delabrouille},
  \citenamefont {Pan}, \citenamefont {Pellejero-Ibanez}, \citenamefont
  {Percival}, \citenamefont {Petitjean}, \citenamefont {Prada}, \citenamefont
  {Price-Whelan}, \citenamefont {Reid}, \citenamefont
  {Rodr{\'{\i}}guez-Torres}, \citenamefont {Roe}, \citenamefont {Ross},
  \citenamefont {Ross}, \citenamefont {Rossi}, \citenamefont
  {Rubi{\~{n}}o-Mart{\'{\i}}n}, \citenamefont {Saito}, \citenamefont
  {Salazar-Albornoz}, \citenamefont {Samushia}, \citenamefont {S{\'{a}}nchez},
  \citenamefont {Satpathy}, \citenamefont {Schlegel}, \citenamefont
  {Schneider}, \citenamefont {Sc{\'{o}}ccola}, \citenamefont {Seo},
  \citenamefont {Sheldon}, \citenamefont {Simmons}, \citenamefont {Slosar},
  \citenamefont {Strauss}, \citenamefont {Swanson}, \citenamefont {Thomas},
  \citenamefont {Tinker}, \citenamefont {Tojeiro}, \citenamefont
  {Maga{\~{n}}a}, \citenamefont {Vazquez}, \citenamefont {Verde}, \citenamefont
  {Wake}, \citenamefont {Wang}, \citenamefont {Weinberg}, \citenamefont
  {White}, \citenamefont {Wood-Vasey}, \citenamefont {Y{\`{e}}che},
  \citenamefont {Zehavi}, \citenamefont {Zhai},\ and\ \citenamefont
  {Zhao}}]{Alam_2017}%
  \BibitemOpen
  \bibfield  {author} {\bibinfo {author} {\bibfnamefont {S.}~\bibnamefont
  {Alam}}, \bibinfo {author} {\bibfnamefont {M.}~\bibnamefont {Ata}}, \bibinfo
  {author} {\bibfnamefont {S.}~\bibnamefont {Bailey}}, \bibinfo {author}
  {\bibfnamefont {F.}~\bibnamefont {Beutler}}, \bibinfo {author} {\bibfnamefont
  {D.}~\bibnamefont {Bizyaev}}, \bibinfo {author} {\bibfnamefont {J.~A.}\
  \bibnamefont {Blazek}}, \bibinfo {author} {\bibfnamefont {A.~S.}\
  \bibnamefont {Bolton}}, \bibinfo {author} {\bibfnamefont {J.~R.}\
  \bibnamefont {Brownstein}}, \bibinfo {author} {\bibfnamefont
  {A.}~\bibnamefont {Burden}}, \bibinfo {author} {\bibfnamefont {C.-H.}\
  \bibnamefont {Chuang}}, \bibinfo {author} {\bibfnamefont {J.}~\bibnamefont
  {Comparat}}, \bibinfo {author} {\bibfnamefont {A.~J.}\ \bibnamefont
  {Cuesta}}, \bibinfo {author} {\bibfnamefont {K.~S.}\ \bibnamefont {Dawson}},
  \bibinfo {author} {\bibfnamefont {D.~J.}\ \bibnamefont {Eisenstein}},
  \bibinfo {author} {\bibfnamefont {S.}~\bibnamefont {Escoffier}}, \bibinfo
  {author} {\bibfnamefont {H.}~\bibnamefont {Gil-Mar{\'{\i}}n}}, \bibinfo
  {author} {\bibfnamefont {J.~N.}\ \bibnamefont {Grieb}}, \bibinfo {author}
  {\bibfnamefont {N.}~\bibnamefont {Hand}}, \bibinfo {author} {\bibfnamefont
  {S.}~\bibnamefont {Ho}}, \bibinfo {author} {\bibfnamefont {K.}~\bibnamefont
  {Kinemuchi}}, \bibinfo {author} {\bibfnamefont {D.}~\bibnamefont {Kirkby}},
  \bibinfo {author} {\bibfnamefont {F.}~\bibnamefont {Kitaura}}, \bibinfo
  {author} {\bibfnamefont {E.}~\bibnamefont {Malanushenko}}, \bibinfo {author}
  {\bibfnamefont {V.}~\bibnamefont {Malanushenko}}, \bibinfo {author}
  {\bibfnamefont {C.}~\bibnamefont {Maraston}}, \bibinfo {author}
  {\bibfnamefont {C.~K.}\ \bibnamefont {McBride}}, \bibinfo {author}
  {\bibfnamefont {R.~C.}\ \bibnamefont {Nichol}}, \bibinfo {author}
  {\bibfnamefont {M.~D.}\ \bibnamefont {Olmstead}}, \bibinfo {author}
  {\bibfnamefont {D.}~\bibnamefont {Oravetz}}, \bibinfo {author} {\bibfnamefont
  {N.}~\bibnamefont {Padmanabhan}}, \bibinfo {author} {\bibfnamefont
  {N.}~\bibnamefont {Palanque-Delabrouille}}, \bibinfo {author} {\bibfnamefont
  {K.}~\bibnamefont {Pan}}, \bibinfo {author} {\bibfnamefont {M.}~\bibnamefont
  {Pellejero-Ibanez}}, \bibinfo {author} {\bibfnamefont {W.~J.}\ \bibnamefont
  {Percival}}, \bibinfo {author} {\bibfnamefont {P.}~\bibnamefont {Petitjean}},
  \bibinfo {author} {\bibfnamefont {F.}~\bibnamefont {Prada}}, \bibinfo
  {author} {\bibfnamefont {A.~M.}\ \bibnamefont {Price-Whelan}}, \bibinfo
  {author} {\bibfnamefont {B.~A.}\ \bibnamefont {Reid}}, \bibinfo {author}
  {\bibfnamefont {S.~A.}\ \bibnamefont {Rodr{\'{\i}}guez-Torres}}, \bibinfo
  {author} {\bibfnamefont {N.~A.}\ \bibnamefont {Roe}}, \bibinfo {author}
  {\bibfnamefont {A.~J.}\ \bibnamefont {Ross}}, \bibinfo {author}
  {\bibfnamefont {N.~P.}\ \bibnamefont {Ross}}, \bibinfo {author}
  {\bibfnamefont {G.}~\bibnamefont {Rossi}}, \bibinfo {author} {\bibfnamefont
  {J.~A.}\ \bibnamefont {Rubi{\~{n}}o-Mart{\'{\i}}n}}, \bibinfo {author}
  {\bibfnamefont {S.}~\bibnamefont {Saito}}, \bibinfo {author} {\bibfnamefont
  {S.}~\bibnamefont {Salazar-Albornoz}}, \bibinfo {author} {\bibfnamefont
  {L.}~\bibnamefont {Samushia}}, \bibinfo {author} {\bibfnamefont {A.~G.}\
  \bibnamefont {S{\'{a}}nchez}}, \bibinfo {author} {\bibfnamefont
  {S.}~\bibnamefont {Satpathy}}, \bibinfo {author} {\bibfnamefont {D.~J.}\
  \bibnamefont {Schlegel}}, \bibinfo {author} {\bibfnamefont {D.~P.}\
  \bibnamefont {Schneider}}, \bibinfo {author} {\bibfnamefont {C.~G.}\
  \bibnamefont {Sc{\'{o}}ccola}}, \bibinfo {author} {\bibfnamefont {H.-J.}\
  \bibnamefont {Seo}}, \bibinfo {author} {\bibfnamefont {E.~S.}\ \bibnamefont
  {Sheldon}}, \bibinfo {author} {\bibfnamefont {A.}~\bibnamefont {Simmons}},
  \bibinfo {author} {\bibfnamefont {A.}~\bibnamefont {Slosar}}, \bibinfo
  {author} {\bibfnamefont {M.~A.}\ \bibnamefont {Strauss}}, \bibinfo {author}
  {\bibfnamefont {M.~E.~C.}\ \bibnamefont {Swanson}}, \bibinfo {author}
  {\bibfnamefont {D.}~\bibnamefont {Thomas}}, \bibinfo {author} {\bibfnamefont
  {J.~L.}\ \bibnamefont {Tinker}}, \bibinfo {author} {\bibfnamefont
  {R.}~\bibnamefont {Tojeiro}}, \bibinfo {author} {\bibfnamefont {M.~V.}\
  \bibnamefont {Maga{\~{n}}a}}, \bibinfo {author} {\bibfnamefont {J.~A.}\
  \bibnamefont {Vazquez}}, \bibinfo {author} {\bibfnamefont {L.}~\bibnamefont
  {Verde}}, \bibinfo {author} {\bibfnamefont {D.~A.}\ \bibnamefont {Wake}},
  \bibinfo {author} {\bibfnamefont {Y.}~\bibnamefont {Wang}}, \bibinfo {author}
  {\bibfnamefont {D.~H.}\ \bibnamefont {Weinberg}}, \bibinfo {author}
  {\bibfnamefont {M.}~\bibnamefont {White}}, \bibinfo {author} {\bibfnamefont
  {W.~M.}\ \bibnamefont {Wood-Vasey}}, \bibinfo {author} {\bibfnamefont
  {C.}~\bibnamefont {Y{\`{e}}che}}, \bibinfo {author} {\bibfnamefont
  {I.}~\bibnamefont {Zehavi}}, \bibinfo {author} {\bibfnamefont
  {Z.}~\bibnamefont {Zhai}}, \ and\ \bibinfo {author} {\bibfnamefont {G.-B.}\
  \bibnamefont {Zhao}},\ }\href {\doibase 10.1093/mnras/stx721} {\bibfield
  {journal} {\bibinfo  {journal} {Monthly Notices of the Royal Astronomical
  Society}\ }\textbf {\bibinfo {volume} {470}},\ \bibinfo {pages} {2617}
  (\bibinfo {year} {2017})}\BibitemShut {NoStop}%
\bibitem [{\citenamefont {Blomqvist}\ \emph {et~al.}(2019)\citenamefont
  {Blomqvist}, \citenamefont {du~Mas~des Bourboux}, \citenamefont {Busca},
  \citenamefont {de~Sainte~Agathe}, \citenamefont {Rich}, \citenamefont
  {Balland}, \citenamefont {Bautista}, \citenamefont {Dawson}, \citenamefont
  {Font-Ribera}, \citenamefont {Guy}, \citenamefont {Goff}, \citenamefont
  {Palanque-Delabrouille}, \citenamefont {Percival}, \citenamefont
  {P{\'{e}}rez-R{\`{a}}fols}, \citenamefont {Pieri}, \citenamefont {Schneider},
  \citenamefont {Slosar},\ and\ \citenamefont {Y{\`{e}}che}}]{Blomqvist_2019}%
  \BibitemOpen
  \bibfield  {author} {\bibinfo {author} {\bibfnamefont {M.}~\bibnamefont
  {Blomqvist}}, \bibinfo {author} {\bibfnamefont {H.}~\bibnamefont {du~Mas~des
  Bourboux}}, \bibinfo {author} {\bibfnamefont {N.~G.}\ \bibnamefont {Busca}},
  \bibinfo {author} {\bibfnamefont {V.}~\bibnamefont {de~Sainte~Agathe}},
  \bibinfo {author} {\bibfnamefont {J.}~\bibnamefont {Rich}}, \bibinfo {author}
  {\bibfnamefont {C.}~\bibnamefont {Balland}}, \bibinfo {author} {\bibfnamefont
  {J.~E.}\ \bibnamefont {Bautista}}, \bibinfo {author} {\bibfnamefont
  {K.}~\bibnamefont {Dawson}}, \bibinfo {author} {\bibfnamefont
  {A.}~\bibnamefont {Font-Ribera}}, \bibinfo {author} {\bibfnamefont
  {J.}~\bibnamefont {Guy}}, \bibinfo {author} {\bibfnamefont {J.-M.~L.}\
  \bibnamefont {Goff}}, \bibinfo {author} {\bibfnamefont {N.}~\bibnamefont
  {Palanque-Delabrouille}}, \bibinfo {author} {\bibfnamefont {W.~J.}\
  \bibnamefont {Percival}}, \bibinfo {author} {\bibfnamefont {I.}~\bibnamefont
  {P{\'{e}}rez-R{\`{a}}fols}}, \bibinfo {author} {\bibfnamefont {M.~M.}\
  \bibnamefont {Pieri}}, \bibinfo {author} {\bibfnamefont {D.~P.}\ \bibnamefont
  {Schneider}}, \bibinfo {author} {\bibfnamefont {A.}~\bibnamefont {Slosar}}, \
  and\ \bibinfo {author} {\bibfnamefont {C.}~\bibnamefont {Y{\`{e}}che}},\
  }\href {\doibase 10.1051/0004-6361/201935641} {\bibfield  {journal} {\bibinfo
   {journal} {Astronomy {\&} Astrophysics}\ }\textbf {\bibinfo {volume}
  {629}},\ \bibinfo {pages} {A86} (\bibinfo {year} {2019})}\BibitemShut
  {NoStop}%
\bibitem [{\citenamefont {{Ch{\'a}vez}}\ \emph {et~al.}(2014)\citenamefont
  {{Ch{\'a}vez}}, \citenamefont {{Terlevich}}, \citenamefont {{Terlevich}},
  \citenamefont {{Bresolin}}, \citenamefont {{Melnick}}, \citenamefont
  {{Plionis}},\ and\ \citenamefont {{Basilakos}}}]{Chavez2014}%
  \BibitemOpen
  \bibfield  {author} {\bibinfo {author} {\bibfnamefont {R.}~\bibnamefont
  {{Ch{\'a}vez}}}, \bibinfo {author} {\bibfnamefont {R.}~\bibnamefont
  {{Terlevich}}}, \bibinfo {author} {\bibfnamefont {E.}~\bibnamefont
  {{Terlevich}}}, \bibinfo {author} {\bibfnamefont {F.}~\bibnamefont
  {{Bresolin}}}, \bibinfo {author} {\bibfnamefont {J.}~\bibnamefont
  {{Melnick}}}, \bibinfo {author} {\bibfnamefont {M.}~\bibnamefont
  {{Plionis}}}, \ and\ \bibinfo {author} {\bibfnamefont {S.}~\bibnamefont
  {{Basilakos}}},\ }\href {\doibase 10.1093/mnras/stu987} {\bibfield  {journal}
  {\bibinfo  {journal} {Monthly Notices of the Royal Astronomical Society}\
  }\textbf {\bibinfo {volume} {442}},\ \bibinfo {pages} {3565} (\bibinfo {year}
  {2014})},\ \Eprint {http://arxiv.org/abs/1405.4010} {arXiv:1405.4010
  [astro-ph.GA]} \BibitemShut {NoStop}%
\bibitem [{\citenamefont {{Gonz{\'a}lez-Mor{\'a}n}}\ \emph
  {et~al.}(2019)\citenamefont {{Gonz{\'a}lez-Mor{\'a}n}}, \citenamefont
  {{Ch{\'a}vez}}, \citenamefont {{Terlevich}}, \citenamefont {{Terlevich}},
  \citenamefont {{Bresolin}}, \citenamefont {{Fern{\'a}ndez-Arenas}},
  \citenamefont {{Plionis}}, \citenamefont {{Basilakos}}, \citenamefont
  {{Melnick}},\ and\ \citenamefont {{Telles}}}]{GonzalezMoran2019}%
  \BibitemOpen
  \bibfield  {author} {\bibinfo {author} {\bibfnamefont {A.~L.}\ \bibnamefont
  {{Gonz{\'a}lez-Mor{\'a}n}}}, \bibinfo {author} {\bibfnamefont
  {R.}~\bibnamefont {{Ch{\'a}vez}}}, \bibinfo {author} {\bibfnamefont
  {R.}~\bibnamefont {{Terlevich}}}, \bibinfo {author} {\bibfnamefont
  {E.}~\bibnamefont {{Terlevich}}}, \bibinfo {author} {\bibfnamefont
  {F.}~\bibnamefont {{Bresolin}}}, \bibinfo {author} {\bibfnamefont
  {D.}~\bibnamefont {{Fern{\'a}ndez-Arenas}}}, \bibinfo {author} {\bibfnamefont
  {M.}~\bibnamefont {{Plionis}}}, \bibinfo {author} {\bibfnamefont
  {S.}~\bibnamefont {{Basilakos}}}, \bibinfo {author} {\bibfnamefont
  {J.}~\bibnamefont {{Melnick}}}, \ and\ \bibinfo {author} {\bibfnamefont
  {E.}~\bibnamefont {{Telles}}},\ }\href {\doibase 10.1093/mnras/stz1577}
  {\bibfield  {journal} {\bibinfo  {journal} {Monthly Notices of the Royal
  Astronomical Society}\ }\textbf {\bibinfo {volume} {487}},\ \bibinfo {pages}
  {4669} (\bibinfo {year} {2019})},\ \Eprint {http://arxiv.org/abs/1906.02195}
  {arXiv:1906.02195 [astro-ph.GA]} \BibitemShut {NoStop}%
\bibitem [{\citenamefont {Gonz\'alez-Mor\'an}\ \emph
  {et~al.}(2021)\citenamefont {Gonz\'alez-Mor\'an}, \citenamefont {Ch\'avez},
  \citenamefont {Terlevich}, \citenamefont {Terlevich}, \citenamefont
  {Fern\'andez-Arenas}, \citenamefont {Bresolin}, \citenamefont {Plionis},
  \citenamefont {Melnick}, \citenamefont {Basilakos},\ and\ \citenamefont
  {Telles}}]{Gonzalez-Moran:2021drc}%
  \BibitemOpen
  \bibfield  {author} {\bibinfo {author} {\bibfnamefont {A.~L.}\ \bibnamefont
  {Gonz\'alez-Mor\'an}}, \bibinfo {author} {\bibfnamefont {R.}~\bibnamefont
  {Ch\'avez}}, \bibinfo {author} {\bibfnamefont {E.}~\bibnamefont {Terlevich}},
  \bibinfo {author} {\bibfnamefont {R.}~\bibnamefont {Terlevich}}, \bibinfo
  {author} {\bibfnamefont {D.}~\bibnamefont {Fern\'andez-Arenas}}, \bibinfo
  {author} {\bibfnamefont {F.}~\bibnamefont {Bresolin}}, \bibinfo {author}
  {\bibfnamefont {M.}~\bibnamefont {Plionis}}, \bibinfo {author} {\bibfnamefont
  {J.}~\bibnamefont {Melnick}}, \bibinfo {author} {\bibfnamefont
  {S.}~\bibnamefont {Basilakos}}, \ and\ \bibinfo {author} {\bibfnamefont
  {E.}~\bibnamefont {Telles}},\ }\href {\doibase 10.1093/mnras/stab1385}
  {\bibfield  {journal} {\bibinfo  {journal} {MNRAS}\ }\textbf {\bibinfo
  {volume} {505}},\ \bibinfo {pages} {1441} (\bibinfo {year} {2021})},\ \Eprint
  {http://arxiv.org/abs/2105.04025} {arXiv:2105.04025 [astro-ph.CO]}
  \BibitemShut {NoStop}%
\bibitem [{\citenamefont {Scolnic}\ \emph {et~al.}(2018)\citenamefont {Scolnic}
  \emph {et~al.}}]{Scolnic:2018}%
  \BibitemOpen
  \bibfield  {author} {\bibinfo {author} {\bibfnamefont {D.~M.}\ \bibnamefont
  {Scolnic}} \emph {et~al.},\ }\href {\doibase 10.3847/1538-4357/aab9bb}
  {\bibfield  {journal} {\bibinfo  {journal} {Astrophys. J.}\ }\textbf
  {\bibinfo {volume} {859}},\ \bibinfo {pages} {101} (\bibinfo {year}
  {2018})},\ \Eprint {http://arxiv.org/abs/1710.00845} {arXiv:1710.00845
  [astro-ph.CO]} \BibitemShut {NoStop}%
\bibitem [{\citenamefont {Peacock}(1998)}]{peacock_1998}%
  \BibitemOpen
  \bibfield  {author} {\bibinfo {author} {\bibfnamefont {J.~A.}\ \bibnamefont
  {Peacock}},\ }\href {\doibase 10.1017/CBO9780511804533} {\emph {\bibinfo
  {title} {Cosmological Physics}}}\ (\bibinfo  {publisher} {Cambridge
  University Press},\ \bibinfo {year} {1998})\BibitemShut {NoStop}%
\bibitem [{\citenamefont {Yang}\ \emph
  {et~al.}(2019{\natexlab{b}})\citenamefont {Yang}, \citenamefont {Pan},
  \citenamefont {Di~Valentino}, \citenamefont {Saridakis},\ and\ \citenamefont
  {Chakraborty}}]{Yang:2019}%
  \BibitemOpen
  \bibfield  {author} {\bibinfo {author} {\bibfnamefont {W.}~\bibnamefont
  {Yang}}, \bibinfo {author} {\bibfnamefont {S.}~\bibnamefont {Pan}}, \bibinfo
  {author} {\bibfnamefont {E.}~\bibnamefont {Di~Valentino}}, \bibinfo {author}
  {\bibfnamefont {E.~N.}\ \bibnamefont {Saridakis}}, \ and\ \bibinfo {author}
  {\bibfnamefont {S.}~\bibnamefont {Chakraborty}},\ }\href {\doibase
  10.1103/PhysRevD.99.043543} {\bibfield  {journal} {\bibinfo  {journal} {Phys.
  Rev. D}\ }\textbf {\bibinfo {volume} {99}},\ \bibinfo {pages} {043543}
  (\bibinfo {year} {2019}{\natexlab{b}})}\BibitemShut {NoStop}%
\bibitem [{\citenamefont {Kass}\ and\ \citenamefont
  {Raftery}(1995)}]{BayesFactor:1995}%
  \BibitemOpen
  \bibfield  {author} {\bibinfo {author} {\bibfnamefont {R.~E.}\ \bibnamefont
  {Kass}}\ and\ \bibinfo {author} {\bibfnamefont {A.~E.}\ \bibnamefont
  {Raftery}},\ }\href {\doibase 10.1080/01621459.1995.10476572} {\bibfield
  {journal} {\bibinfo  {journal} {Journal of the American Statistical
  Association}\ }\textbf {\bibinfo {volume} {90}},\ \bibinfo {pages} {773}
  (\bibinfo {year} {1995})}\BibitemShut {NoStop}%
\bibitem [{\citenamefont {Heavens}\ \emph
  {et~al.}(2017{\natexlab{a}})\citenamefont {Heavens}, \citenamefont {Fantaye},
  \citenamefont {Sellentin}, \citenamefont {Eggers}, \citenamefont {Hosenie},
  \citenamefont {Kroon},\ and\ \citenamefont {Mootoovaloo}}]{Heavens:2017}%
  \BibitemOpen
  \bibfield  {author} {\bibinfo {author} {\bibfnamefont {A.}~\bibnamefont
  {Heavens}}, \bibinfo {author} {\bibfnamefont {Y.}~\bibnamefont {Fantaye}},
  \bibinfo {author} {\bibfnamefont {E.}~\bibnamefont {Sellentin}}, \bibinfo
  {author} {\bibfnamefont {H.}~\bibnamefont {Eggers}}, \bibinfo {author}
  {\bibfnamefont {Z.}~\bibnamefont {Hosenie}}, \bibinfo {author} {\bibfnamefont
  {S.}~\bibnamefont {Kroon}}, \ and\ \bibinfo {author} {\bibfnamefont
  {A.}~\bibnamefont {Mootoovaloo}},\ }\href {\doibase
  10.1103/PhysRevLett.119.101301} {\bibfield  {journal} {\bibinfo  {journal}
  {Phys. Rev. Lett.}\ }\textbf {\bibinfo {volume} {119}},\ \bibinfo {pages}
  {101301} (\bibinfo {year} {2017}{\natexlab{a}})}\BibitemShut {NoStop}%
\bibitem [{\citenamefont {Heavens}\ \emph
  {et~al.}(2017{\natexlab{b}})\citenamefont {Heavens}, \citenamefont {Fantaye},
  \citenamefont {Mootoovaloo}, \citenamefont {Eggers}, \citenamefont {Hosenie},
  \citenamefont {Kroon},\ and\ \citenamefont {Sellentin}}]{Heavens:2017arxiv}%
  \BibitemOpen
  \bibfield  {author} {\bibinfo {author} {\bibfnamefont {A.}~\bibnamefont
  {Heavens}}, \bibinfo {author} {\bibfnamefont {Y.}~\bibnamefont {Fantaye}},
  \bibinfo {author} {\bibfnamefont {A.}~\bibnamefont {Mootoovaloo}}, \bibinfo
  {author} {\bibfnamefont {H.}~\bibnamefont {Eggers}}, \bibinfo {author}
  {\bibfnamefont {Z.}~\bibnamefont {Hosenie}}, \bibinfo {author} {\bibfnamefont
  {S.}~\bibnamefont {Kroon}}, \ and\ \bibinfo {author} {\bibfnamefont
  {E.}~\bibnamefont {Sellentin}},\ }\href {\doibase 10.48550/ARXIV.1704.03472}
  {\enquote {\bibinfo {title} {Marginal likelihoods from monte carlo markov
  chains},}\ } (\bibinfo {year} {2017}{\natexlab{b}})\BibitemShut {NoStop}%
\bibitem [{\citenamefont {Cardenas}\ and\ \citenamefont
  {Rivera}(2012)}]{Cardenas:2012mw}%
  \BibitemOpen
  \bibfield  {author} {\bibinfo {author} {\bibfnamefont {V.~H.}\ \bibnamefont
  {Cardenas}}\ and\ \bibinfo {author} {\bibfnamefont {M.}~\bibnamefont
  {Rivera}},\ }\href {\doibase 10.1016/j.physletb.2012.03.004} {\bibfield
  {journal} {\bibinfo  {journal} {Phys. Lett. B}\ }\textbf {\bibinfo {volume}
  {710}},\ \bibinfo {pages} {251} (\bibinfo {year} {2012})},\ \Eprint
  {http://arxiv.org/abs/1203.0984} {arXiv:1203.0984 [astro-ph.CO]} \BibitemShut
  {NoStop}%
\bibitem [{\citenamefont {Cardenas}\ \emph {et~al.}(2013)\citenamefont
  {Cardenas}, \citenamefont {Bernal},\ and\ \citenamefont
  {Bonilla}}]{Cardenas:2013roa}%
  \BibitemOpen
  \bibfield  {author} {\bibinfo {author} {\bibfnamefont {V.~H.}\ \bibnamefont
  {Cardenas}}, \bibinfo {author} {\bibfnamefont {C.}~\bibnamefont {Bernal}}, \
  and\ \bibinfo {author} {\bibfnamefont {A.}~\bibnamefont {Bonilla}},\ }\href
  {\doibase 10.1093/mnras/stt983} {\bibfield  {journal} {\bibinfo  {journal}
  {Mon. Not. Roy. Astron. Soc.}\ }\textbf {\bibinfo {volume} {433}},\ \bibinfo
  {pages} {3534} (\bibinfo {year} {2013})},\ \Eprint
  {http://arxiv.org/abs/1306.0779} {arXiv:1306.0779 [astro-ph.CO]} \BibitemShut
  {NoStop}%
\bibitem [{\citenamefont {Magana}\ \emph {et~al.}(2017)\citenamefont {Magana},
  \citenamefont {Motta}, \citenamefont {Cardenas},\ and\ \citenamefont
  {Foex}}]{ElBoss}%
  \BibitemOpen
  \bibfield  {author} {\bibinfo {author} {\bibfnamefont {J.}~\bibnamefont
  {Magana}}, \bibinfo {author} {\bibfnamefont {V.}~\bibnamefont {Motta}},
  \bibinfo {author} {\bibfnamefont {V.~H.}\ \bibnamefont {Cardenas}}, \ and\
  \bibinfo {author} {\bibfnamefont {G.}~\bibnamefont {Foex}},\ }\href {\doibase
  10.1093/mnras/stx750} {\bibfield  {journal} {\bibinfo  {journal} {Mon. Not.
  Roy. Astron. Soc.}\ }\textbf {\bibinfo {volume} {469}},\ \bibinfo {pages}
  {47} (\bibinfo {year} {2017})},\ \Eprint {http://arxiv.org/abs/1703.08521}
  {arXiv:1703.08521 [astro-ph.CO]} \BibitemShut {NoStop}%
\bibitem [{\citenamefont {Zhang}\ and\ \citenamefont
  {Xia}(2018)}]{Zhang:2017jvo}%
  \BibitemOpen
  \bibfield  {author} {\bibinfo {author} {\bibfnamefont {M.-J.}\ \bibnamefont
  {Zhang}}\ and\ \bibinfo {author} {\bibfnamefont {J.-Q.}\ \bibnamefont
  {Xia}},\ }\href {\doibase 10.1016/j.nuclphysb.2018.02.020} {\bibfield
  {journal} {\bibinfo  {journal} {Nucl. Phys. B}\ }\textbf {\bibinfo {volume}
  {929}},\ \bibinfo {pages} {438} (\bibinfo {year} {2018})},\ \Eprint
  {http://arxiv.org/abs/1701.04973} {arXiv:1701.04973 [astro-ph.CO]}
  \BibitemShut {NoStop}%
\bibitem [{\citenamefont {Bolotin}\ \emph {et~al.}(2022)\citenamefont
  {Bolotin}, \citenamefont {Cherkaskiy}, \citenamefont {Konchatnyi},
  \citenamefont {Pan},\ and\ \citenamefont {Yang}}]{Bolotin:2020qbx}%
  \BibitemOpen
  \bibfield  {author} {\bibinfo {author} {\bibfnamefont {Y.~L.}\ \bibnamefont
  {Bolotin}}, \bibinfo {author} {\bibfnamefont {V.~A.}\ \bibnamefont
  {Cherkaskiy}}, \bibinfo {author} {\bibfnamefont {M.~I.}\ \bibnamefont
  {Konchatnyi}}, \bibinfo {author} {\bibfnamefont {S.}~\bibnamefont {Pan}}, \
  and\ \bibinfo {author} {\bibfnamefont {W.}~\bibnamefont {Yang}},\ }\href
  {\doibase 10.1142/S0218271822500365} {\bibfield  {journal} {\bibinfo
  {journal} {Int. J. Mod. Phys. D}\ }\textbf {\bibinfo {volume} {31}},\
  \bibinfo {pages} {2250036} (\bibinfo {year} {2022})},\ \Eprint
  {http://arxiv.org/abs/2008.09602} {arXiv:2008.09602 [gr-qc]} \BibitemShut
  {NoStop}%
\bibitem [{\citenamefont {Escobal}\ \emph {et~al.}(2023)\citenamefont
  {Escobal}, \citenamefont {Jesus}, \citenamefont {Pereira},\ and\
  \citenamefont {Lima}}]{Escobal:2023saa}%
  \BibitemOpen
  \bibfield  {author} {\bibinfo {author} {\bibfnamefont {A.~A.}\ \bibnamefont
  {Escobal}}, \bibinfo {author} {\bibfnamefont {J.~F.}\ \bibnamefont {Jesus}},
  \bibinfo {author} {\bibfnamefont {S.~H.}\ \bibnamefont {Pereira}}, \ and\
  \bibinfo {author} {\bibfnamefont {J.~A.~S.}\ \bibnamefont {Lima}},\
  }\href@noop {} {\  (\bibinfo {year} {2023})},\ \Eprint
  {http://arxiv.org/abs/2302.01946} {arXiv:2302.01946 [astro-ph.CO]}
  \BibitemShut {NoStop}%
\bibitem [{\citenamefont {Mamon}(2018)}]{Mamon}%
  \BibitemOpen
  \bibfield  {author} {\bibinfo {author} {\bibfnamefont {A.}~\bibnamefont
  {Mamon}},\ }\href {\doibase doi.org/10.1142/S0217732318501134} {\bibfield
  {journal} {\bibinfo  {journal} {Mod. Phys. Lett. A}\ }\textbf {\bibinfo
  {volume} {33}} (\bibinfo {year} {2018}),\
  doi.org/10.1142/S0217732318501134}\BibitemShut {NoStop}%
\bibitem [{\citenamefont {Riess}\ \emph {et~al.}(2004)\citenamefont {Riess},
  \citenamefont {Strolger}, \citenamefont {Tonry}, \citenamefont {Casertano},
  \citenamefont {Ferguson}, \citenamefont {Mobasher}, \citenamefont {Challis},
  \citenamefont {Filippenko}, \citenamefont {Jha}, \citenamefont {Li},
  \citenamefont {Chornock}, \citenamefont {Kirshner}, \citenamefont
  {Leibundgut}, \citenamefont {Dickinson}, \citenamefont {Livio}, \citenamefont
  {Giavalisco}, \citenamefont {Steidel}, \citenamefont {Benitez},\ and\
  \citenamefont {Tsvetanov}}]{Riess_2004}%
  \BibitemOpen
  \bibfield  {author} {\bibinfo {author} {\bibfnamefont {A.~G.}\ \bibnamefont
  {Riess}}, \bibinfo {author} {\bibfnamefont {L.-G.}\ \bibnamefont {Strolger}},
  \bibinfo {author} {\bibfnamefont {J.}~\bibnamefont {Tonry}}, \bibinfo
  {author} {\bibfnamefont {S.}~\bibnamefont {Casertano}}, \bibinfo {author}
  {\bibfnamefont {H.~C.}\ \bibnamefont {Ferguson}}, \bibinfo {author}
  {\bibfnamefont {B.}~\bibnamefont {Mobasher}}, \bibinfo {author}
  {\bibfnamefont {P.}~\bibnamefont {Challis}}, \bibinfo {author} {\bibfnamefont
  {A.~V.}\ \bibnamefont {Filippenko}}, \bibinfo {author} {\bibfnamefont
  {S.}~\bibnamefont {Jha}}, \bibinfo {author} {\bibfnamefont {W.}~\bibnamefont
  {Li}}, \bibinfo {author} {\bibfnamefont {R.}~\bibnamefont {Chornock}},
  \bibinfo {author} {\bibfnamefont {R.~P.}\ \bibnamefont {Kirshner}}, \bibinfo
  {author} {\bibfnamefont {B.}~\bibnamefont {Leibundgut}}, \bibinfo {author}
  {\bibfnamefont {M.}~\bibnamefont {Dickinson}}, \bibinfo {author}
  {\bibfnamefont {M.}~\bibnamefont {Livio}}, \bibinfo {author} {\bibfnamefont
  {M.}~\bibnamefont {Giavalisco}}, \bibinfo {author} {\bibfnamefont {C.~C.}\
  \bibnamefont {Steidel}}, \bibinfo {author} {\bibfnamefont {T.}~\bibnamefont
  {Benitez}}, \ and\ \bibinfo {author} {\bibfnamefont {Z.}~\bibnamefont
  {Tsvetanov}},\ }\href {\doibase 10.1086/383612} {\bibfield  {journal}
  {\bibinfo  {journal} {The Astrophysical Journal}\ }\textbf {\bibinfo {volume}
  {607}},\ \bibinfo {pages} {665} (\bibinfo {year} {2004})}\BibitemShut
  {NoStop}%
\bibitem [{\citenamefont {Visser}(2004)}]{Visser_2004}%
  \BibitemOpen
  \bibfield  {author} {\bibinfo {author} {\bibfnamefont {M.}~\bibnamefont
  {Visser}},\ }\href {\doibase 10.1088/0264-9381/21/11/006} {\bibfield
  {journal} {\bibinfo  {journal} {Classical and Quantum Gravity}\ }\textbf
  {\bibinfo {volume} {21}},\ \bibinfo {pages} {2603} (\bibinfo {year}
  {2004})}\BibitemShut {NoStop}%
\bibitem [{\citenamefont {Valcin}\ \emph {et~al.}(2021)\citenamefont {Valcin},
  \citenamefont {Jimenez}, \citenamefont {Verde}, \citenamefont {Bernal},\ and\
  \citenamefont {Wandelt}}]{Valcin:2021}%
  \BibitemOpen
  \bibfield  {author} {\bibinfo {author} {\bibfnamefont {D.}~\bibnamefont
  {Valcin}}, \bibinfo {author} {\bibfnamefont {R.}~\bibnamefont {Jimenez}},
  \bibinfo {author} {\bibfnamefont {L.}~\bibnamefont {Verde}}, \bibinfo
  {author} {\bibfnamefont {J.~L.}\ \bibnamefont {Bernal}}, \ and\ \bibinfo
  {author} {\bibfnamefont {B.~D.}\ \bibnamefont {Wandelt}},\ }\href {\doibase
  10.1088/1475-7516/2021/08/017} {\bibfield  {journal} {\bibinfo  {journal}
  {Journal of Cosmology and Astroparticle Physics}\ }\textbf {\bibinfo {volume}
  {2021}},\ \bibinfo {pages} {017} (\bibinfo {year} {2021})}\BibitemShut
  {NoStop}%
\bibitem [{\citenamefont {Valcin}\ \emph {et~al.}(2020)\citenamefont {Valcin},
  \citenamefont {Bernal}, \citenamefont {Jimenez}, \citenamefont {Verde},\ and\
  \citenamefont {Wandelt}}]{Valcin_2020}%
  \BibitemOpen
  \bibfield  {author} {\bibinfo {author} {\bibfnamefont {D.}~\bibnamefont
  {Valcin}}, \bibinfo {author} {\bibfnamefont {J.~L.}\ \bibnamefont {Bernal}},
  \bibinfo {author} {\bibfnamefont {R.}~\bibnamefont {Jimenez}}, \bibinfo
  {author} {\bibfnamefont {L.}~\bibnamefont {Verde}}, \ and\ \bibinfo {author}
  {\bibfnamefont {B.~D.}\ \bibnamefont {Wandelt}},\ }\href {\doibase
  10.1088/1475-7516/2020/12/002} {\bibfield  {journal} {\bibinfo  {journal}
  {Journal of Cosmology and Astroparticle Physics}\ }\textbf {\bibinfo {volume}
  {2020}},\ \bibinfo {pages} {002} (\bibinfo {year} {2020})}\BibitemShut
  {NoStop}%
\bibitem [{\citenamefont {Bernal}\ \emph {et~al.}(2021)\citenamefont {Bernal},
  \citenamefont {Verde}, \citenamefont {Jimenez}, \citenamefont {Kamionkowski},
  \citenamefont {Valcin},\ and\ \citenamefont {Wandelt}}]{Bernal:2021}%
  \BibitemOpen
  \bibfield  {author} {\bibinfo {author} {\bibfnamefont {J.~L.}\ \bibnamefont
  {Bernal}}, \bibinfo {author} {\bibfnamefont {L.}~\bibnamefont {Verde}},
  \bibinfo {author} {\bibfnamefont {R.}~\bibnamefont {Jimenez}}, \bibinfo
  {author} {\bibfnamefont {M.}~\bibnamefont {Kamionkowski}}, \bibinfo {author}
  {\bibfnamefont {D.}~\bibnamefont {Valcin}}, \ and\ \bibinfo {author}
  {\bibfnamefont {B.~D.}\ \bibnamefont {Wandelt}},\ }\href {\doibase
  10.1103/physrevd.103.103533} {\bibfield  {journal} {\bibinfo  {journal}
  {Physical Review D}\ }\textbf {\bibinfo {volume} {103}} (\bibinfo {year}
  {2021}),\ 10.1103/physrevd.103.103533}\BibitemShut {NoStop}%
\bibitem [{\citenamefont {Krishnan}\ \emph {et~al.}(2021)\citenamefont
  {Krishnan}, \citenamefont {Ó~Colgáin}, \citenamefont {Sheikh-Jabbari},\
  and\ \citenamefont {Yang}}]{H0diagnostic:2021}%
  \BibitemOpen
  \bibfield  {author} {\bibinfo {author} {\bibfnamefont {C.}~\bibnamefont
  {Krishnan}}, \bibinfo {author} {\bibfnamefont {E.}~\bibnamefont
  {Ó~Colgáin}}, \bibinfo {author} {\bibfnamefont {M.}~\bibnamefont
  {Sheikh-Jabbari}}, \ and\ \bibinfo {author} {\bibfnamefont {T.}~\bibnamefont
  {Yang}},\ }\href {\doibase 10.1103/physrevd.103.103509} {\bibfield  {journal}
  {\bibinfo  {journal} {Phys. Rev. D}\ }\textbf {\bibinfo {volume} {103}}
  (\bibinfo {year} {2021}),\ 10.1103/physrevd.103.103509}\BibitemShut {NoStop}%
\bibitem [{\citenamefont {Di~Valentino}\ \emph {et~al.}(2018)\citenamefont
  {Di~Valentino}, \citenamefont {Linder},\ and\ \citenamefont
  {Melchiorri}}]{Valentino:2008tsh0}%
  \BibitemOpen
  \bibfield  {author} {\bibinfo {author} {\bibfnamefont {E.}~\bibnamefont
  {Di~Valentino}}, \bibinfo {author} {\bibfnamefont {E.~V.}\ \bibnamefont
  {Linder}}, \ and\ \bibinfo {author} {\bibfnamefont {A.}~\bibnamefont
  {Melchiorri}},\ }\href {\doibase 10.1103/PhysRevD.97.043528} {\bibfield
  {journal} {\bibinfo  {journal} {Phys. Rev. D}\ }\textbf {\bibinfo {volume}
  {97}},\ \bibinfo {pages} {043528} (\bibinfo {year} {2018})}\BibitemShut
  {NoStop}%
\end{thebibliography}%

\end{document}